\documentclass[useAMS,usenatbib,usegraphicx]{mn2e}

\title[The long term X--ray spectral variability of AGN]{The long term X--ray spectral variability of AGN}
\author[M. A. Sobolewska and I. E. Papadakis]{M. A.
Sobolewska$^{1,2}$\thanks{E-mail: malgosia@physics.uoc.gr} and I. E.
Papadakis$^{1,2}$\\
$^1$Foundation for Research and Technology - Hellas, IESL, Voutes, 71110
Heraklion, Crete, Greece\\
$^2$University of Crete, Department of Physics, Voutes, 71003 Heraklion, Crete,
Greece}

\newcommand{\apj}{ApJ}
\newcommand{\apjl}{ApJL}
\newcommand{\mnras}{MNRAS}
\newcommand{\aap}{A\&A}

\newcommand{\pasj}{PASJ}
\newcommand{\nat}{Nature}
\newcommand{\chjaa}{CHJAA}
\newcommand{\aj}{AJ}

\newcommand{\aarv}{A\&ARv}
\newcommand{\procspie}{Proc. SPIE}
\newcommand{\rxte}{{\it RXTE}}

\begin{document}

\date{}

\pagerange{\pageref{firstpage}--\pageref{lastpage}} \pubyear{2009}

\maketitle

\label{firstpage}

\begin{abstract}

We present the results from the spectral analysis of more than 7,500 {\it RXTE}
spectra of 10 AGN. Our main goal was to study their long term X--ray spectral
variability. The sources in the sample are nearby, X-ray bright, and they have
been observed by {\it RXTE} regularly over a long period of time ($\sim$7--11
years). High frequency breaks have been detected  in their power spectra, and
these  characteristic frequencies imply time scales of the order of a few days
or weeks. Consequently, the {\it RXTE} observations we have used most probably
sample most of the flux and spectral variations that these objects exhibit.
Thus, the {\it RXTE} data are ideal for our purpose. Fits to the individual
spectra were performed in the 3--20 keV energy band. We modeled the data in a
uniform way using simple phenomenological models (a power-law with the addition
of Gaussian line and/or edge to model the iron K$\alpha$ emission/absorption
features, if needed) to consistently parametrize the shape of the observed
X--ray {\it continuum} of the sources in the sample. We found that the average
spectral slope does not correlate with source luminosity or black hole mass,
while it correlates positively with the average accretion rate. We have also
determined the (positive) ``spectral slope -- flux" relation for each object,
over a flux range larger than before. We found that this correlation is  {\it
similar} in all objects, except for NGC~5548  which displays limited spectral
variations for its flux variability. We discuss this global  ``spectral slope --
flux'' trend in the light of current models for spectral variability. We
consider (i) intrinsic variability, expected e.g. from Comptonization processes,
(ii) variability caused by absorption of X-rays by a single absorber whose
ionization parameter varies proportionally to the continuum flux variations,
(iii) variability resulting from the superposition of  a constant reflection
component and  an intrinsic power-law which is variable in flux but constant in
shape,  and, (iv) variability resulting from the superposition of a constant
reflection component and an intrinsic power-law which is  variable both in flux
and shape. Our final conclusion is that scenario (iv) provides the best fit to
the data of all objects, except for NGC~5548. 

\end{abstract}


\section{Introduction}

The current paradigm for the central source in AGN postulates a black hole (BH)
with a mass of $10^6-10^9$ M$_{\odot}$, and a geometrically thin, optically
thick accretion disc that presumably extends to the innermost stable circular
orbit around the black hole. This disc  is supposed to be  responsible for the
broad, quasi-thermal emission component in the optical-UV spectrum of AGN (the
so-called  Big Blue Bump).  At energies above $\sim$2 keV, a power-law like
component is observed. This is attributed to emission by a hot corona ($T \sim
10^8 - 10^9$ K) overlying the thin disc. The corona up-Comptonizes the disc soft
photons to produce the hard ($E \sim 2 - 100$ keV) X-ray emission. An   Fe
K$\alpha$ line is also detected, produced either by fluorescence on cold matter
at $E = 6.4$ keV or by photoionization of H-like or He-like Fe at 6.9 keV and
6.7 keV, respectively. This feature is believed to be emitted on the surface of
the accretion disc, from the reprocessing of the X-rays. In many cases the line
is resolved, and its width can, in theory, constrain the position of the inner
radius of the disc and the spin of the central BH.

The AGN X--ray emission is strongly variable, even on time scales as short as a
few hundred seconds, both in flux and in the shape of the observed energy
spectrum. These variations can provide information on the  physical conditions,
the size, and the geometry of the X--ray emitting region. Regarding the flux
variability, although the broad band power spectra of AGN are not well
understood, the identification of characteristic time scales in them, combined
with the large range in the masses of the BHs that have been studied so far, can
constrain (to some degree) the size and the geometry of the X--ray source
\citep[e.g.][]{mchardyea:2006}. On the other hand, recent detailed studies of the
spectral variations observed in a few objects provide valuable details of the
physical conditions of the X--ray reprocessing \citep[see e.g.]{miniuttiea:2007}
and/or on the physical properties of absorbing material which may exist close to
the central source \citep*[see e.g.][]{millerea:2008}.

Progress in settling alternative interpretations of the observed spectral
variability can be made by correlating the spectral and timing properties in
AGN. Along this line, in a recent paper \citep[][]{papadakisea:2009} we studied 
11 nearby, X--ray bright Seyferts, which have been intensively observed over the
last two decades by the {\it Rossi X-ray Timing Explorer} ({\it RXTE}). We
detected a positive correlation between their mean spectral slope and the
characteristic break frequency, $\nu_{\rm br}$, at which the power spectrum
changes its slope from -1 to -2. To the extent that this frequency corresponds
to a (still unknown) characteristic  time scale of the X--ray source, this
correlation suggests that the mean X--ray spectral slope in AGN is determined by
a fundamental parameter of the system, most probably its accretion rate. If
true, this result has important implications on current Comptonization models. 

In this work we use the results from the model fitting of thousands of {\it
RXTE} spectra of 10 AGN that were included in the sample of
\citet[][]{papadakisea:2009}. Our prime goal is to study in detail the X--ray
spectral variability of each object. Although the spectral resolution of {\it
RXTE} is not as good as that of {\it Chandra}, {\it XMM-Newton} or {\it Suzaku},
the {\it RXTE} data sets that we used are ideal for our purpose. First, the
sources in the sample are nearby and X-ray bright. As a result, their X--ray
spectral shape can be accurately determined even from relatively short
($\sim$1--2 ksec) {\it RXTE} observations. Second, each source in the sample was
observed regularly (more than 200--300 times, and in a few cases more than a
thousand times) over a long  period of time (of the order of 7--11 years). This
is contrary to typical {\it Chandra} and {\it XMM-Newton} observations of
individual AGN, which usually span a period of a few days (maximum). Third,  the
characteristic frequencies of the objects imply time scales of the order of a
few days or weeks. Consequently, the {\it RXTE} observations we have used most
probably sample most of the flux and spectral variations that they exhibit.

A subsample of the observations we study have already  been studied by
\citet[][]{papadakisea:2002}. These authors used hardness ratios in various
energy bands to study the spectral variability of four objects, namely NGC~5548,
NGC~5506, NGC~4051 and MCG~-6-30-15. In the present work we studied many more
observations of these objects and we have also analysed data for six more
Seyferts.  Furthermore, instead of computing hardness ratios, we fitted each
spectrum with the same phenomenological model of a  ``power law plus a Gaussian
line"  to derive the observed photon index, $\Gamma_{\rm obs}$. In this way, we 
parametrized all available 3--20 keV {\it RXTE} spectra of the sources in a
uniform way. We then used the resulting best model fitting values to: i) 
determine the mean spectral slope for each object and investigate its
correlation with the main source parameters such as BH mass, accretion rate and
luminosity, and ii)  construct their long term ``spectral slope vs. flux"
relations. For the reasons explained above, we believe that we  determined the
flux related spectral variability in these sources more accurately than before. 

Several possible reasons for the observed X--ray spectral variability in AGN 
have been proposed in the past. The simplest possibility is that the
$\Gamma_{\rm obs}$ variations correspond to intrinsic variations in continuum
slope   \citep*[e.g.][]{haardtea:1997,coppi:1999,beloborodov:1999}. For
example, \citet{nandrapapadakis:2001} have showed that  the X-ray spectral changes in
NGC~7469 are intrinsic and  respond to the UV-soft photon input variations,
exactly as one would expect in the case of thermal Comptonization models.
However, there has also been  certain observational evidence that the
$\Gamma_{\rm obs}$ variations originate as a result of superposition of
different spectral components with different variability properties. One such
possibility is a spectrum composed of a constant reflection component and a
variable, in normalization only, power-law like continuum of constant slope
\citep*[e.g.][and references
therein]{taylorea:2003,pontiea:2006,miniuttiea:2007}. Another  possibility is a
constant slope power law which varies in flux and complex variations in an
absorber, e.g. its ionization state and/or the covering factor of the source
\citep[see recent review by][]{turnermiller:2009}. 

One of our main result is that the X--ray spectral slope steepens with
increasing flux in a {\it similar} way for all the objects in the sample. This
similarity implies that, although all the above mentioned mechanisms may 
operate in AGN, it is just one of them which is mainly responsible for the
observed spectral variability in AGN. 

The paper is organized as follows. In Section 2, we discuss the data selection
and reduction, and we also describe the  phenomenological models that we used to
parametrize the shape of AGN continuum. In Section 3, we report the spectral
fitting results, and the results from the spectral variability study.  In
Section 4, we discuss our findings and compare them with predictions of several
spectral variability models. We give our conclusions in Section 5.

\section{Data Reduction and Analysis}


\begin{table*}
\caption{Summary of the {\it RXTE} observations and properties of the AGN in the
sample. In column 2, $N$ is the total number of observations considered for a given AGN.
Dates of the first and last observation are listed in column 3. The luminosity
distances, $D$, and redshifts, $z$, listed in columns 4 and 5, were
taken form the NED database. In column 6 we list black hole mass
estimates.}

\centering
\begin{tabular}{l l c c c c}
\hline
Source      & $N$ & Start/End & Lum. Distance & Redshift & $M_{\rm BH}$\\
 & & & {\it D} / Mpc & {\it z} & $\times 10^6$ M$_{\odot}$\\
(1) & (2) & (3) & (4) & (5) & (6) \\
\hline
Fairall 9   & 671  & 1996-11-03/2003-03-01 & 199  & 0.047 & 255 \\
Akn 564     & 505  & 1996-12-23/2003-03-04 & 98.6 & 0.025 & 1.9 \\
Mrk 766     & 219  & 1997-03-05/2006-11-07 & 57.7 & 0.013 & 3.6 \\
NGC 4051    & 1257 & 1996-04-23/2006-10-01 & 12.7 & 0.002 & 1.9\\
NGC 3227    & 1021 & 1996-11-19/2005-12-04 & 20.4 & 0.004 & 42.2\\
NGC 5548    & 866  & 1996-05-05/2006-11-06 & 74.5 & 0.017 & 67.1\\
NGC 3783    & 874  & 1996-01-31/2006-11-08 & 44.7 & 0.010 & 29.8\\
NGC 5506    & 627  & 1996-03-17/2004-08-08 & 29.1 & 0.006 & 28.8 \\
MCG-6-30-15 & 1214 & 1996-03-17/2006-12-24 & 35.8 & 0.008 & 6.8 \\
NGC 3516    & 250  & 1997-03-16/2006-10-13 & 37.5 & 0.009 & 42.7\\
\hline
\end{tabular}
\label{tab:tab1}
\end{table*}

\begin{table*}
\caption{Quality of the various model fits to the data. The number of model-fit
``successes", $n_s$, and global propabilities, P$_{c}$ (see Sec. 2.2 for details),
are listed for various models.}
\centering
\begin{tabular}{l c c c c c c c c c c c}
\hline
            & \multicolumn{2}{c}{PL} & \multicolumn{2}{c}{PLG} & \multicolumn{2}{c}{ePLG$^a$} & \multicolumn{2}{c}{ePLG$^b$} & \multicolumn{2}{c}{ePLG$^c$} \\
Source      &  $n_s$ & P$_c$ &  $n_s$ & P$_c$ & $n_s$ & P$_c$ & $n_s$ & P$_c$ & $n_s$ & P$_c$ \\
(1) & (2) & (3) & (4) & (5) & (6) & (7) & (8) & (9) & (10) & (11)\\
\hline
Fairall 9   & 642 	& 0.81       		&            	&              	&               &               &            &            &   &  \\
Akn 564     & 486 	& 0.92       		&            	&              	&               &               &            &            &   &  \\
Mrk 766     & 208 	& 0.54       		&           	&              	&               &               &            &            &   &  \\
NGC 4051    & 1167      & $5\times10^{-4}$ 	& 1199 		& 0.75   	&               &               &            &            &   &  \\
NGC 3227    & 919       & $<10^{-5}$       	& 980  		& 0.94   	&               &               &            &            &   &  \\
NGC 5548    & 802       & $1\times10^{-3}$ 	& 827  		& 0.77   	&               &               &            &            &   &  \\
NGC 3783    & 635       & $<10^{-5}$       	& 799        	& $<10^{-5}$    & 821    	& 0.09     	&            &            &   &  \\
NGC 5506    & 382       & $<10^{-5}$       	& 556$^d$    	& $<10^{-5,d}$ 	& 586$^d$ 	& 0.05$^d$ 	&            &            &   &  \\
MCG-6-30-15 & 925       & $<10^{-5}$       	& 1076       	& $<10^{-5}$   	& 1136          & 0.02          & 1147 	     & 0.22 	  &   &  \\
NGC 3516    & 160       & $<10^{-5}$       	& 221        	& $<10^{-5}$   	& 227           & 0.004         & 227        & 0.004      & 231  & 0.05  \\
\hline
\end{tabular}\\
\label{tab:tab2}
\begin{minipage}[b]{0.75\linewidth}
$^a$ the edge energy fixed at 7.1 keV\\
$^b$ the edge energy allowed to vary\\
$^c$ both the line and edge energies allowed to vary\\
$^d$ PLG or ePLG model plus cold absorption with $N_{\rm H}=3.7\times10^{22}$~cm$^{-2}$
\end{minipage}
\end{table*}


\subsection{Data selection and reduction}

In Tab.~\ref{tab:tab1}, we list the 10 AGN in our sample, some of their properties
(distance, redshift and BH mass estimates), and details of the \rxte\ observations we
used in this work. We used the \citet[][]{petersonea:2004} estimates, from
reverbaration mapping, for the BH mass (M$_{\rm BH})$ in Fairall 9, NGC~4051, NGC~3227,
NGC~5548, NGC~3783, and NGC~3516. In the case of Mrk 766 and MCG-6-30-15 we used the
stellar velocity dispecrsion measurements of \citet[][]{botteea:2004} and
\citet[][]{mchardyea:2005}, respectively, and the \citet[][]{tremaineea:2002} $M_{\rm
BH}-\sigma$ relation to estimate BH mass. The BH mass estimates for Ark 564 and
NGC~5506 were taken from \citet[][]{zhangwang:2006}  and \citet[][]{wangzhang:2007},
respectively. They were also based on the \citet[][]{tremaineea:2002} M$_{\rm
BH}-\sigma$ relation, although in this case the stellar velocity dispecrsion was
obtained from the FWHM of the O$_{\rm III}$ line, using the FWHM O$_{\rm III}-\sigma$
relation of \citet[][]{greeneho:2005}. Since the reverbaration mapping estimates have
been obtained by ``forcing" the AGN $M_{\rm BH}-\sigma$ relationship to have the same
normalization as the  $M_{\rm BH}-\sigma$ relationship for quiescent galaxies
\citep[][]{petersonea:2004}, we belive that the usege of two different M$_{\rm BH}$
determination methods should not add significant scatter to the correlations we present
below. 

For each object we considered all the \rxte\ observations that were performed
until the end of 2006. Some of these objects were monitored with {\it RXTE} for
as long as 10 years. The reason we chose to study these AGN is precisely because
they have been observed frequently by {\it RXTE}. They are all nearby, X--ray
bright, Type I objects (except for NGC~5506, which is optically classified as
Type II Seyfert).  Their central BH mass ranges from $\sim$2$\times 10^6$
M$_{\odot}$ to $\sim$2$\times 10^8$ M$_{\odot}$, and their accretion rate from a
few percent of the Eddington limit (in NGC~3227) to almost close to this limit
(in Ark~564; see e.g. M$^{\rm c}$Hardy et al. 2006). Consequently, they could be
considered as representative of the Seyfert population in the local Universe.

We used data from the Proportional Counter Array \citep[PCA;][]{jahodaea:1996}.
The typical duration of each observation was $\sim$1--2 ksec. The data were
reduced using {\tt FTOOLS} v.6.3. The PCA data were screened according to the
following criteria: the satellite was out of the South Atlantic Anomaly (SAA)
for at least 30 min, the Earth elevation angle was $\geq 10^{\circ}$, the offset
from the nominal optical position was $\leq 0^{\circ}\!\!.02$, and the parameter
ELECTRON-2 was $\leq 0.1$.

We extracted STANDARD-2 mode, 3--20 keV, Layer 1, energy spectra from PCU2 only.
For background subtraction we used the appropriate background model for the faint
objects\footnote{\tt pca\_bkgd\_cmfaintl7\_eMv20051128.mdl} available from {\it
RXTE} data analysis web-pages. The background subtracted spectra were rebinned using {\tt grppha} so that each bin contained more
than 15 photons for the $\chi^2$ statistic to be valid.  We considered only
these data sets for which binning resulted in at least 15 PHA channels. We used PCA response matrices and effective area
curves created specifically for the individual observations by {\tt pcarsp}.

\subsection{Modeling the spectral shape of the AGN}

Fits to the individual spectra were performed in the energy range 3--20 keV, where the
PCA is most sensitive, using the {\tt XSPEC} v.11.3.2 software package
\citep[][]{arnaud:1996}. Our prime  aim was to parametrize in a uniform way (i.e. by
fiting the same model to all the observations) the shape of the observed X--ray {\it
continuum} of the sources in our sample. We first fitted a simple power-law model
(hereafter PL) to all the individual spectra of all sources. In the cases when the
global goodness of the PL fit was unacceptable, we added additional spectral
components.  The form of these components was chosen based on visual examination of the
PL fit residuals in the cases when the model did not fit the data well.

We did not consider the Galactic absorption effects as the Galactic column to all
sources is so low that cannot affect the spectrum above 3 keV. In the case of NGC~5506,
all models were modified by cold absorption using {\tt wabs} in {\tt XSPEC},  with the
hydrogen column density fixed at N$_{\rm H} = 3.7\times 10^{22}$ cm$^{-2}$
\citep[][]{perolaea:2002}.

We assumed that a model fits well the spectrum of an individual observation if
the probability of accepting the null hypothesis, $p_{\rm null}$, is larger than
0.05. Given that the number of observations for each AGN is quite large, it is
possible that $p_{\rm null}<0.05$ in some cases, even if the model is the
correct one. Therefore, the global goodness of a model fit to all the
spectra of each source had to be judged in a statistical way. One method to
achieve this is to consider the fit to the individual spectra as a ``success"
when $p_{\rm null}>0.05$, and as a ``failure" otherwise. Under the hypothesis
that the model is the correct one, the probability of a ``success" (``failure")
is $p_{\rm s} = 0.95$ (1-$p_{\rm s}$ = 0.05), by definition. In this approach,
the model fit to each spectrum can be considered as a Bernoulli trial.

For a given source and a given model we recorded the total number of spectra,
$N$,  the number of model-fit ``successes", $n_{\rm s}$, and the number of
model-fit ``failures", $n_{\rm f}$. Then, we used the binomial distribution to
estimate the probability, P$_{\rm c}$, that the number of ``failures"  will be
larger than $n_{\rm f}$ in $N$ trials by chance. This is equal to the
probability that the numbers of ``successes" will be smaller than $n_{\rm s}$.
Therefore, P$_{\rm c}$ can be estimated as

\begin{equation} P_{\rm c} =
\sum_{j=0}^{n_{\rm s}} {N \choose j } p_{\rm s}^{j} (1-p_{\rm s})^{N-j}.
\end{equation} 

\noindent We accepted that a model fits {\it globally} the spectra of a source
if $0.05<P_{\rm c}< 0.95$. We report below the results from the various model
fits  to the data of all sources. 

\subsubsection{``Power-law" model fits}

\begin{figure}
\includegraphics[height=8.3cm,bb=214 57 355 591,clip,angle=-90]{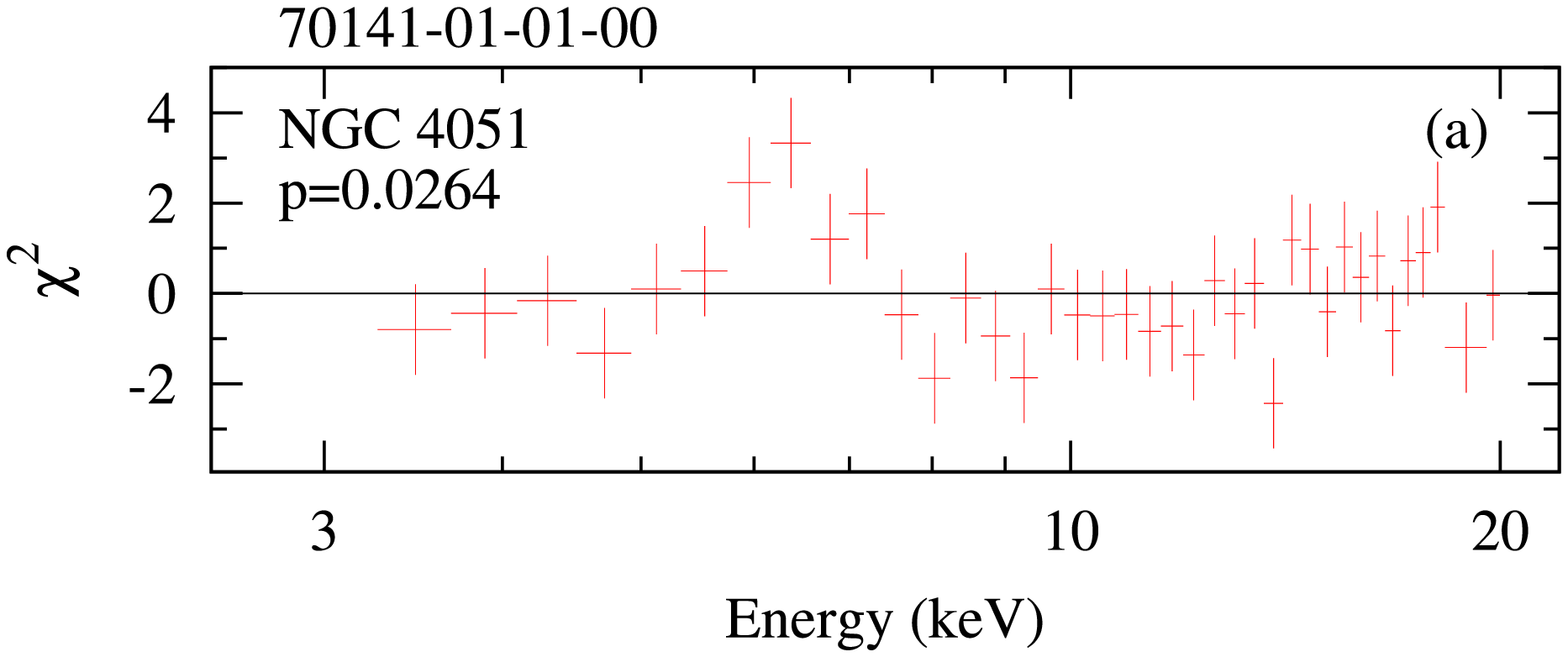}\\
\includegraphics[height=8.3cm,bb=214 57 355 591,clip,angle=-90]{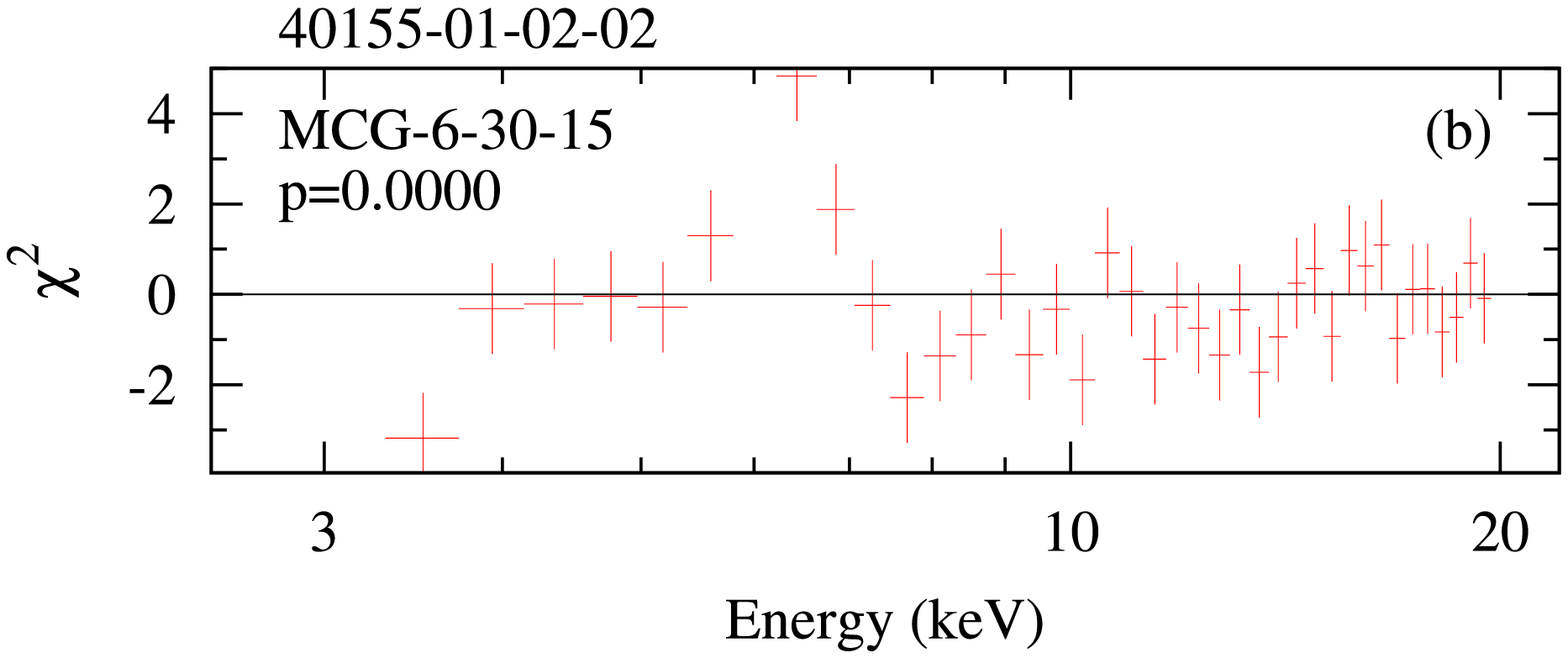}\\
\includegraphics[height=8.3cm,bb=214 57 355 591,clip,angle=-90]{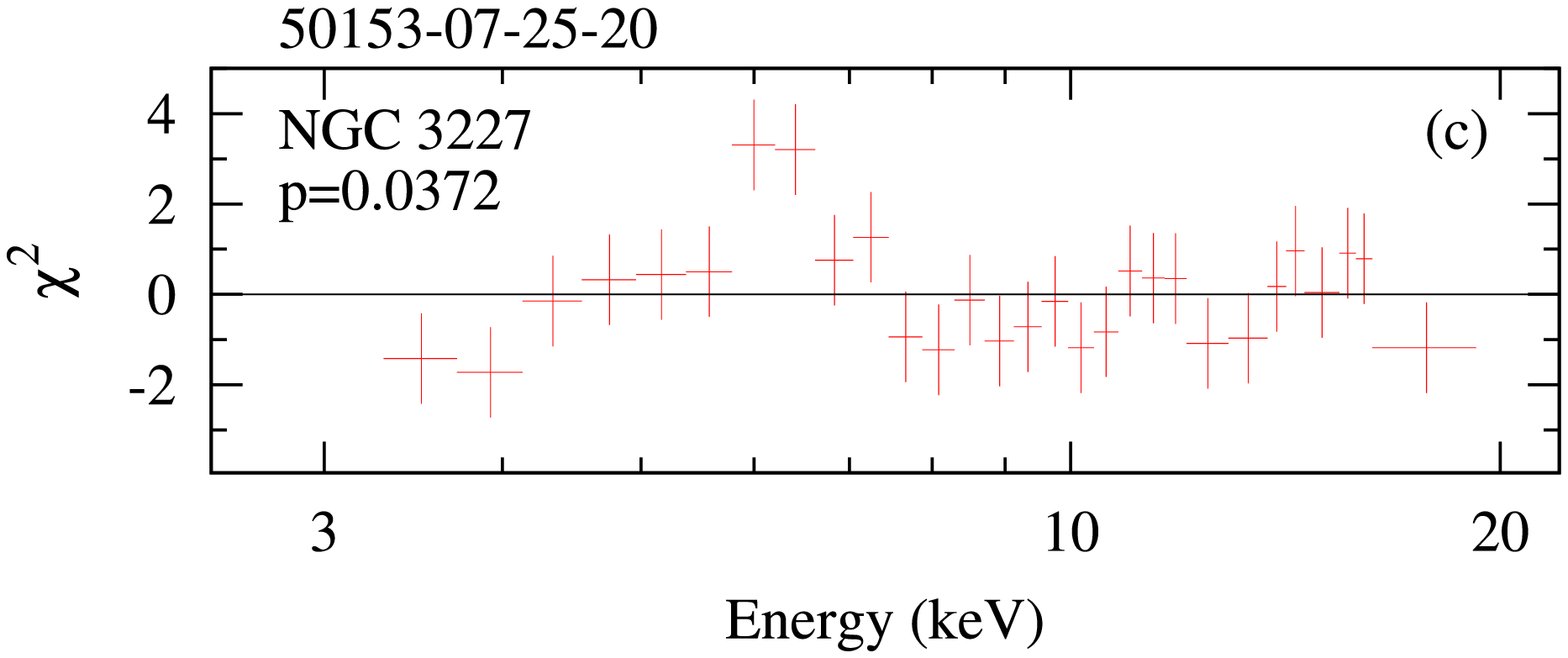}\\
\includegraphics[height=8.3cm,bb=214 57 355 591,clip,angle=-90]{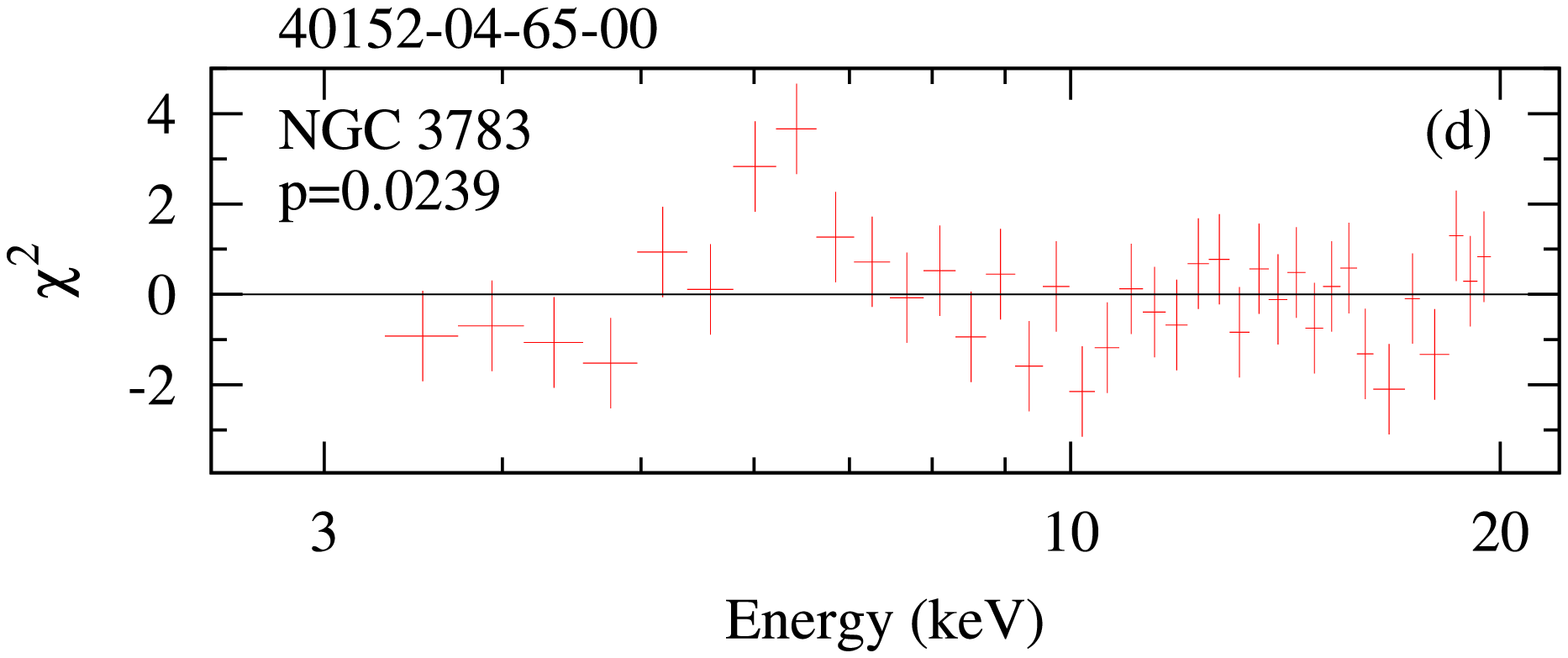}\\
\includegraphics[height=8.3cm,bb=214 57 355 591,clip,angle=-90]{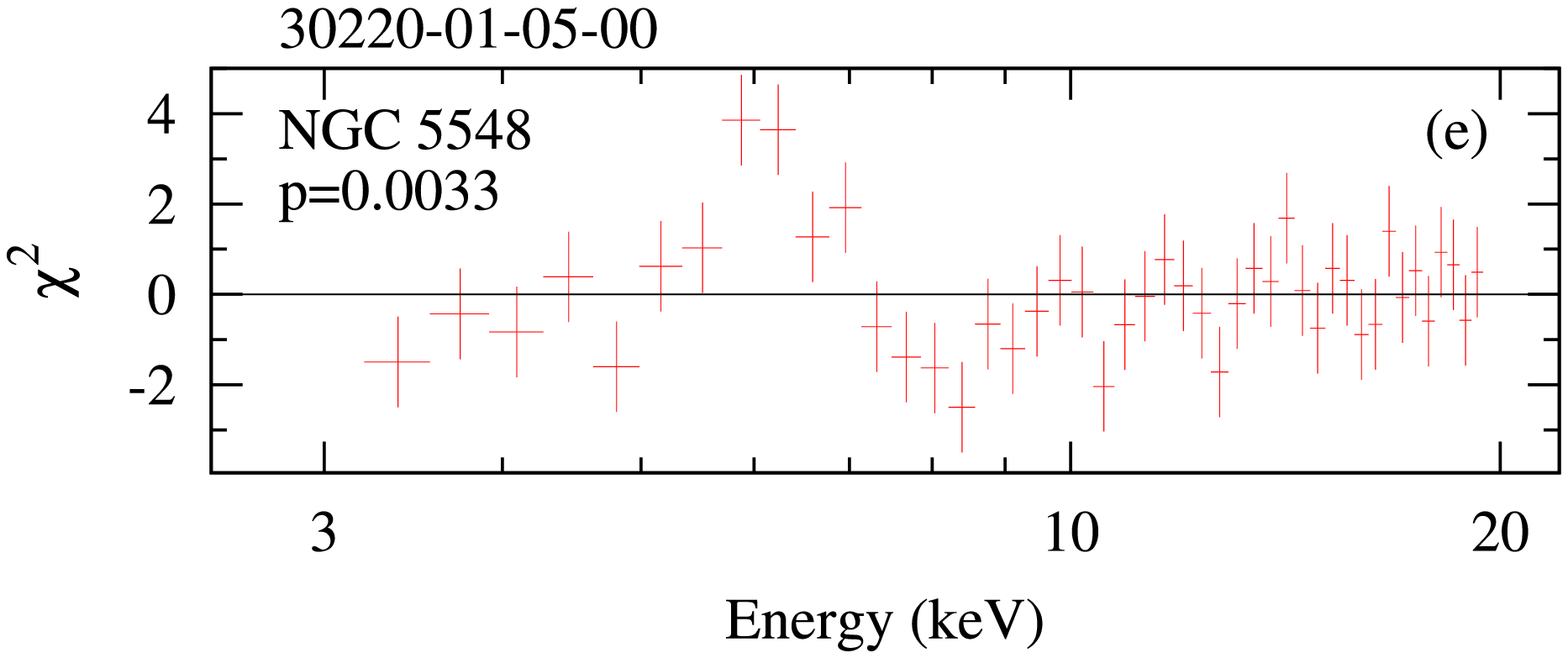}\\
\includegraphics[height=8.3cm,bb=214 57 355 591,clip,angle=-90]{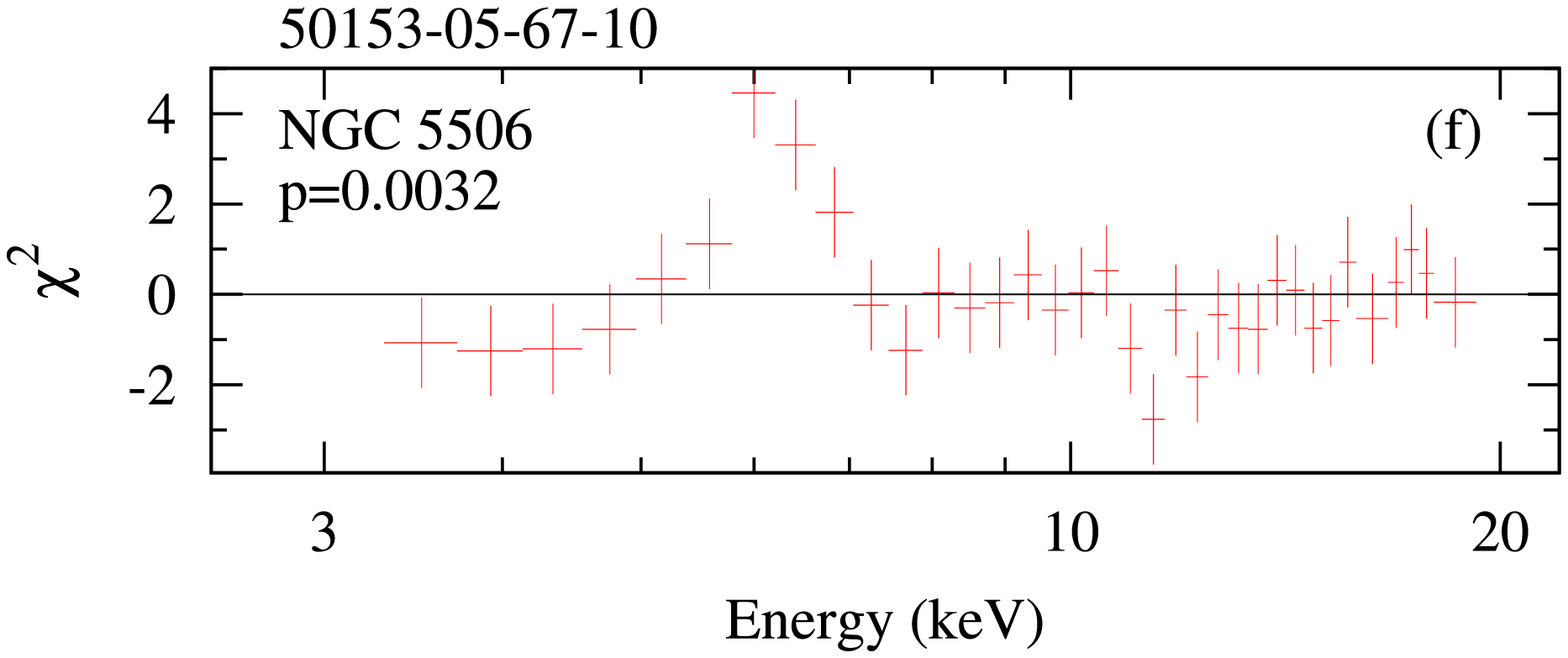}\\
\includegraphics[height=8.3cm,bb=214 57 420 591,clip,angle=-90]{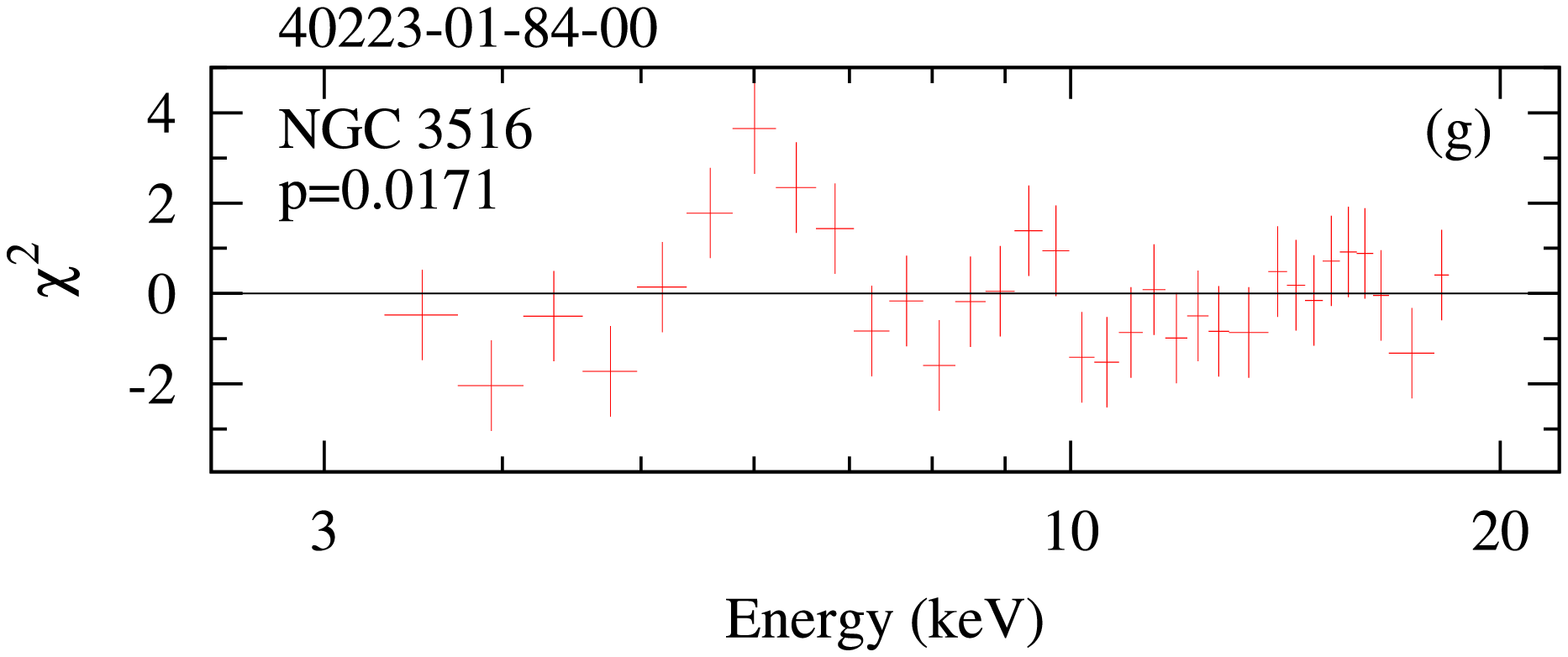}
\caption{Contribution to $\chi^2$ for the PL model fits to  the data of (a) NGC
4051 (obsid 70141-01-01-00), (b) MCG-6-30-15 (40155-01-02-02), (c) NGC 3227
(50153-07-25-20), (d) NGC3783 (40152-04-65-00), (e) NGC 5548 (30220-01-05-00),
(f) NGC 5506 (50153-05-67-10), and (g) NGC 3516 (40223-01-84-00). The fit
probabilities are indicated on each panel. Residuals indicative of an iron K$\alpha$ line are visible in the 5--7 keV band.}
\label{fig:fig1}
\end{figure}

The PL model fited well the spectra of only three sources: Fairall 9, Akn 564
and Mkn 766 ($n_{\rm s}$ and P$_{\rm c}$ values are listed in the second and
third columns of Tab.~\ref{tab:tab2}). In the remaining 7 objects we clearly
detected 6--7 keV residua indicative of an iron K$_{\alpha}$ line (see Fig.
\ref{fig:fig1}).

\subsubsection{``Power-law plus line" model fits}

Motivated by shape of the PL best model fit  residua around 6--7 keV, we added a
Gaussian component to account for the K$\alpha$ iron line in these spectra ({\tt
powerlaw+gaussian} in {\tt XSPEC}, hereafter PLG). In all cases, the iron line
energy and width were fixed at 6.4 keV and 0.1 keV, respectively. The goodness
of the new model fits improved and became acceptable in NGC 4051, NGC 3227 and
NGC 5548  ($n_{\rm s}$ and P$_{\rm c}$ values for the PLG model are listed in
columns 4 and 5  of Tab.~\ref{tab:tab2}). 

\subsubsection{``Power-law plus line plus edge" model fits }

Figure~\ref{fig:fig2} shows the best PLG model fit residua in the case of an
unacceptable model fit  to a spectrum of (a) NGC~5506, (b)  NGC~3783, (c) MCG~-6-30-15,
and (d--e) NGC~3516. The best-fit model residuals indicate the presence of an edge-like
feature around $\sim$7 keV. We therefore considered  a {\tt edge*(powerlaw+gauss)}
model in {\tt XSPEC} (ePLG hereafter). We kept the edge energy fixed at 7.1 keV (the
threshold energy for neutral iron) and we re-fitted the individual spectra of these
sources. 

The global fit became acceptable  in the case of NGC~5506 and NGC~3783. The optical
depth of the edge, $\tau$, was detected at the 2$\sigma$ significance level in 298 out
of 627 observations, in the case of NGC~5506, and in 319 out of 874 observations in
NGC~3783. The respective  $n_{\rm s}$ and P$_{\rm c}$ values listed in columns 6 and 7
of Tab.~\ref{tab:tab2}.

In MCG-6-30-15 the model fit improved significantly, and the model became acceptable,
when we allowed the energy of the edge to vary between 6 and 10 keV (results are listed
in columns 8 and 9 of Tab.~\ref{tab:tab2}). The edge was detected at the 2$\sigma$
significance level in 863 out of 1214 data sets. Using the best fit results for these
spectra only, we estimated that $\overline{E}_{\rm edge}\sim7.4$ keV, an energy that is
indicative of absorption by ionized iron. However, the standard deviation of the best
fit edge energy values is substantial ($\sigma \sim 1$ keV), which suggests that one
should treat this result with caution (see discussion in Section 2.2.5).

In the case of NGC 3516 we did not record an improvement to the global fit of
the ePLG model, even when the edge energy was allowed to vary (compare columns 6
and 7, and 8 and 9 of Tab.~\ref{tab:tab2}). Thus, we thawed the energy of the
iron line and allowed it to vary between 5.5 and 7.5 keV. This extra
modification provided a statistically acceptable global fit (the  $n_{\rm s}$
and P$_{\rm c}$ values listed in  columns 10 and 11 in Tab.~\ref{tab:tab2}
correspond to this model). The average line energy, and its standard deviation,
calculated for the data sets where the line is detected at the 2$\sigma$
significance level (117 of 250) were $6.1$ and 0.3 keV, respectively. The
absorption edge was detected at the 2$\sigma$ significance level in 122 out of
the 250 spectra of this source. Using the best fit results of these observations
only, we calculated that the average energy of the absorption edge, and its
standard deviation, were $\overline{E}_{\rm edge} = 7.9$ and $\sigma = 1.1$ keV,
respectively.

\subsubsection{Summary of the acceptable model fits}


\begin{figure}
\includegraphics[height=8.3cm,bb=214 57 355 591,clip,angle=-90]{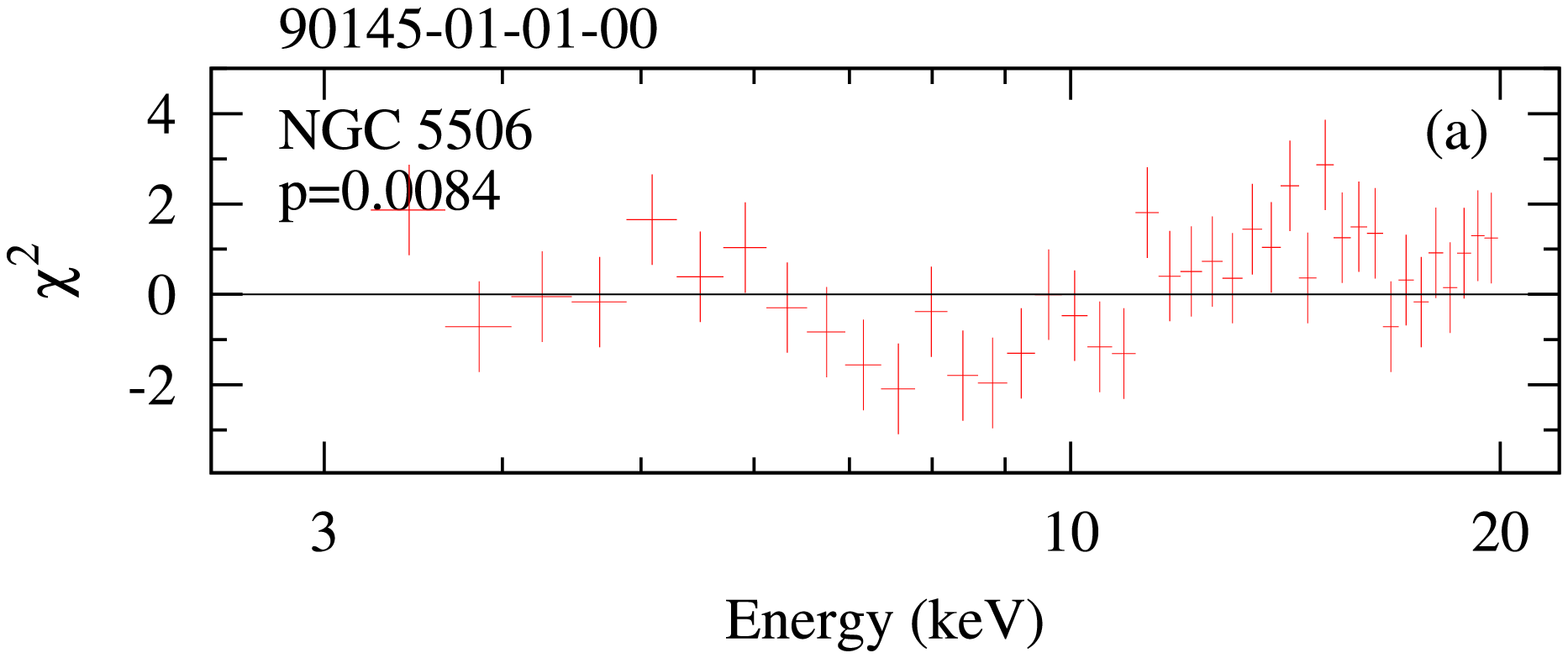}\\
\includegraphics[height=8.3cm,bb=214 57 355 591,clip,angle=-90]{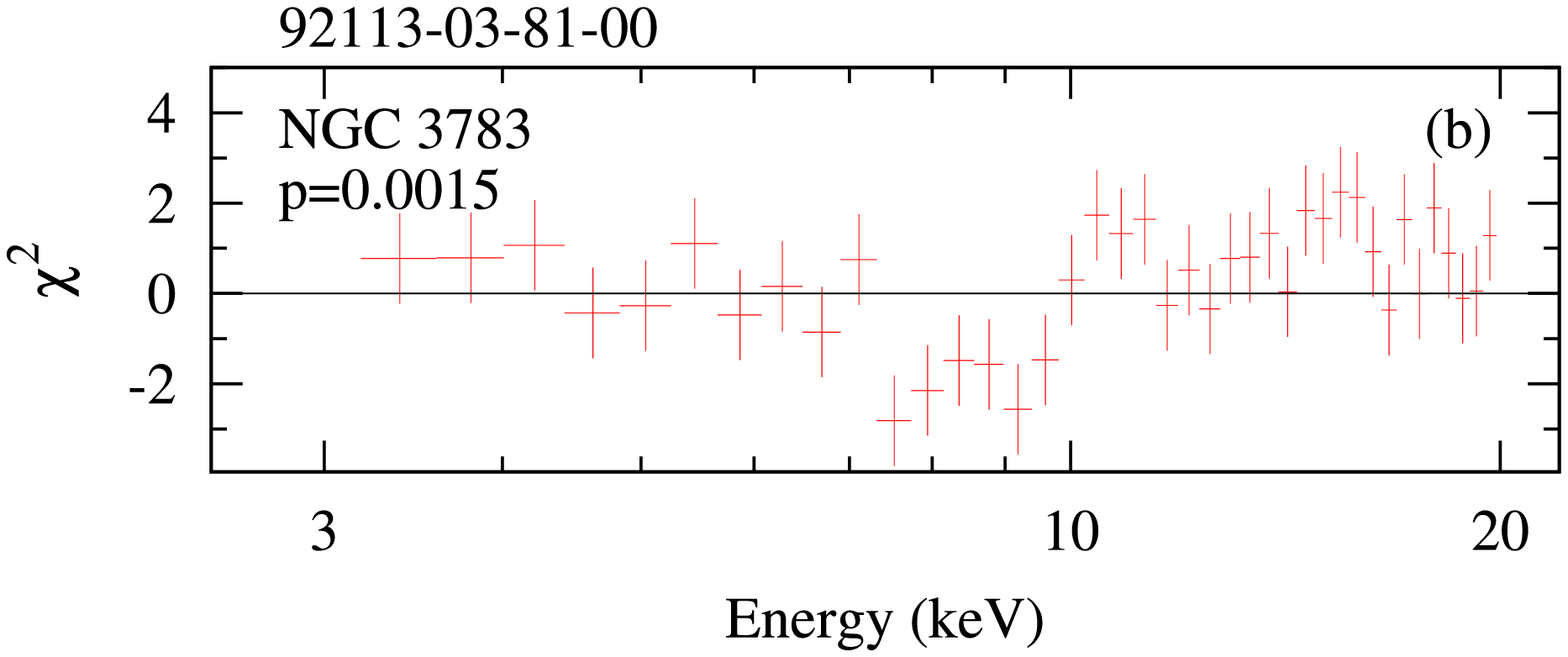}\\
\includegraphics[height=8.3cm,bb=214 57 355 591,clip,angle=-90]{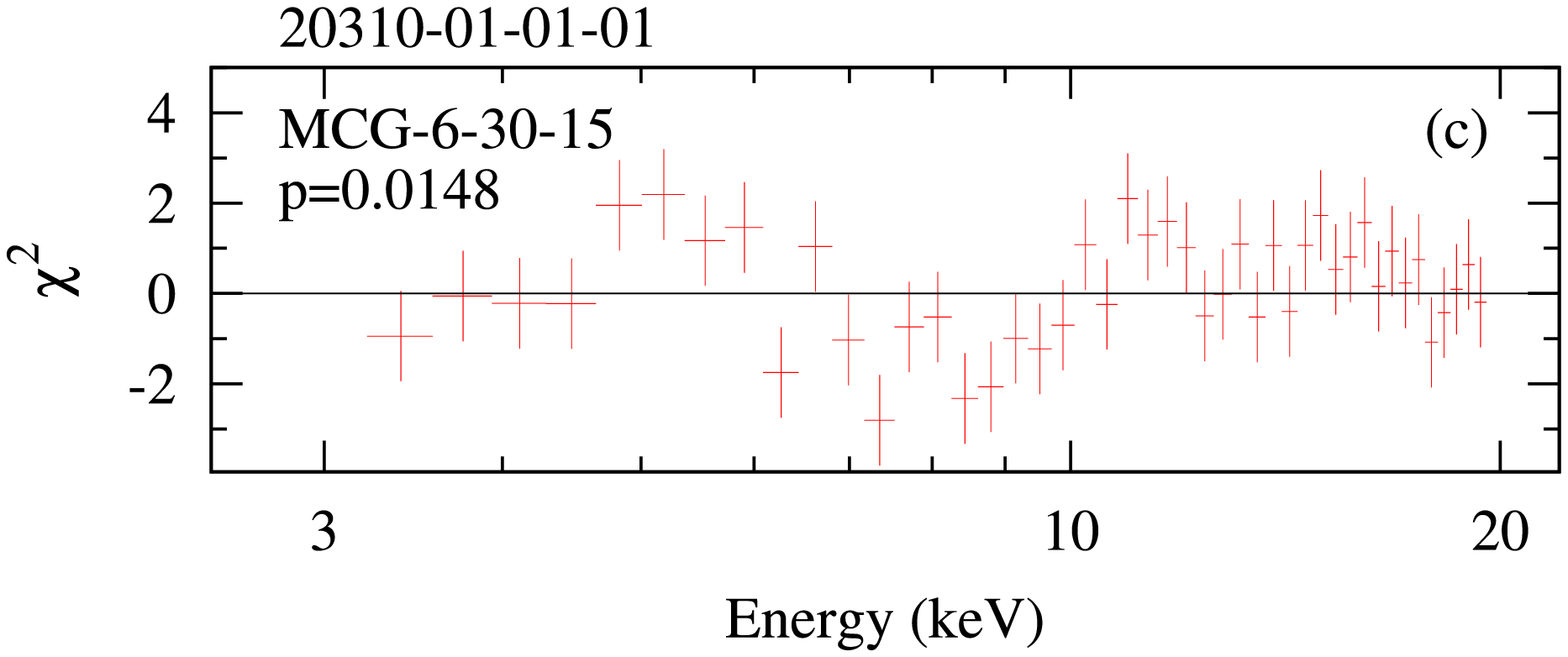}\\
\includegraphics[height=8.3cm,bb=214 57 355 591,clip,angle=-90]{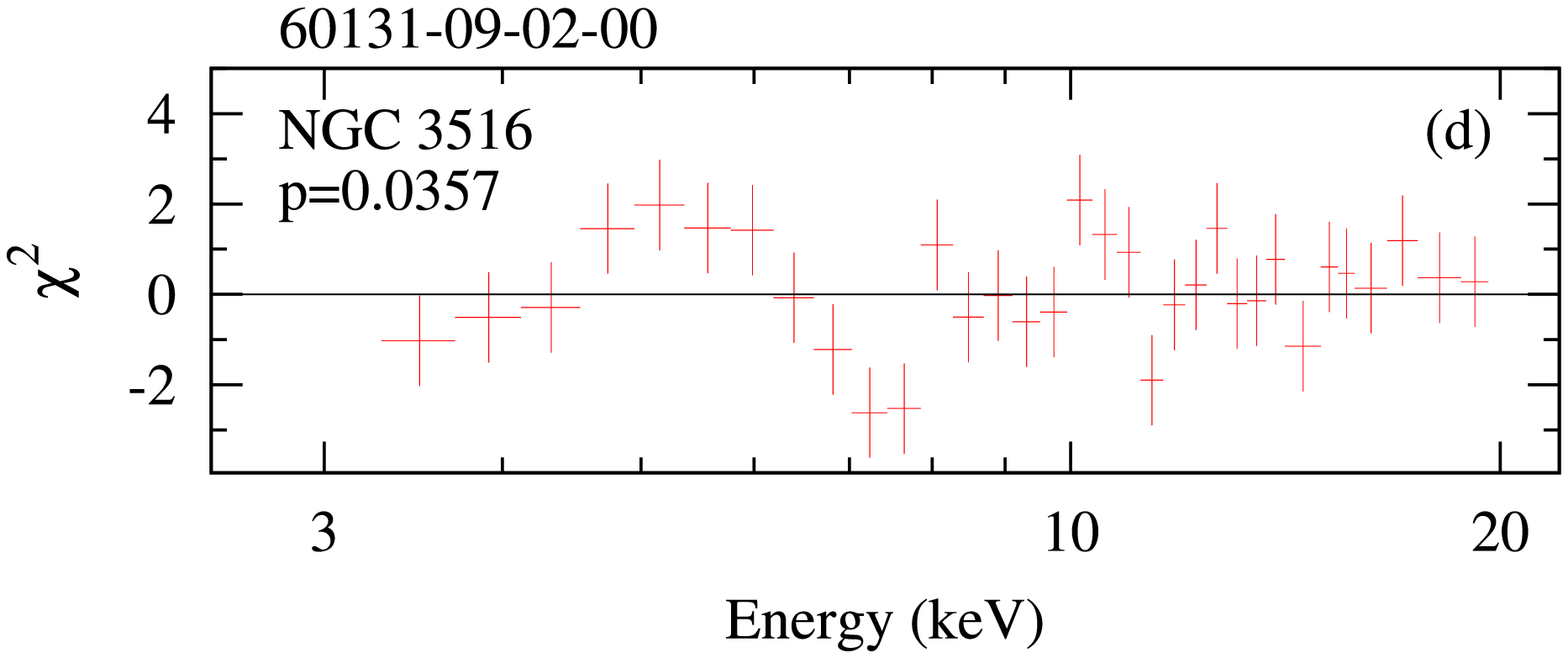}
\includegraphics[height=8.3cm,bb=214 57 420 591,clip,angle=-90]{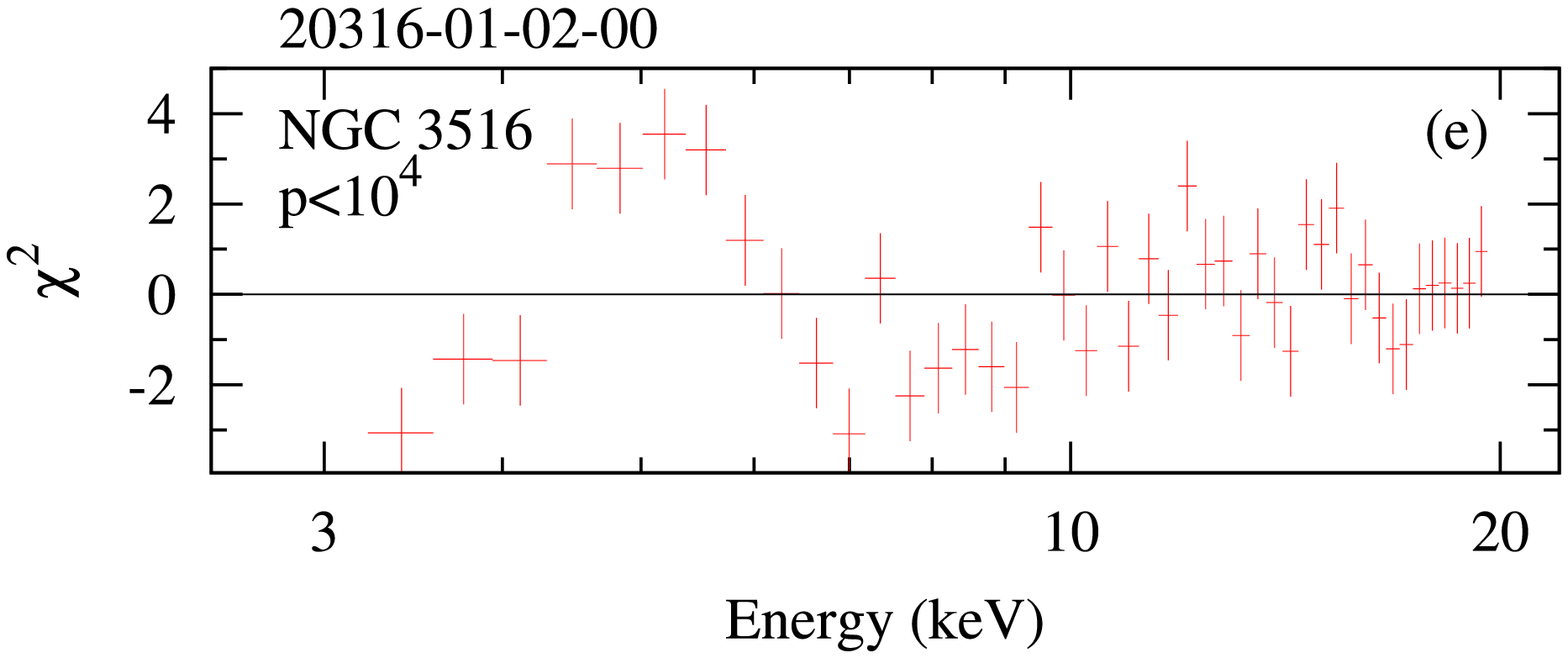}
\caption{Contributions to $\chi^2$ for the PLG best fit  models applied to
exemplary data of (a) NGC 5506 (obsid 90145-01-01-00), (b) NGC 3783
(obsid 92113-03-81-00), (c) MCG-6-30-15 (obsid 20310-01-01-01) and
(d--e) NGC 3516 (obsids 60131-09-02-00 and 20316-01-02-00). The fit
probabilities are indicated on each panel.}
\label{fig:fig2}
\end{figure}

\begin{figure}
\includegraphics[height=8.3cm,bb=349 110 500 685,clip,angle=-90]{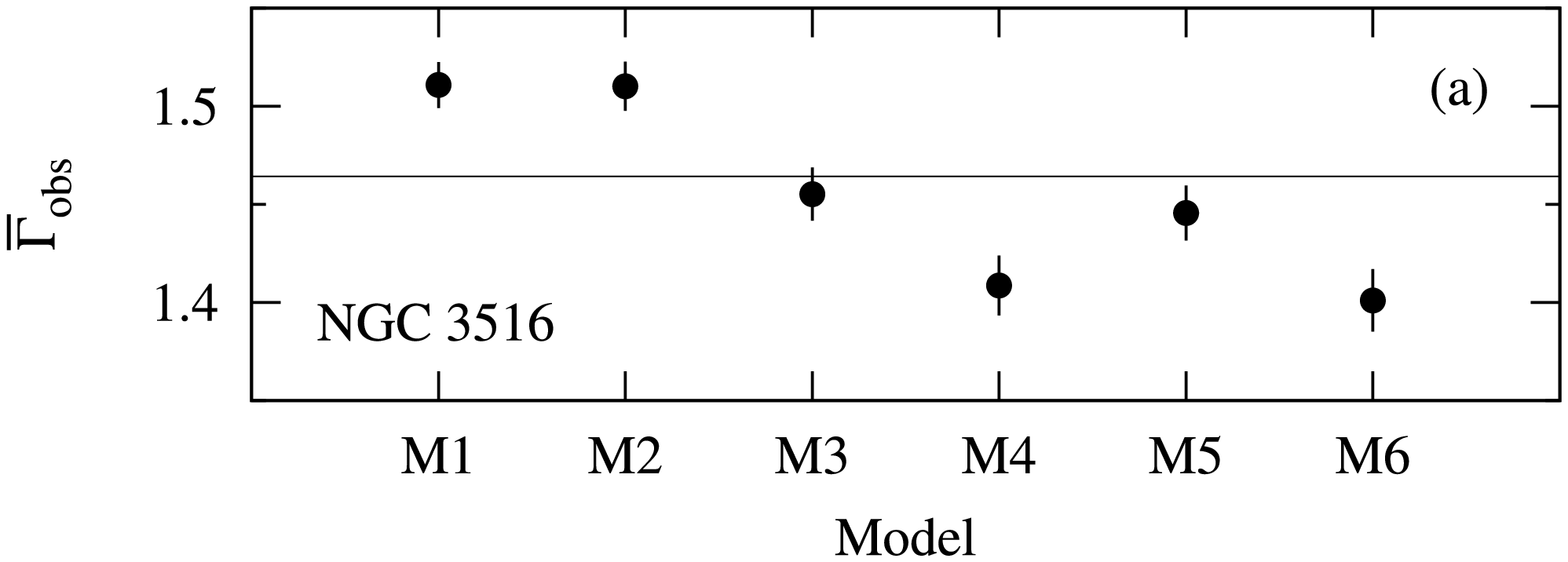}\\
\includegraphics[height=8.3cm,bb=349 109 551 685,clip,angle=-90]{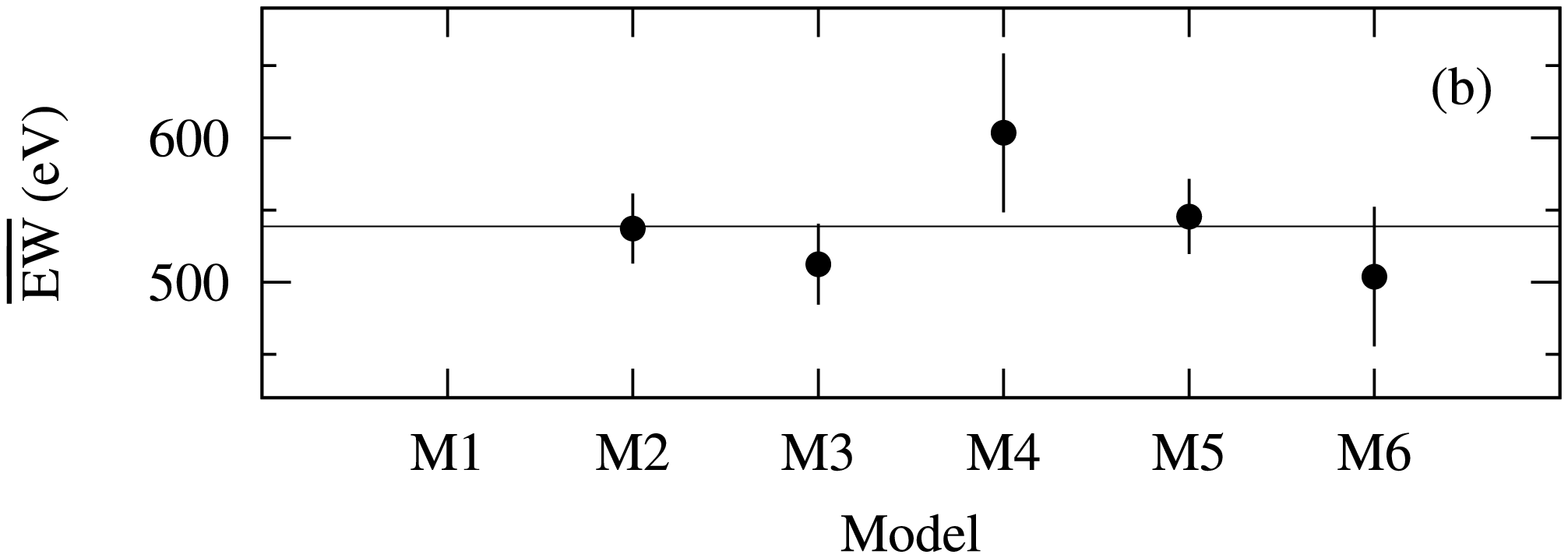}
\caption{The average $\Gamma_{\rm obs}$ (a) and EW of the iron line (b) in NGC
3516 for different spectral models. The model labels are M1 -- PL;  M2 -- PLG;
M3 -- ePLG; M4 -- same as M3 but with variable edge energy, $E_{\rm edge}$; M5
-- same as M3 but with variable line energy, $E_{\rm line}$; M6 -- same as M3
but with variable both $E_{\rm edge}$ and $E_{\rm line}$. The error-bars
represent standard errors on the mean. The horizontal lines indicate the average
$\overline{\Gamma_{\rm obs}}$ and $\overline{EW}$.}
\label{fig:fig3}
\end{figure}

\begin{list}{\labelitemi}{\leftmargin=0em\itemsep=0.5em}
\item[1.] A simple PL model describes well the spectra of Fairall~9, Akn~564 and
Mkn~766.
\item[2.] In NGC~4051, NGC~3227, and NGC~5548 a statistically
acceptable global fit to the data is obtained with the PLG model.
\item[3.] In NGC~3783 and NGC~5506, the addition of  an absorption edge with energy fixed at
7.1 keV is needed to achieve an acceptable model fit (ePLG). 
\item[4.] In MCG~6-30-15 the best global fit is reached by the ePLG model but
only when the edge energy is allowed to vary (the resulting mean edge energy is $\overline{E}_{\rm edge}\sim7.4$
keV).
\item[5.] Finally, the same ePLG model provides an acceptable global fit to the 
NGC~3516 spectra but only when both the energy of the iron line and the energy of the
absorption edge were left free to vary ($\overline{E}_{\rm line}\sim 6.1$, and
$\overline{E}_{\rm edge}\sim 7.95$).
\end{list} 

In Section 3, we report the results from: (a) the study of the correlation of
the average spectral slope with other source parameters, such as luminosity,
accretion rate and black hole mass, (b) the study of the  spectral slope
variations as a function of the source luminosity for each  object in the
sample, and (c) the study of the average line's equivalent width variations with
other source parameters. In each case, we use the appropriate, for each source,
best fitting slope, $\Gamma_{\rm obs}$, and line's EW values.

We do not consider the implications from the ``variable edge energy"  model fits
to MCG~-6-30-15  or from the ``variable iron line and edge energy" model fits to
NGC 3516. We believe that the $E_{\rm edge}$ and $E_{\rm line}$ best fit values
from these model fits  should be treated with caution due to the low energy
resolution of the PCA instrument. The residua plot in Figs.~\ref{fig:fig2}d--e show
this clearly in the case of NGC 3516. It is the line-like positive residual at
$\sim$5 keV that forces the best fit iron line energy to be less than 6.4 keV in
this case.  However, we can also get a good fit to the the spectrum from e.g. the
20316-01-02-00 observation of NGC 3516 (Fig.~\ref{fig:fig2}e) if we consider the
ePLG model with $E_{\rm line}$ and $E_{\rm edge}$ kept fixed at 6.4 keV and 7.1 keV,
respectively, together with intrinsic neutral absorption, i.e. if we add a {\tt
wabs} component to the model and we let N$_{\rm H}$ to be a free model
parameter.

We note that in the case of NGC~5506, NGC~3783, MCG~6-30-15 and NGC~3516, the
$\Gamma_{\rm obs}$ values were not dramatically affected when we added an
absorption edge or when we let $E_{\rm edge}$ or $E_{\rm line}$ to vary freely
during the model fitting process. For example, Fig.~\ref{fig:fig3}a shows the
mean $\Gamma_{\rm obs}$  for NGC~3516 in the case of all 6 models that we
applied to its individual spectra  (PL, PLG, ePLG and its modifications: ePLG
model with variable $E_{\rm edge}$ and/or $E_{\rm line}$). Despite the fact
that, statistically speaking, models 1--5 are not formally acceptable, the mean 
best fit spectral slope values are almost identical. The largest difference is
the one between $\overline{\Gamma}_{\rm obs}$ in the case of the PL model (M1 in
Fig.~\ref{fig:fig3}) and in the case of the ePLG model when both the edge and
line energies are left free to vary (M6 in Fig.~\ref{fig:fig3}). Although the
difference between the mean $\Gamma$'s is statistically significant (the
error-bars show standard errors on $\overline{\Gamma}$), it
is so small ($\sim 0.1$) that it cannot affect any of the results we discuss in
Section 3.

We reached the same conclusion for the mean EW of the iron line. In
Fig.~\ref{fig:fig3}b, we plot $\overline{\rm EW}$ in the case of the 5 models
that include an emission line component. The averages were calculated over the
data-sets in which the line was detected at 2$\sigma$ level. All five
measurements are in agreement with each other (again, the
error-bars show standard errors on $\overline{\rm EW}$). The largest difference is the one
between the average EW in the case of model M4 and model M6 ($\sim$100 eV).
Taking into account their errors, the two measurements are consistent with each
other, and so this difference does not affect any of the conclusions we present
in Section 3.

\section{Spectral variability analysis}
\label{sec:variability}

\begin{table*}

\caption{Variability properties of $\Gamma_{\rm obs}$ and $L_{\rm 2-10}$ based
on the best-fit model parameter values.}

\centering
\begin{tabular}{l c c c | c c c}
\hline
Target      & $\overline{\Gamma}_{\rm obs}$ & $\chi^2/d.o.f.$   & $F_{\rm var}$ &
$\overline{L}_{\rm 2-10}$  & $\chi^2/d.o.f$       & $F_{\rm var}$\\
            &                     &            &        (\%)  &
($\times10^{42}$) erg s$^{-1}$&   & (\%)\\
\hline
Fairall 9   & 1.794$\pm$0.006 & 1.3   &   3.7$\pm$0.3  & 102$\pm$1       & 16  
& 26.0$\pm$0.4 \\
Ark 564     & 2.48 $\pm$0.01  & 1.6   &   5.0$\pm$0.3  & 25.5$\pm$0.4    & 20  
& 30.7$\pm$0.3 \\
Mrk 766     & 2.02 $\pm$0.01  & 2.3   &   4.6$\pm$0.4  & 10.6$\pm$0.2    & 55  
& 30.3$\pm$0.5 \\
NGC 4051    & 1.842$\pm$0.009 & 6.9   &  14.9$\pm$0.3  & 0.493$\pm$0.007 & 83  
& 49.7$\pm$0.2 \\
NGC 3227    & 1.552$\pm$0.008 & 7.1   &  13.7$\pm$0.2  & 1.86$\pm$0.03   & 109 
& 48.0$\pm$0.2 \\
NGC 5548    & 1.735$\pm$0.004 & 2.1   &   2.9$\pm$0.2  & 28.8$\pm$0.4    & 196 
& 39.4$\pm$0.2 \\
NGC 3783    & 1.616$\pm$0.004 & 2.5   &   4.7$\pm$0.1  & 17.0$\pm$0.1    & 111 
& 23.2$\pm$0.1 \\
NGC 5506    & 1.800$\pm$0.003 & 2.4   &  3.1$\pm$0.1   & 10.2$\pm$0.1    & 169 
& 22.79$\pm$0.09\\
MCG-6-30-15 & 1.731$\pm$0.005 & 3.6   &   7.2$\pm$0.2  & 7.48$\pm$0.06   & 93  
& 27.2$\pm$0.1 \\
NGC 3516    & 1.40 $\pm$0.01  & 7.6   &  14.2$\pm$0.5  & 6.5$\pm$0.2     & 179 
& 41.3$\pm$0.3 \\
\hline
\end{tabular}
\label{tab:tab3} 
\end{table*}


\begin{figure*}
\includegraphics[height=8.8cm,bb=80 120 210 815,clip,angle=-90]{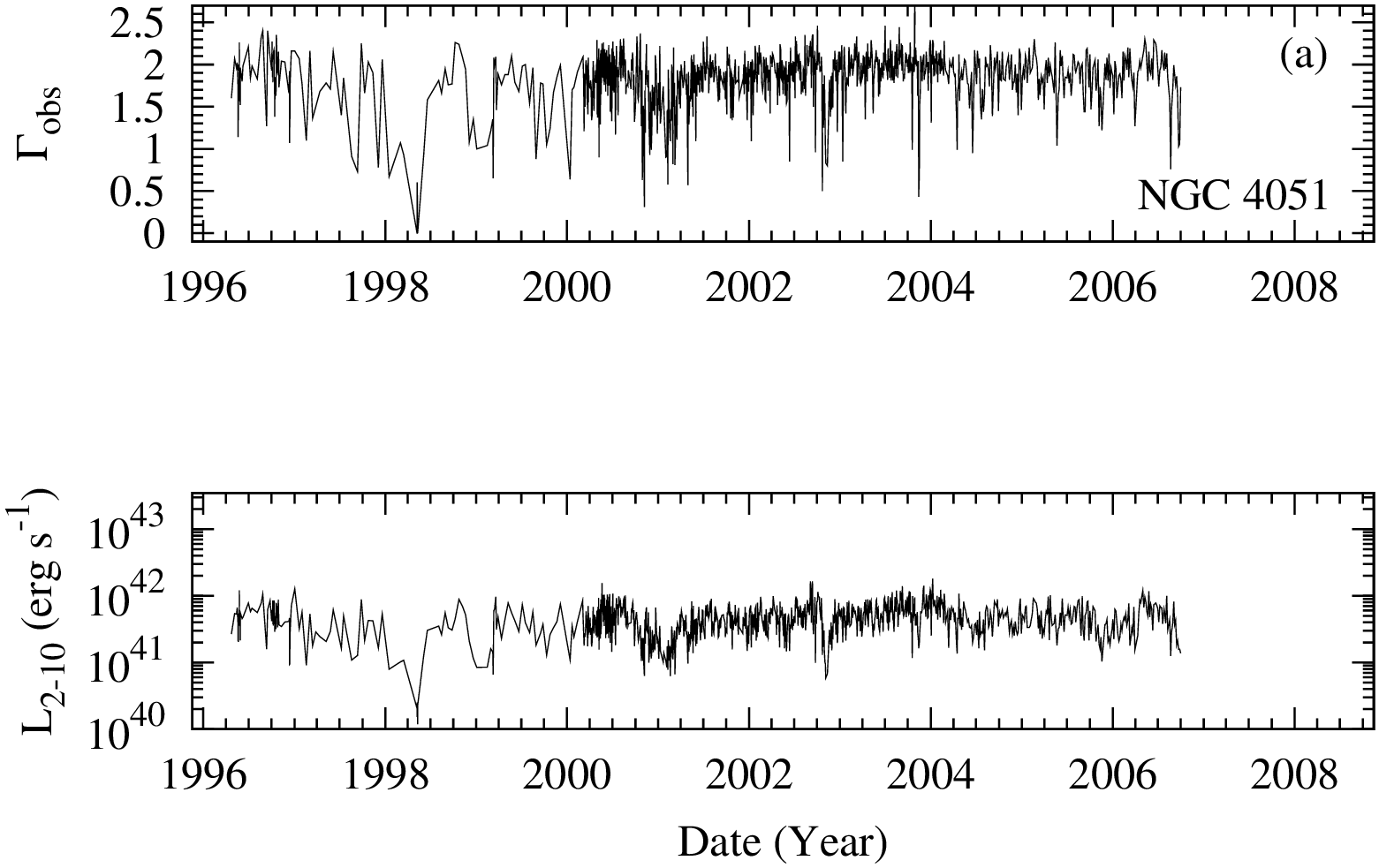}
\includegraphics[height=8.8cm,bb=80 120 210 815,clip,angle=-90]{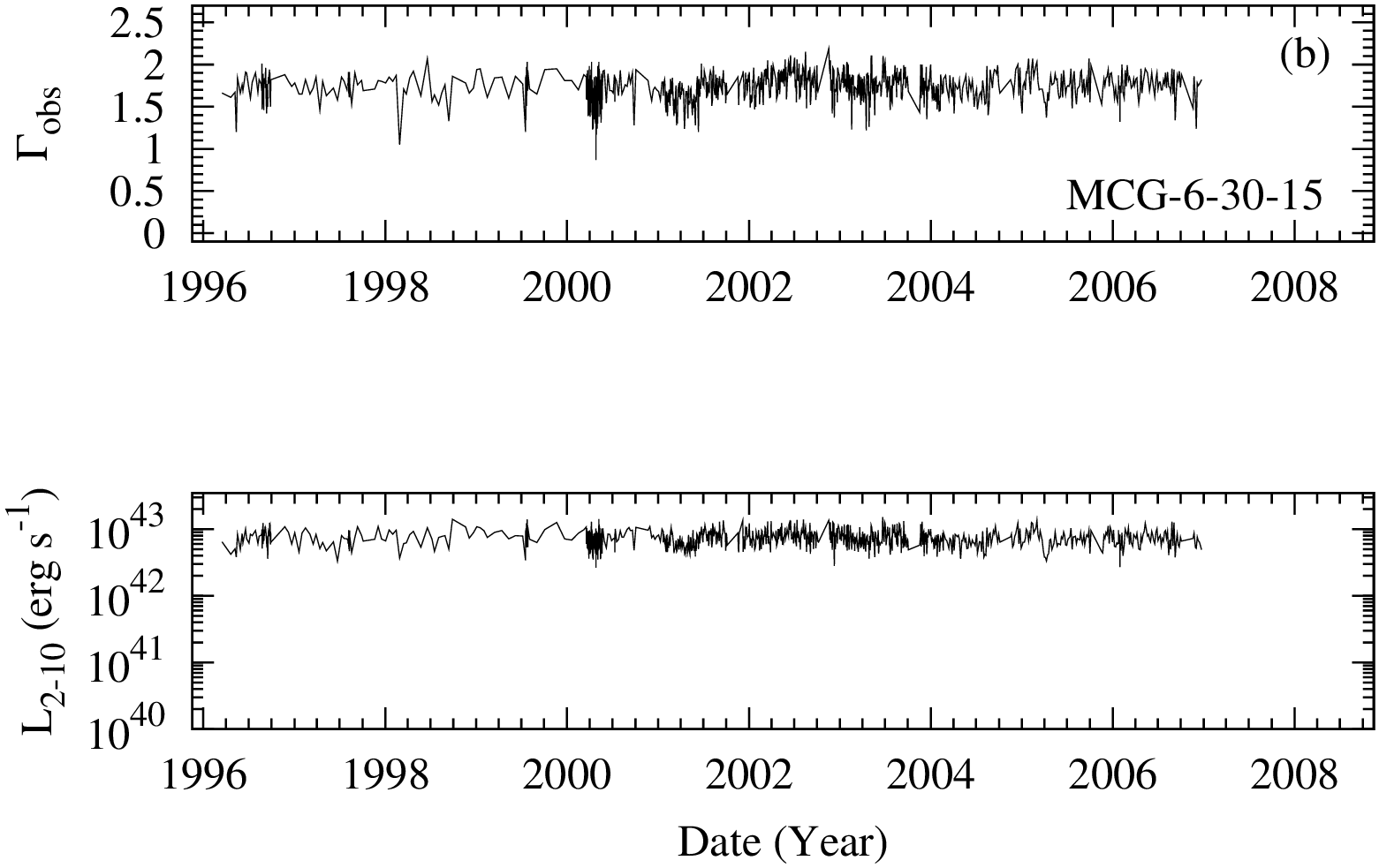}\\
\includegraphics[height=8.8cm,bb=320 120 515 815,clip,angle=-90]{fig4a.ps}
\includegraphics[height=8.8cm,bb=320 120 515 815,clip,angle=-90]{fig4b.ps}\\
\includegraphics[height=8.8cm,bb=80 120 210 815,clip,angle=-90]{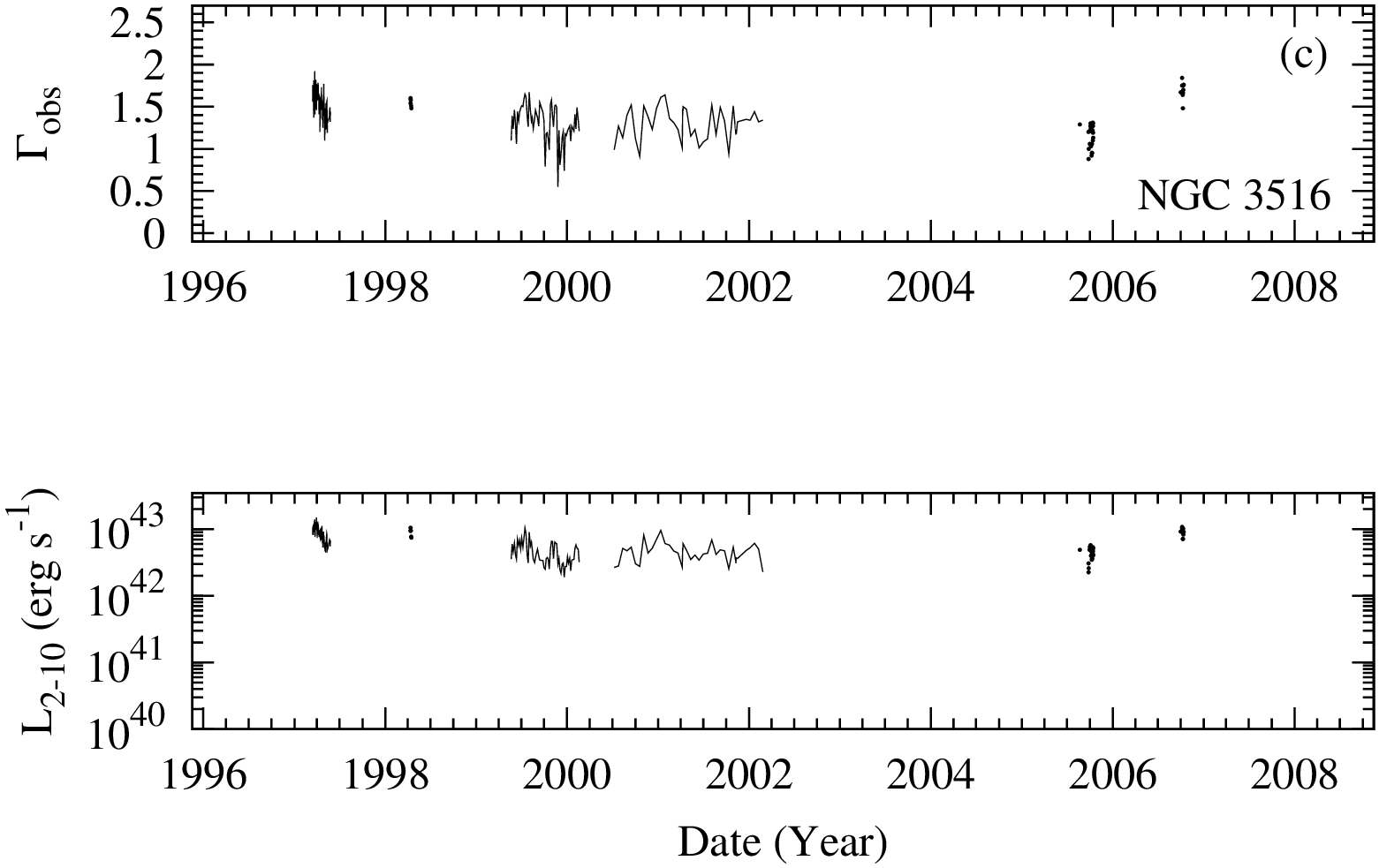}
\includegraphics[height=8.8cm,bb=80 120 210 815,clip,angle=-90]{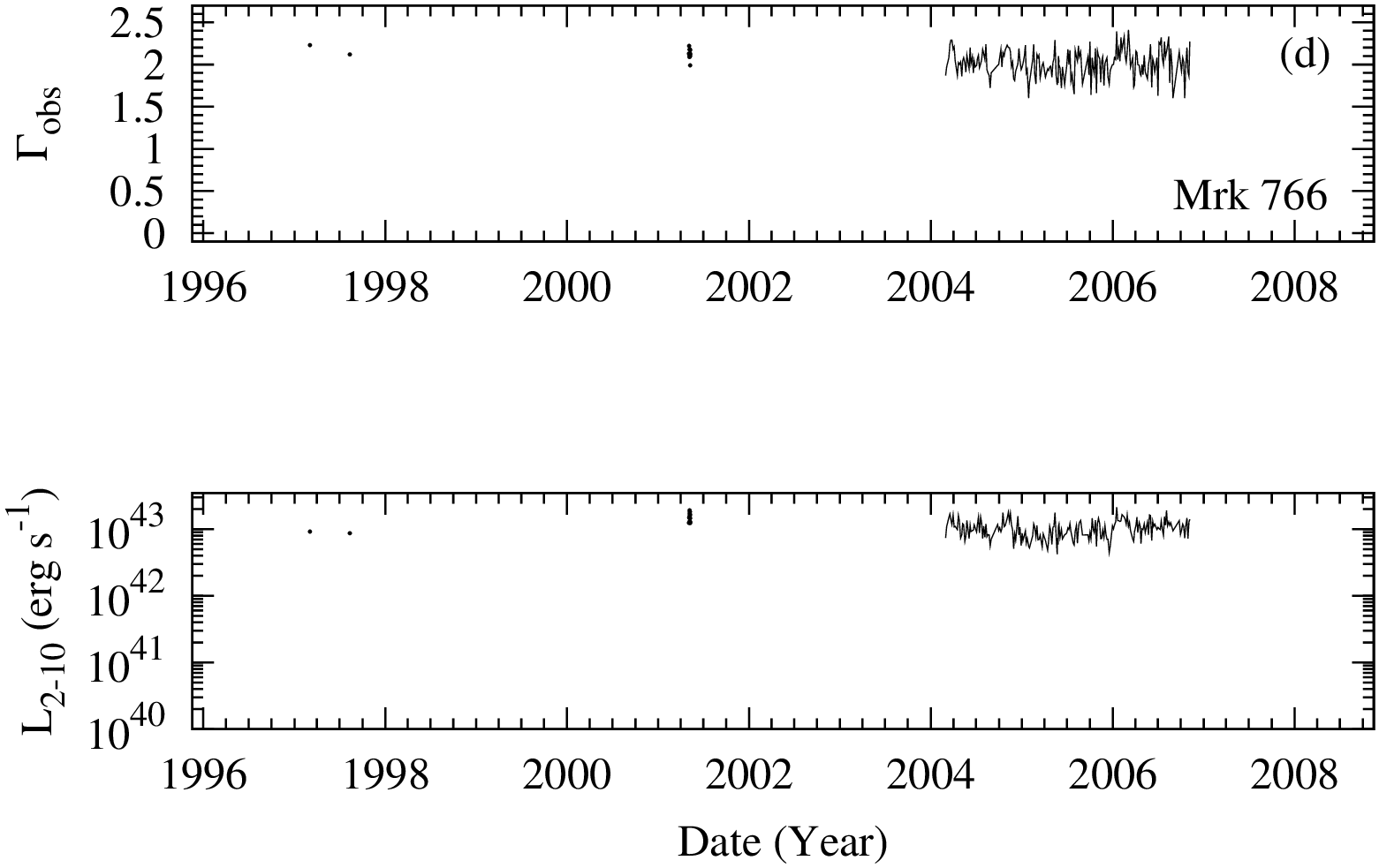}\\
\includegraphics[height=8.8cm,bb=320 120 515 815,clip,angle=-90]{fig4c.ps}
\includegraphics[height=8.8cm,bb=320 120 515 815,clip,angle=-90]{fig4d.ps}
\caption{The {\it RXTE} light curves of $\Gamma_{\rm obs}$ and $L_{\rm 2-10}$ for
objects with the best (NGC 4051 and MCG-6-30-15, a--b) and worst (NGC 3516
and Mrk 766, c--d) time coverage.}
\label{fig:fig4}
\end{figure*}

\begin{figure*}
\includegraphics[width=6.0cm,bb=174 71 545 438,clip,angle=-90]{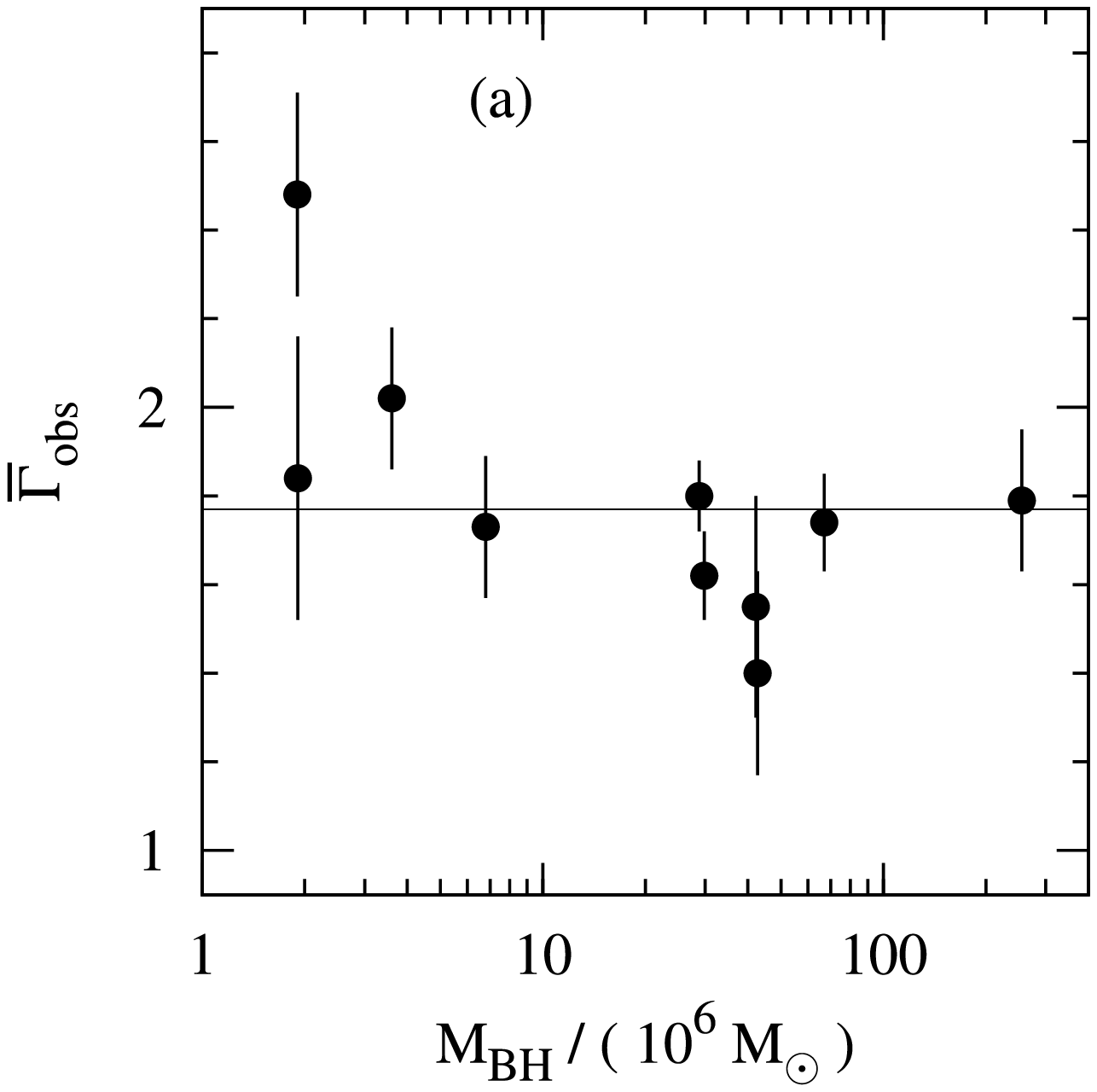}
\includegraphics[width=6.0cm,bb=174 133 545 438,clip,angle=-90]{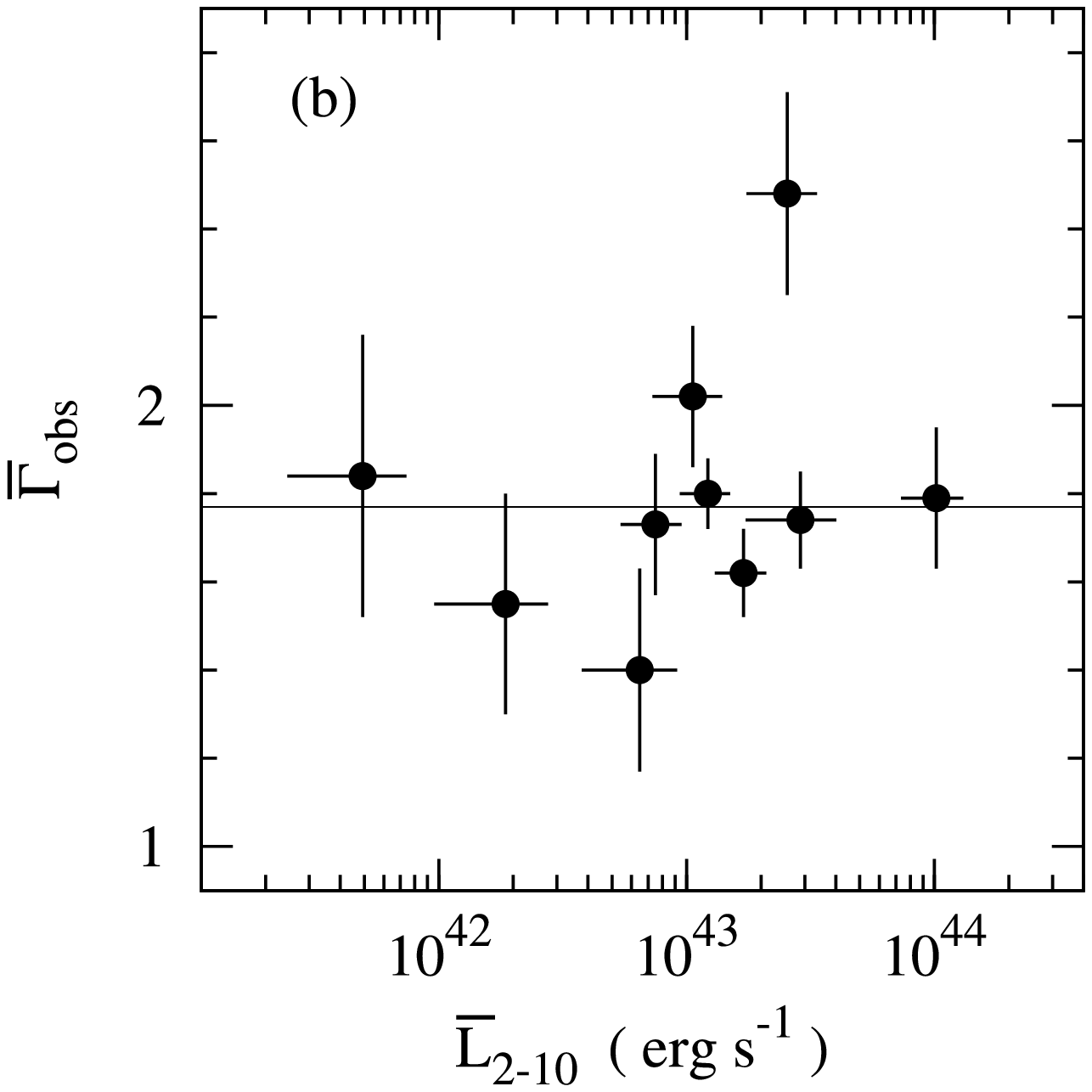}
\includegraphics[width=6.0cm,bb=174 133 545 438,clip,angle=-90]{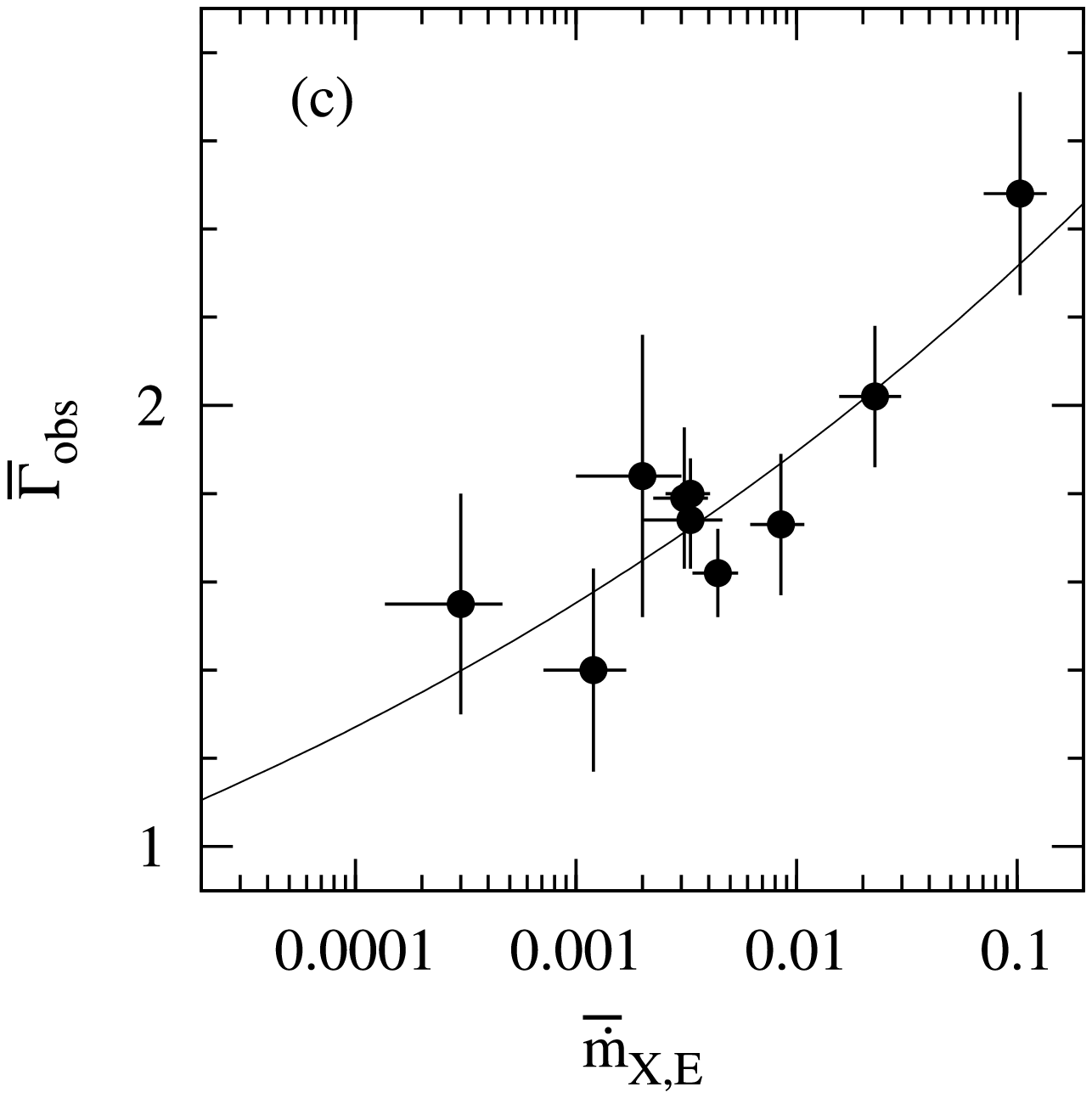}
\caption{Relations between the average photon index, $\overline{\Gamma}_{\rm
obs}$, and (a) BH mass, (b) average 2--10 keV luminosity, and (c) average mass
accretion rate in Eddington units. The correlation in (c) is best described with
a power-law, $\overline{\Gamma}_{\rm obs} \propto \overline{\dot{m}}_{\rm
X,E}^{~0.08\pm0.02}$ (solid curve).}
\label{fig:fig5}
\end{figure*}

\begin{figure*}
\includegraphics[width=6.0cm,bb=174 125 540 523,clip,angle=-90]{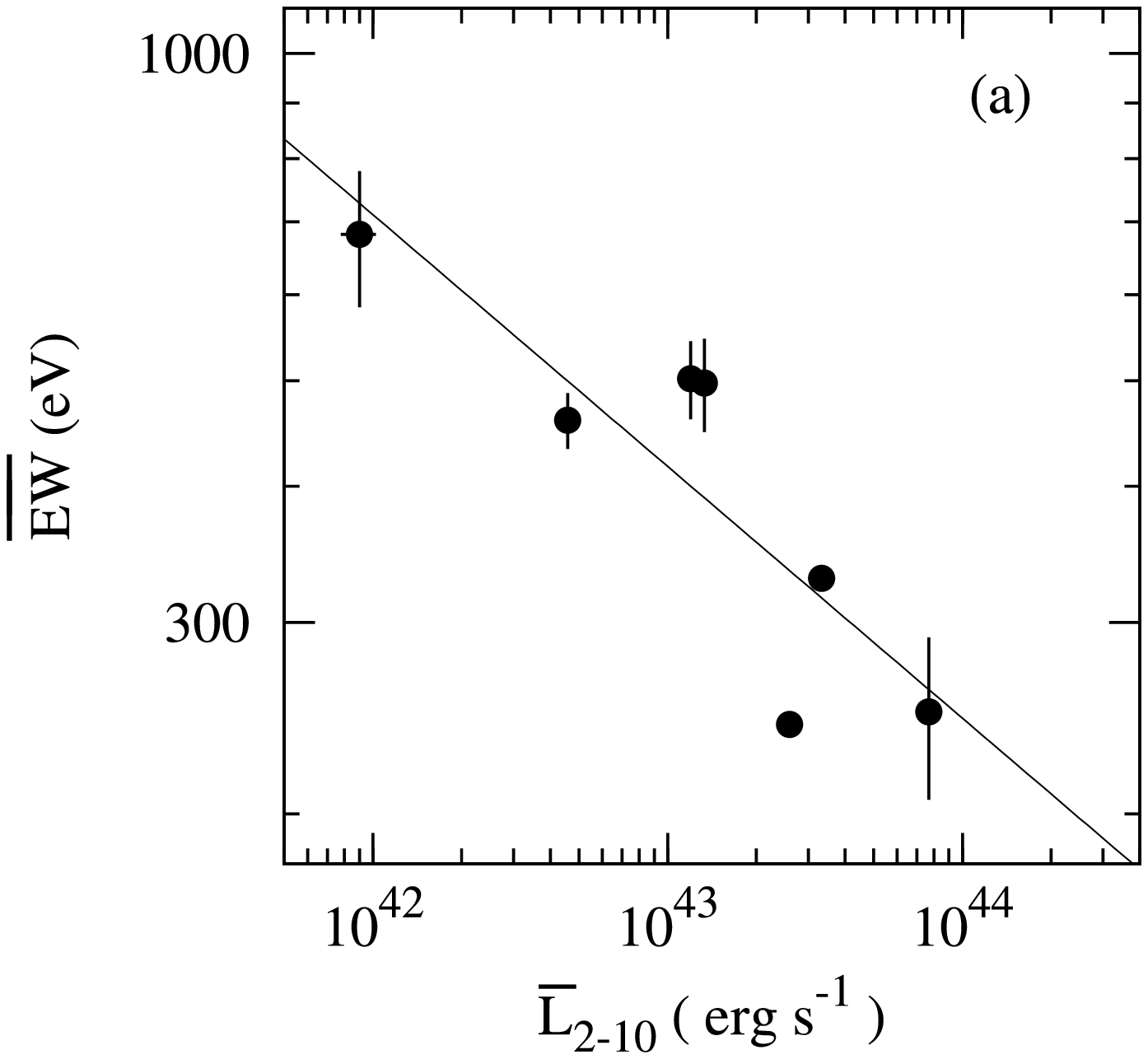}
\includegraphics[width=6.0cm,bb=174 220 540 523,clip,angle=-90]{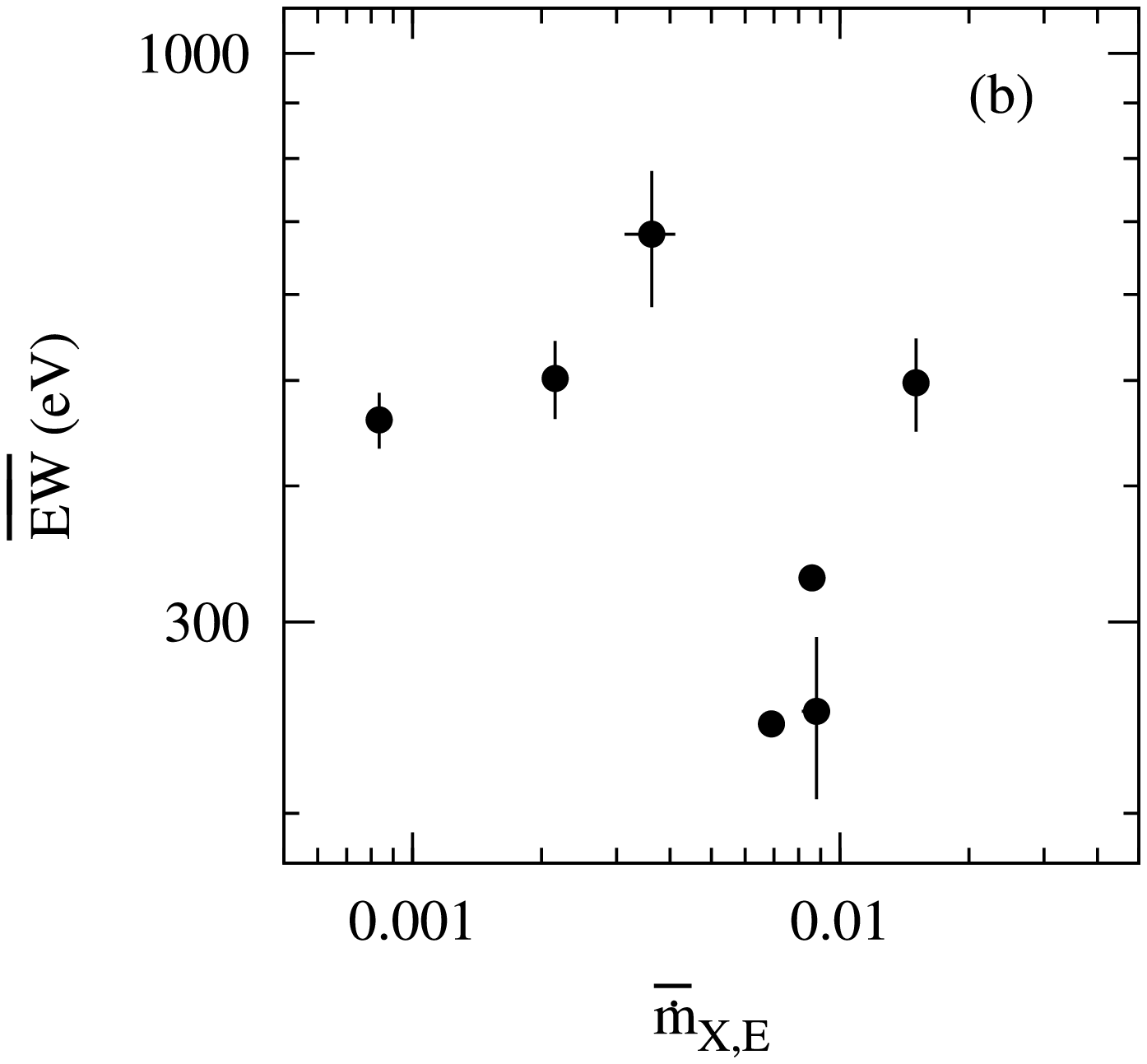}
\includegraphics[width=6.0cm,bb=174 220 540 523,clip,angle=-90]{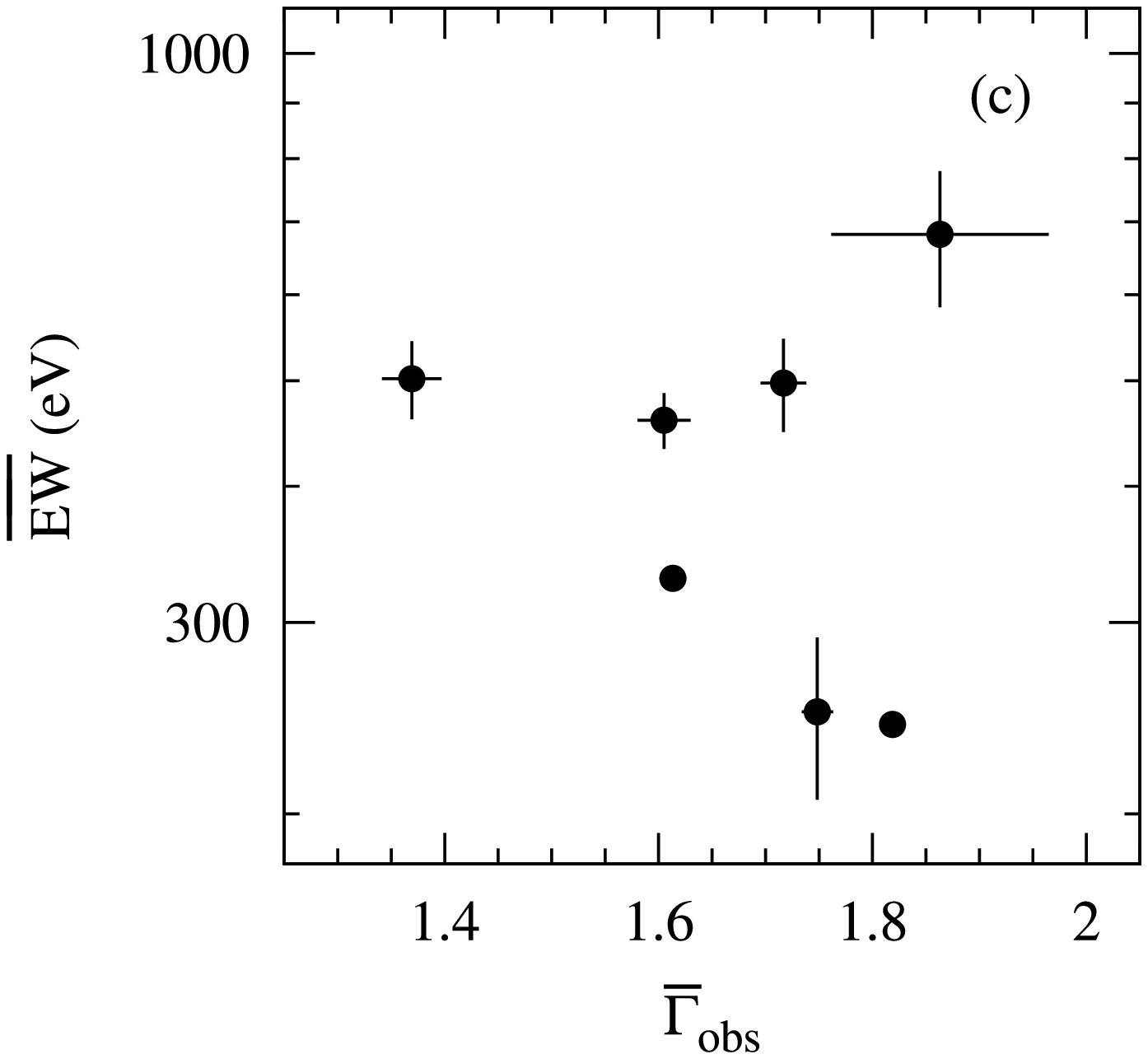}
\caption{Relations between the average EW of the iron line and the average (a)
2--10 luminosity, (b) mass accretion rate, and (c) photon index. The solid line
in panel (a) indicates the best fit relation $\overline{\rm EW} \propto
\overline{L}_{\rm 2-10}^{~-0.23\pm0.06}$.}
\label{fig:fig6}
\end{figure*}

\begin{figure*}
\includegraphics[width=3.85cm,bb=176 63 470 427,clip,angle=-90]{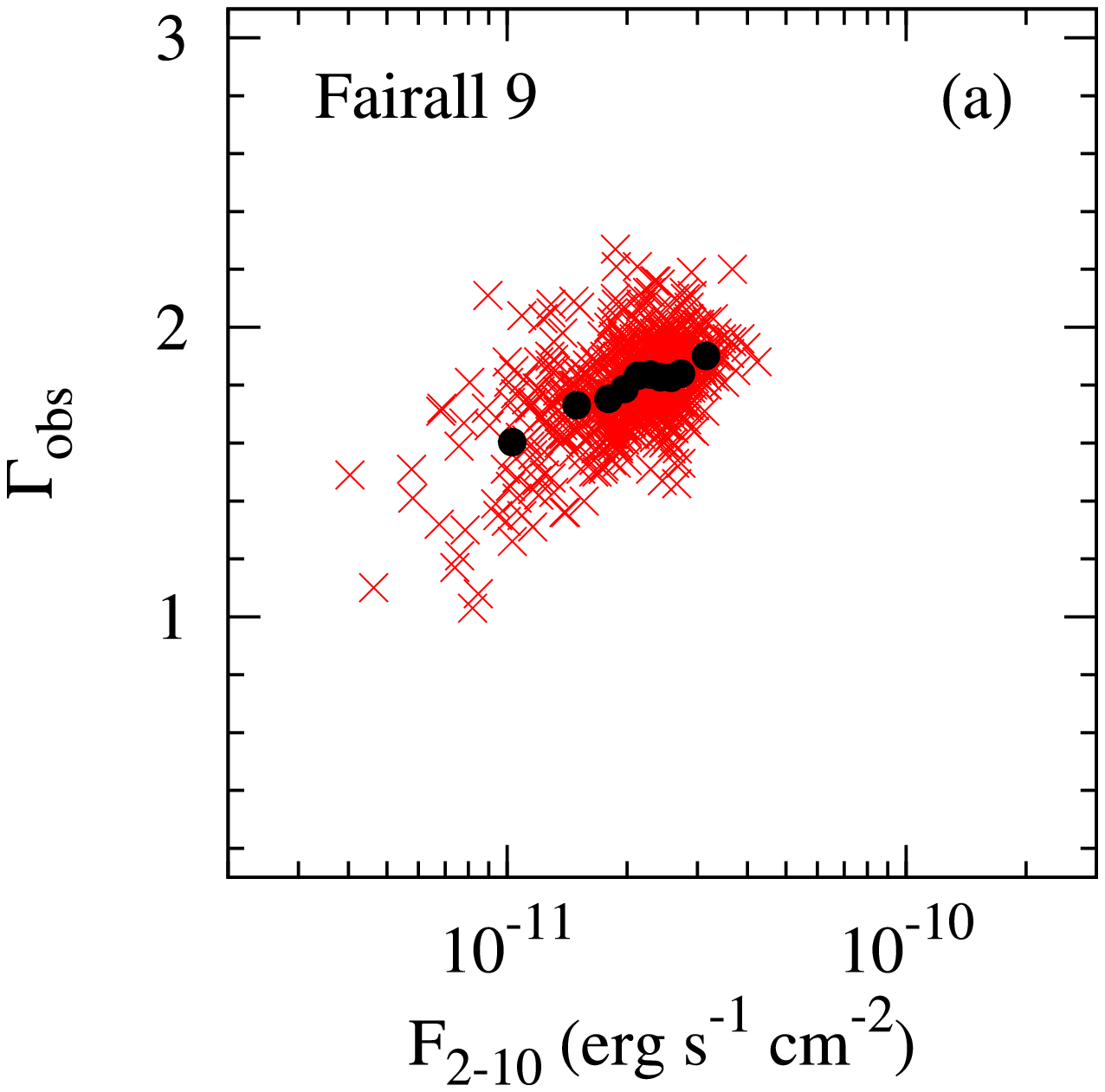}
\includegraphics[width=3.85cm,bb=176 133 470 427,clip,angle=-90]{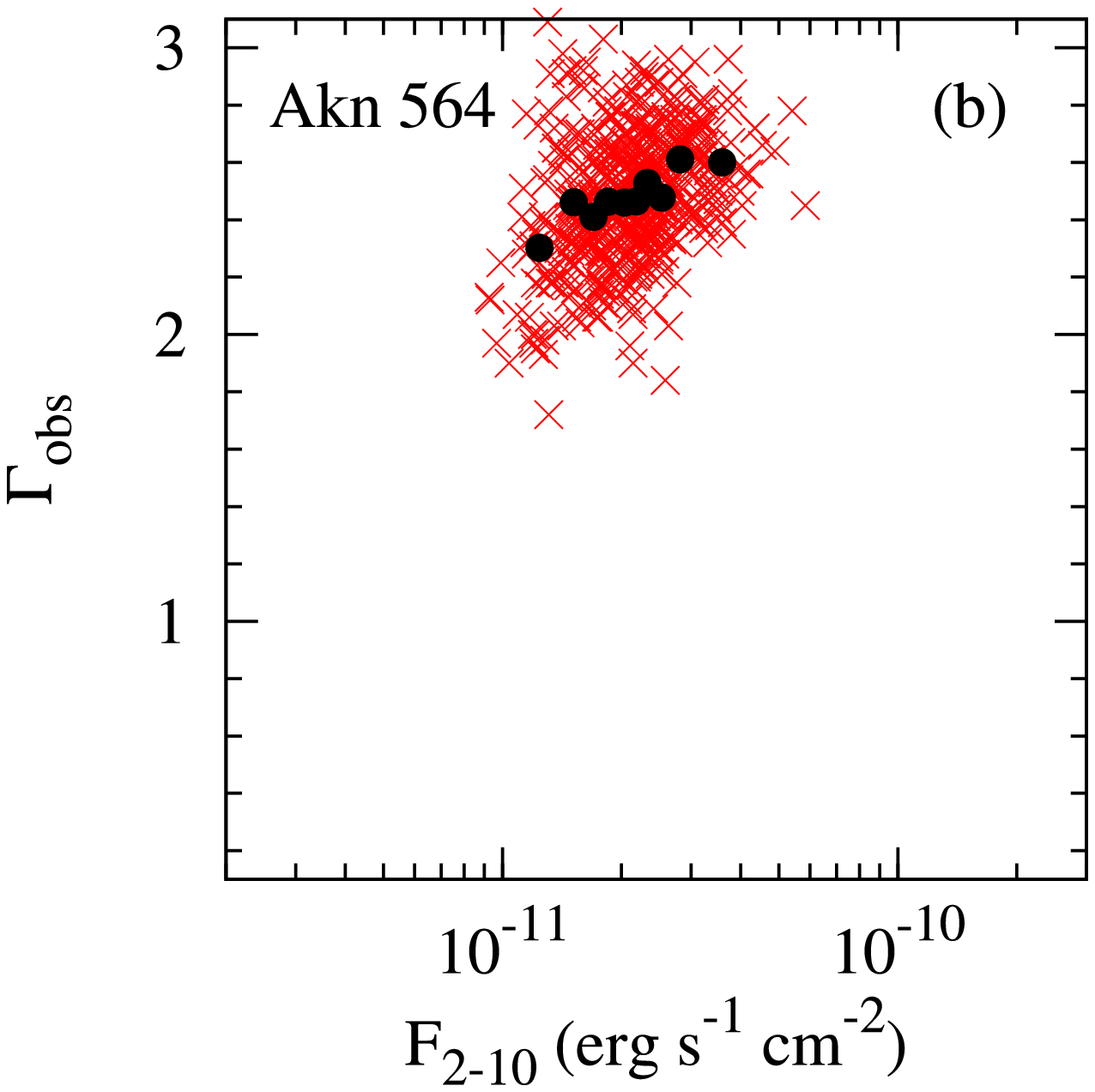}
\includegraphics[width=3.85cm,bb=176 133 470 427,clip,angle=-90]{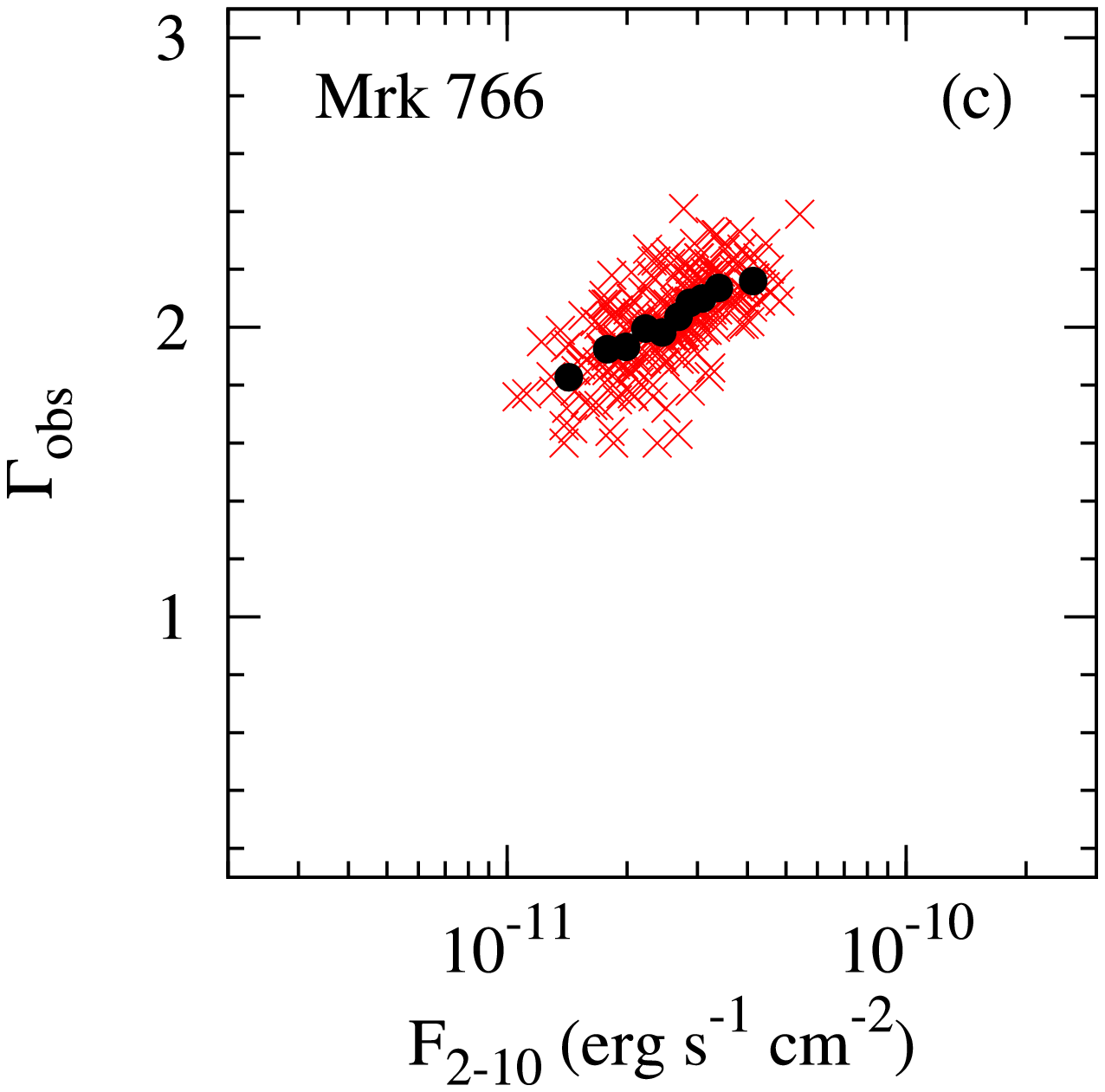}
\includegraphics[width=3.85cm,bb=176 133 470 500,clip,angle=-90]{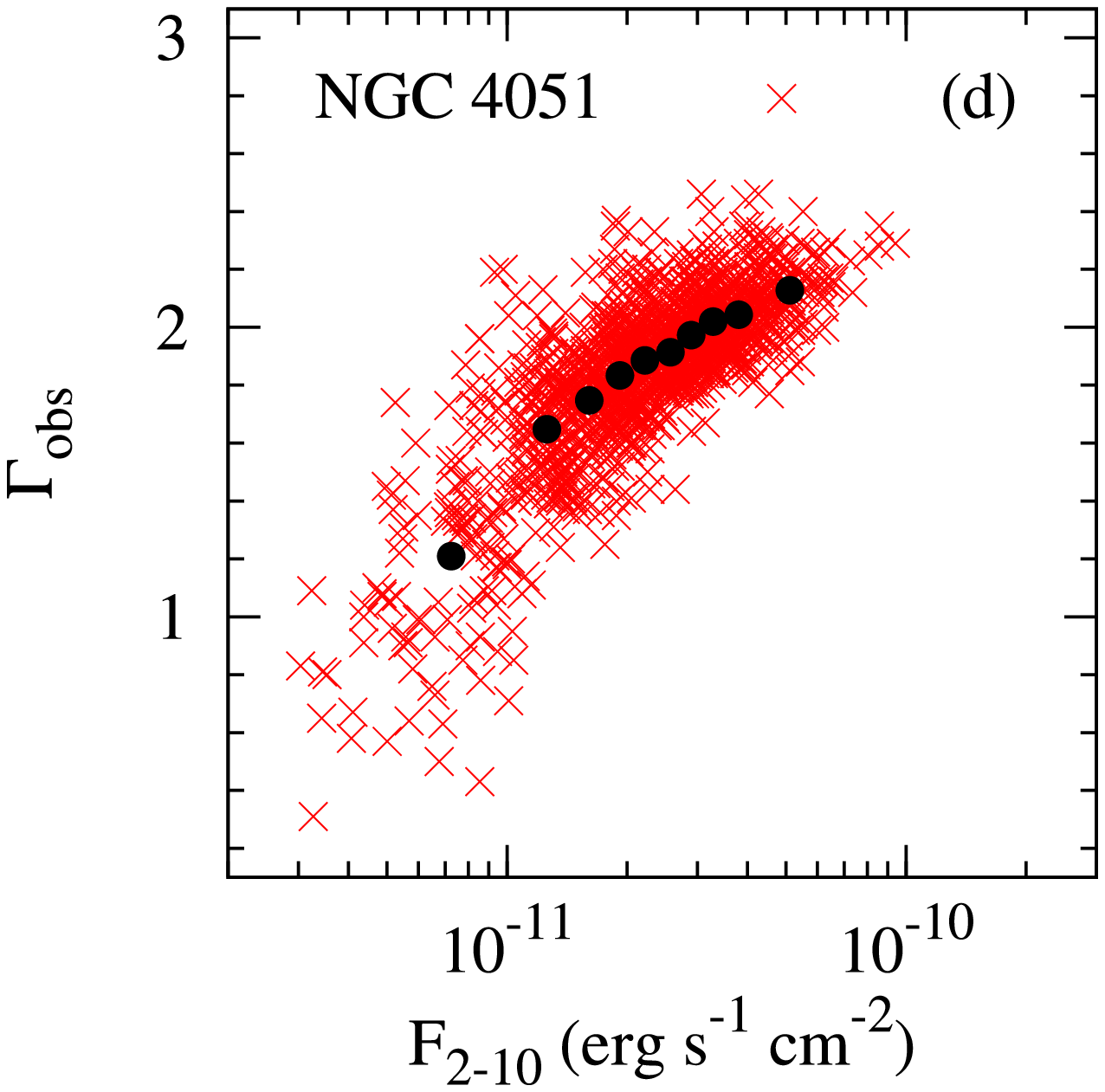}\\
\includegraphics[width=3.85cm,bb=176 63 470 427,clip,angle=-90]{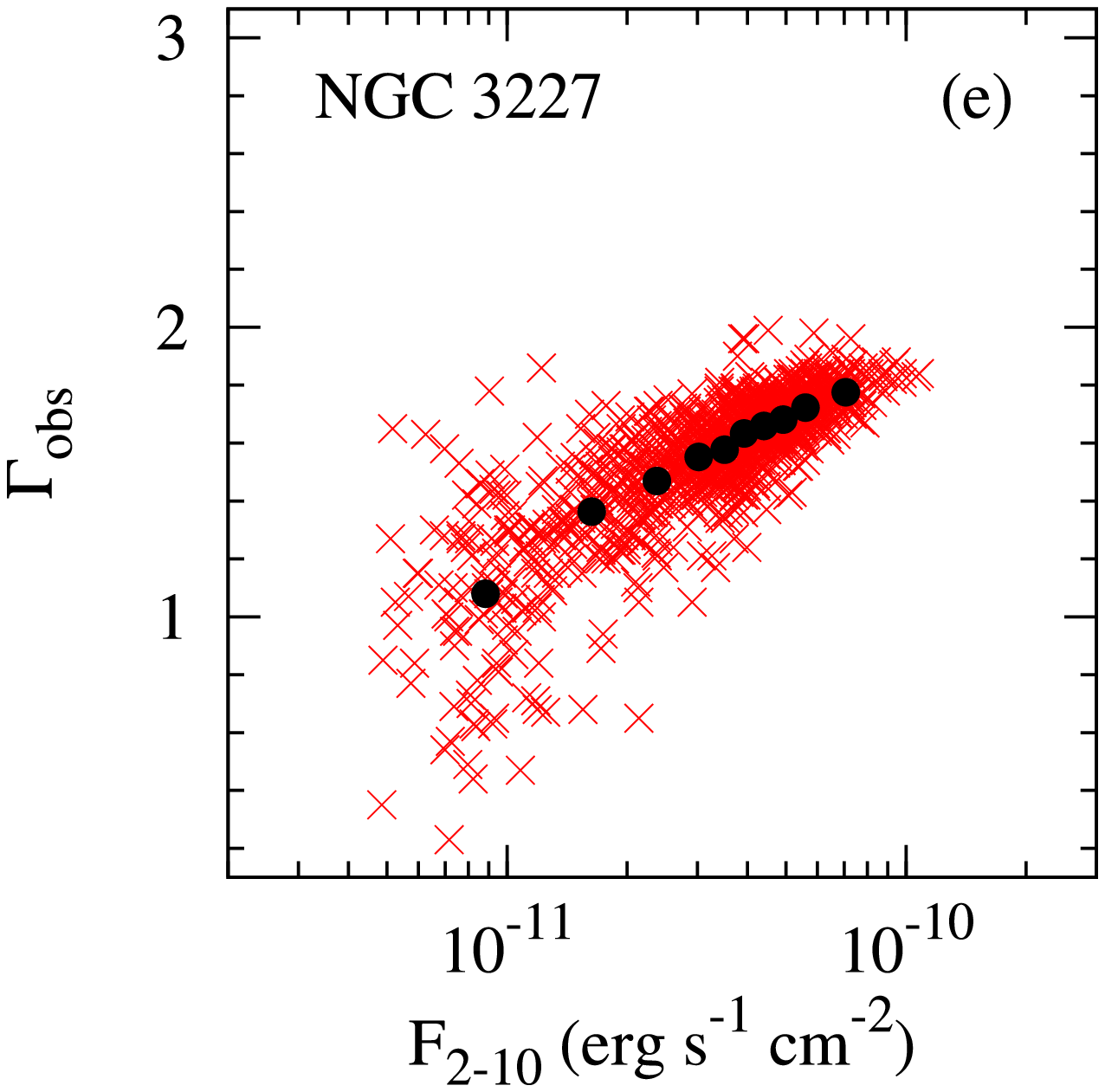}
\includegraphics[width=3.85cm,bb=176 133 470 427,clip,angle=-90]{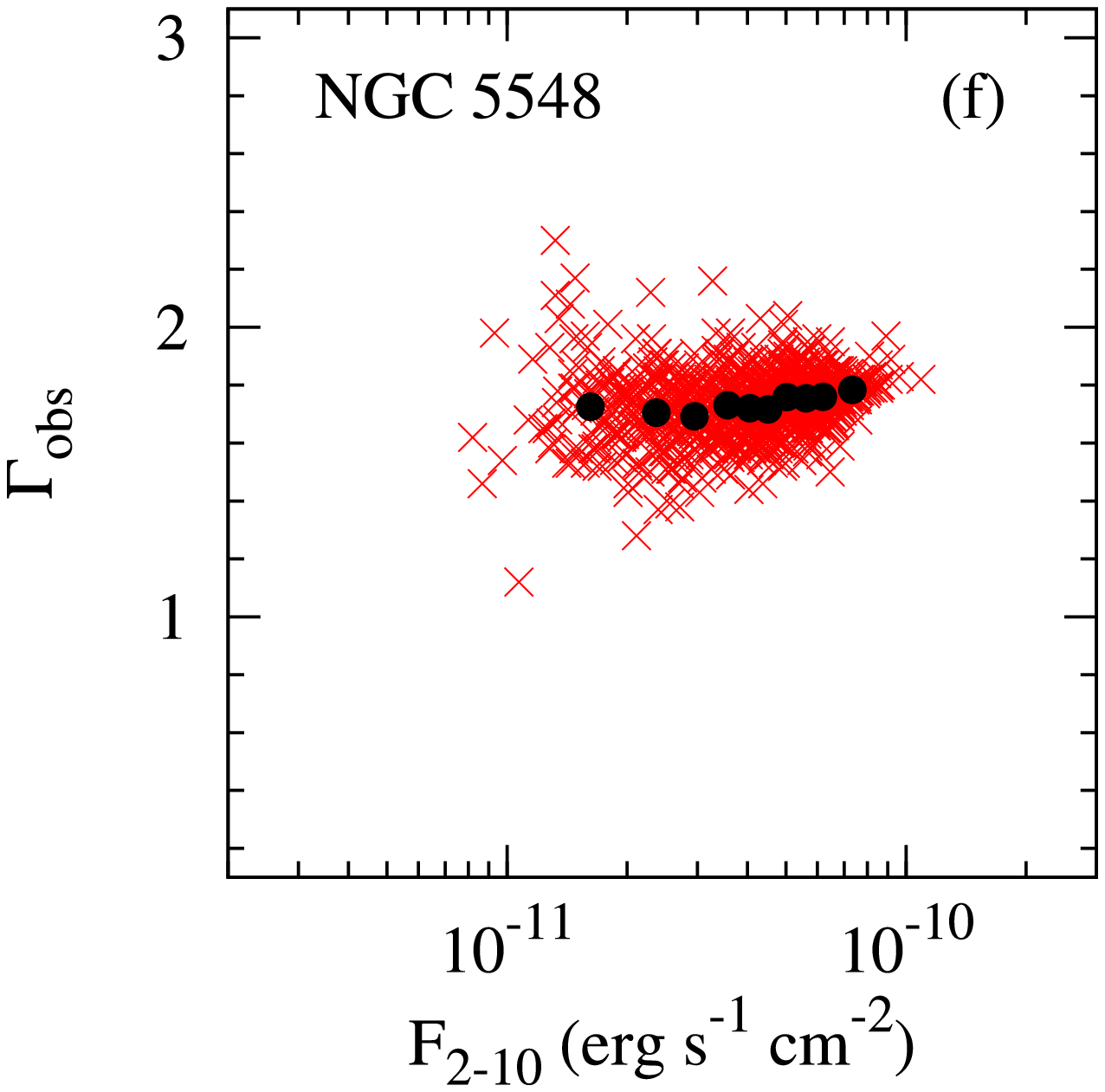}
\includegraphics[width=3.85cm,bb=176 133 470 427,clip,angle=-90]{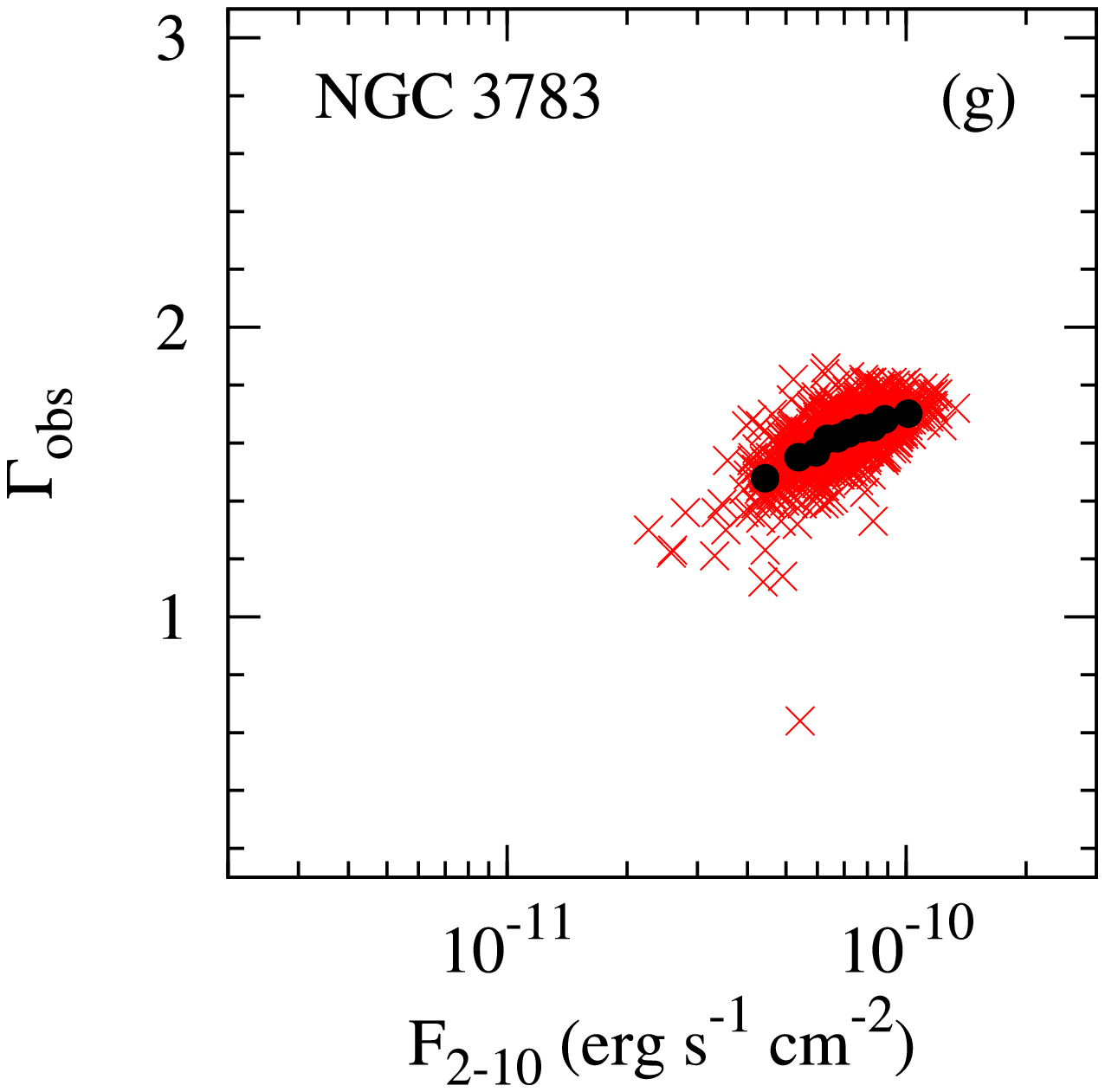}
\includegraphics[width=4.7cm,bb=176 133 536 500,clip,angle=-90]{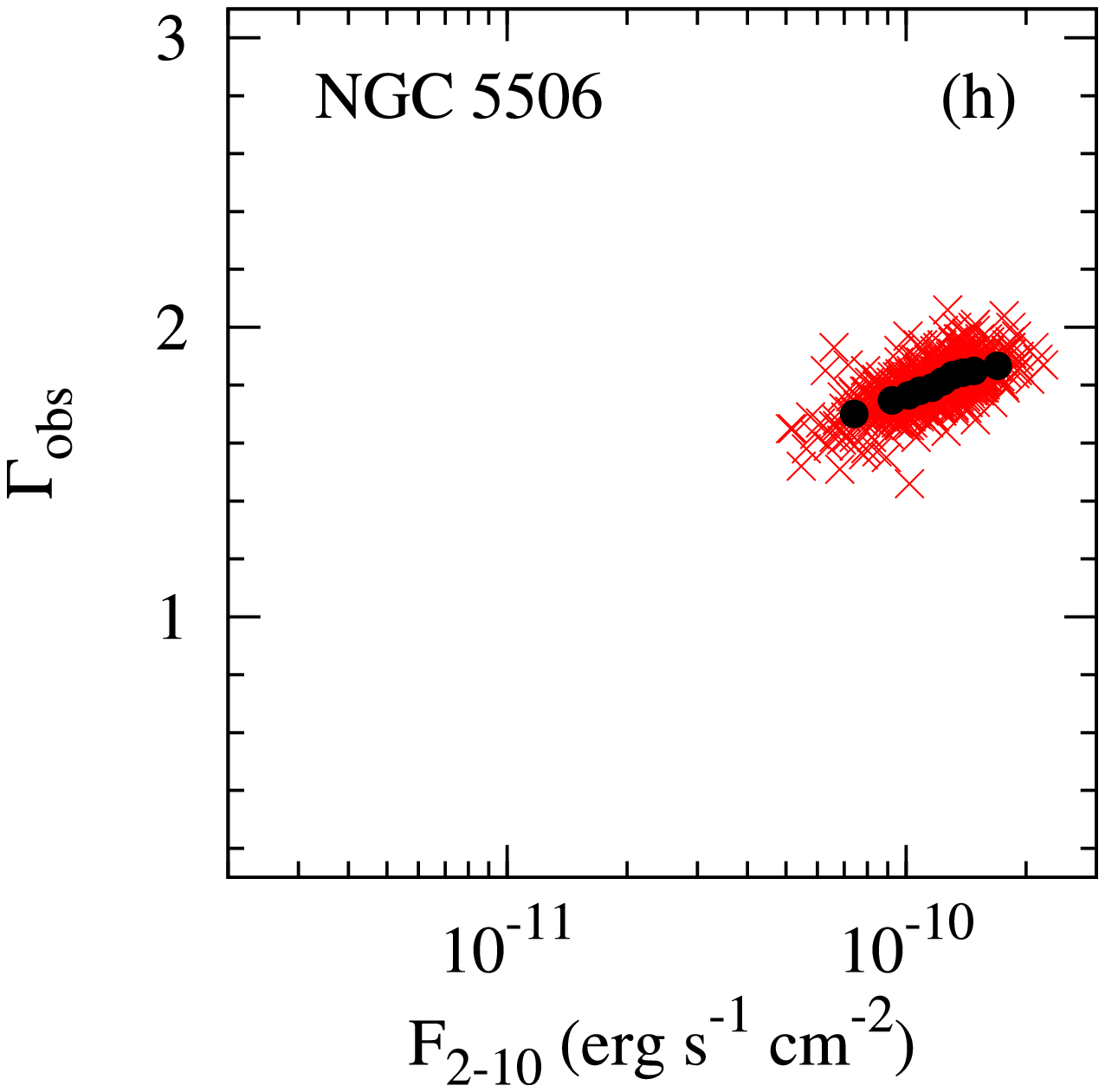}\\
\includegraphics[width=4.7cm,bb=176 63 536 427,clip,angle=-90]{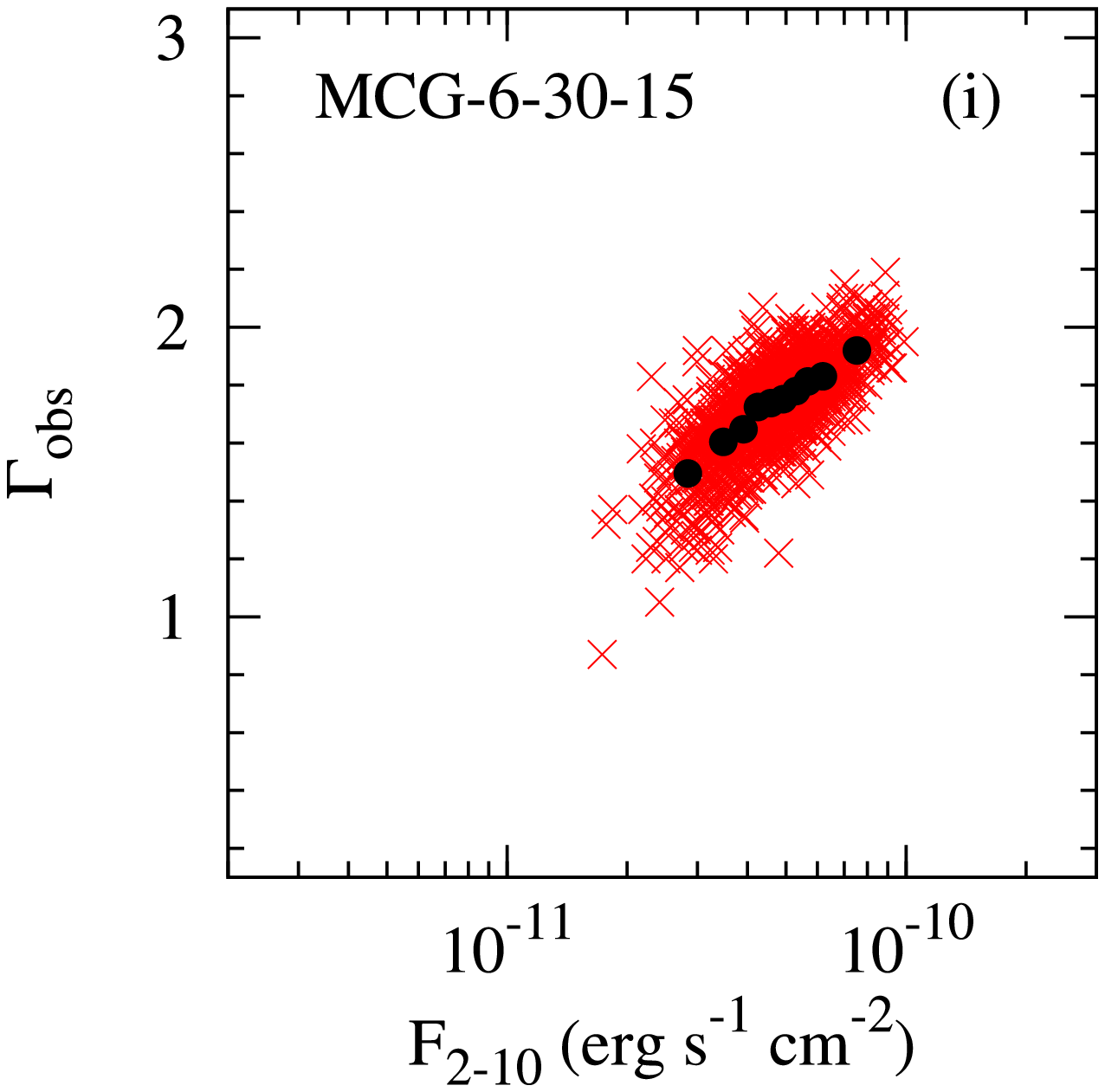}
\includegraphics[width=4.7cm,bb=176 133 536 427,clip,angle=-90]{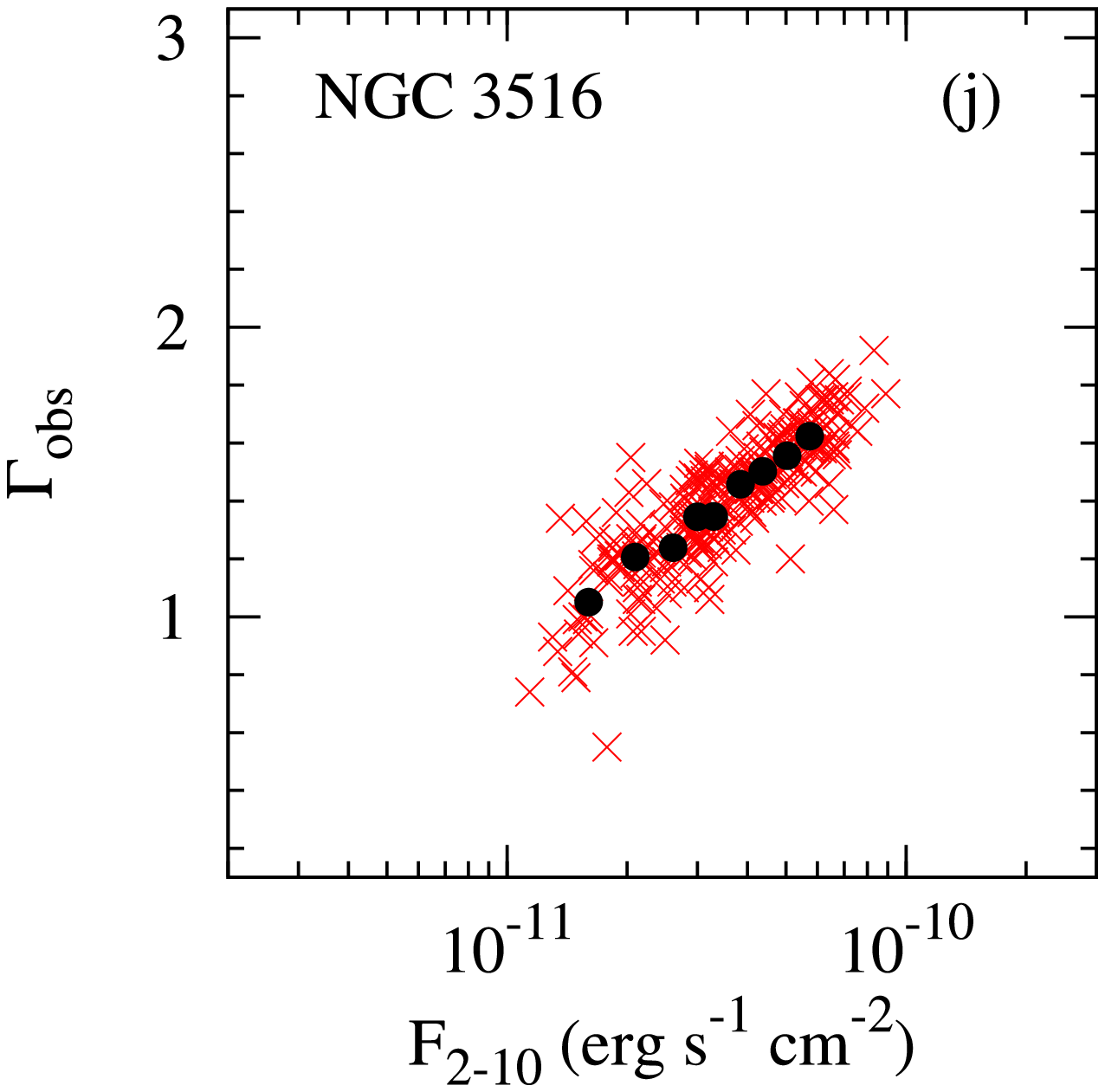}
\includegraphics[width=4.7cm,bb=176 133 536 427,clip,angle=-90]{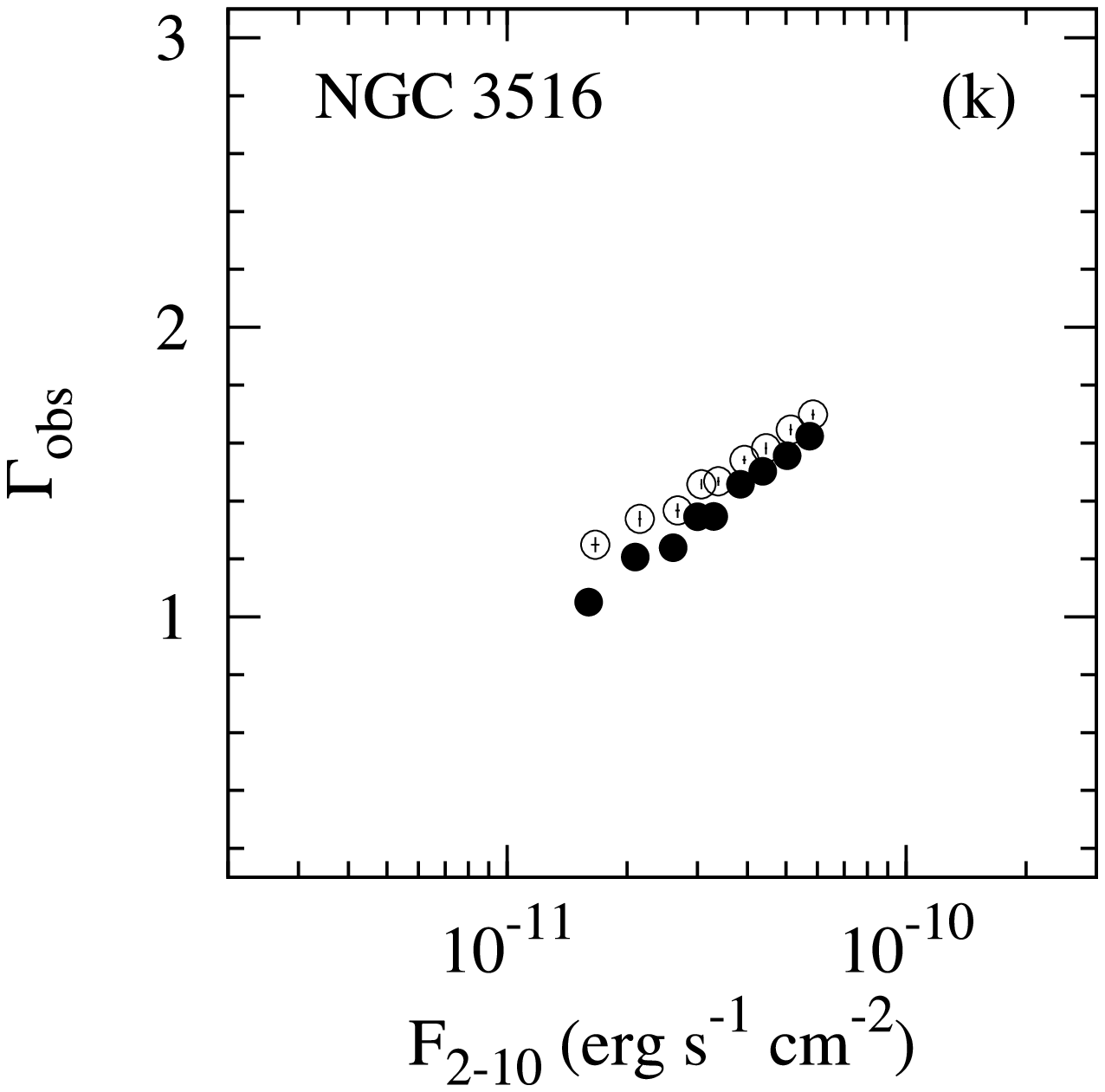}
\includegraphics[width=4.7cm,bb=176 109 538 483,clip,angle=-90]{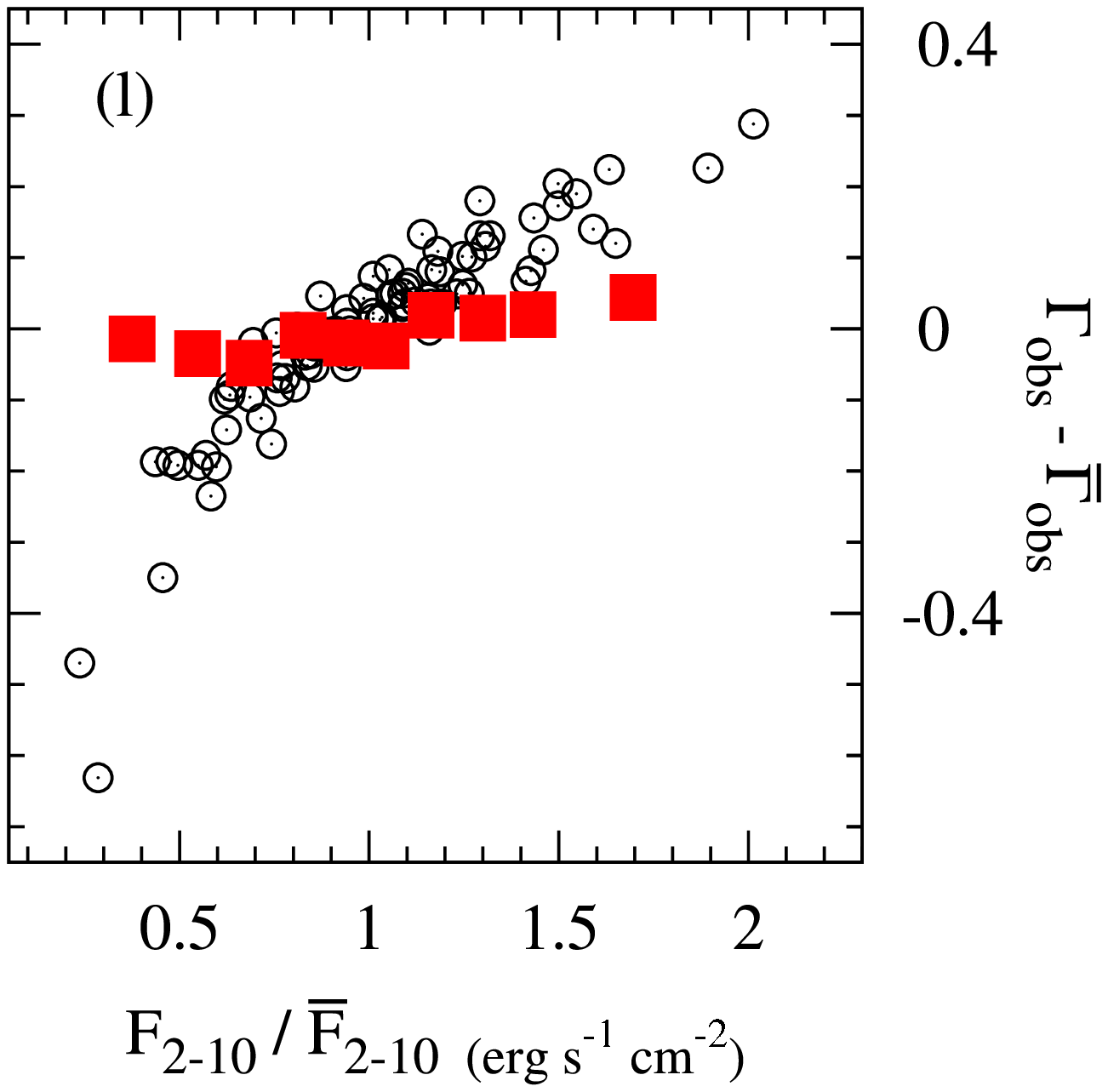}
\caption{Dependence of $\Gamma_{\rm obs}$ on $F_{\rm 2-10}$ for all AGN in the
sample. (a--j) The individual data (crosses) are shown together with the binned
data (filled circles, see Sec. 3.3 for details of binning). Errors on the binned data are plotted but are smaller
than the size of the symbols. (k) Binned data of NGC~3516 resulting from the simple power-law model (filled circles) and power-law model modyfied by a Gaussian line and absorption edge with variable line and edge energies(open circles; models M1 and M6 in Fig. 4, see Sec. 3.3 for details). (l) Binned $\Gamma-F_{\rm 2-10}$  data of all
AGN. Fluxes are  normalized to the respective $\overline{F}_{\rm 2-10}$, and we
have subtracted the respective $\overline{\Gamma}_{\rm obs}$ from the spectral
slope values. NGC~5548 (filled squares) follows a different ``slope vs. flux''
relation than the other AGN (open circles).}
\label{fig:fig7}
\end{figure*}


Figure~\ref{fig:fig4} shows $\Gamma_{\rm obs}$ and the X-ray luminosity, $L_{\rm
2-10}$  plotted as a function of time for NGC~4051, MCG-6-30-15, NGC~3516 and
Mrk~766\footnote{The luminosity is estimated as $L_{\rm 2-10} = 4\pi D^2 F_{\rm
2-10}$, where $D$ is the luminosity distance, listed in Tab.~\ref{tab:tab1}, and
$F_{\rm 2-10}$ is the observed 2--10 keV flux, from the appropriate best model
fitting results for each source. The flux measurements are corrected for
absorption in the case of NGC~5506.}. The first two objects are representative
of sources observed by {\it RXTE} frequently (on average every $2-3$ days),
between 1996 and 2006.  On the other hand,  NGC 3516 and Mkn 766  were
systematically observed by {\it RXTE} only between the years 1999--2002 and
2004--2007, respectively. Outside of these dates data for these sources  are
scarce. Their light curves are representative of sources which were observed
less than $\sim$700 times by {\it RXTE} from the start of the mission until the
end of 2006.

The (unweighted) mean photon indices, $\overline{\Gamma}_{\rm obs}$, and 2--10
keV luminosities, $\overline{L}_{\rm 2-10}$, for the sources in our sample are
listed in  Tab.~\ref{tab:tab3} (the standard error on $\overline{\Gamma}_{\rm
obs}$ and $\overline{L}_{\rm 2-10}$ are very small, mainly because of the large
number of data points). Application of the  standard $\chi^2$ test to the
$\Gamma_{\rm obs}$ and $L_{\rm 2-10}$  light curves suggests that all sources
show significant spectral and flux variations (reduced $\chi^2$ values are
listed in Tab.~\ref{tab:tab3}; the number of data points is large enough that
even in the case when $\chi^2/d.o.f.\sim 1.3$--1.4 the variations we detected
are significant at more than the 3$\sigma$ level). 

To further quantify the variability properties of the AGN in our sample, we
calculated the excess variance, $\sigma^2_{\rm xs}$, of the  $\Gamma_{\rm
obs}$ and $L_{\rm 2-10}$  light curves as follows,

\begin{equation}
\sigma^2_{\rm xs} = \frac{1}{n_s}\sum_{i=1}^{n_s}(x_i - \overline{x})^2-
\frac{1}{n_s}\sum_{i=1}^{n_s}\sigma^2_{{\rm err,}i}.
\end{equation}

The excess variance is a measure of the scatter about the mean in a light curve,
corrected for the contribution expected from the experimental errors. In the
formula above,  $x$ stands either for $\Gamma_{\rm obs}$ or $L_{\rm 2-10}$, and
$\sigma^2_{err,i}$ is the arithmetic mean of the upper and lower 90\% confidence
error on $\Gamma_{\rm obs}$ and $L_{\rm 2-10}$.  We then estimated the
fractional root mean square variability amplitude, $F_{\rm var} =
\sqrt{\sigma^2_{\rm xs}/\overline{x}^2}$ and its error using  equation (B2) in
\citet{vaughanea:2003} (this error accounts only for the uncertainty due to the
experimental error in each point). 

The results are listed in Tab.~\ref{tab:tab3}. In terms  of variability amplitude, the flux
variations are stronger (i.e. of larger amplitude) than the spectral
variations.  Furthermore, there seems to be a tendency for the objects with
stronger flux variations to show stronger spectral variations as well.  To
quantify this correlation,  we used the Kendall's $\tau$ test. We found that
$\tau=0.51$, which implies that the positive correlation between  $F_{\rm var,
\overline{\Gamma}}$ and  $F_{\rm var, \overline{L}(2-10)}$ is significant at
the 4\% level.

\subsection{Correlation between the average spectral slope and other source
parameters}

Figure~\ref{fig:fig5} shows the average observed photon index,
$\overline{\Gamma}_{\rm obs}$, plotted as a function of (a) BH mass, (b) average
2--10 luminosity, and (c) average X--ray mass accretion rate in Eddington units,
$\overline{\dot{m}}_{\rm X,E} = L_{\rm 2-10}/L_{\rm E}$, where $L_{\rm E} =
1.3\times 10^{38}$ M/M$_{\odot}$ ergs s$^{-1}$ is the Eddington luminosity of an
AGN with a black hole mass M. This quantity should be representative of the
total accretion rate (i.e. $L_{\rm bol}/L_{\rm E}$) of each source. For
simplicity we will  refere to it as the accretion rate hereafter. The errors on 
$\overline{\Gamma}_{\rm obs}$ plotted in Fig.~\ref{fig:fig5} are the standard
deviation of $\Gamma_{\rm obs}$ and we use them, instead of the standard error
of the mean, to account for the scatter of the individual $\Gamma_{\rm obs}$
values about $\overline{\Gamma}_{\rm obs}$. 

The solid lines plotted in Figs.~\ref{fig:fig5}a and \ref{fig:fig5}b indicate
the average spectral slope of all AGN in our sample. Individual points appear to
scatter randomly about this line. These plots suggest that the average spectral
shape does not correlate with either the average luminosity ($\tau=0.2$, $P_{\rm
null}=0.42$) or BH mass  ($\tau=-0.38$, $P_{\rm null}=0.13$).  On the other
hand, panel (c) in Fig.~\ref{fig:fig5} implies a positive correlation between
$\overline{\Gamma}_{\rm obs}$ and $\overline{\dot{m}}_{\rm X,E}$ ($\tau=0.51$,
$P_{\rm null}=0.04$).  Sources with a higher accretion rate appear to have
steeper average spectra as well. We found that a power-law model of the form
$y=\alpha x^{\beta}$ can fit well the data plotted in Fig.~\ref{fig:fig5}c
($\chi^2=8.14/8~d.o.f.$, $p_{\rm null}=0.42$). The best fitting parameters are:
$\alpha=2.7\pm0.3$, and $\beta=0.08\pm0.02$ (the error on the slope indicates
that the correlation between $\overline{\Gamma}_{\rm obs}$ and
$\overline{\dot{m}}_{\rm X,E}$ is significant at the $4\sigma$ level).

\subsection{Variations of the EW of the iron line}

An iron $K_{\alpha}$ line is detected in many spectra of all sources (except
for Akn~564, Fairall 9 and Mrk~766). To investigate the line's EW variations,
we considered the spectral fit results in the case when the line is  detected at
the 2$\sigma$ level in a spectrum of a source. The number of spectra which do
show such a strong line, and the average equivalent width, $\overline{\rm EW}$,
using the individual EW values from these spectra only, are listed in 
Tab.~\ref{tab:tab4}. The fourth column in Tab.~\ref{tab:tab4} lists the $\chi^2$
values when we fit a constant to the EW light curves. The result is that we
do not detect significant EW variations, except for NGC 5548, MCG~-6-30-15, and
NGC 3516. This is mainly due to the fact that, although the line is detected,
the small exposure time of the individual {\it RXTE} observations did not allow
the determination of the line's EW with a high accuracy. We will present a more
detailed study of the variability of the EW of the iron line in a forthcomming
paper.

Panels (a), (b) and (c) in Fig.~\ref{fig:fig6} show $\overline{\rm EW}$ plotted
as a function of $\overline{L}_{\rm 2-10}$, $\overline{\dot{m}}_{\rm X,E}$, and
$\overline{\Gamma}_{\rm obs}$.  Fig.~\ref{fig:fig6}b and Fig.~\ref{fig:fig6}c
imply that $\overline{\rm EW}$ does not correlate strongly with either accretion
rate ($\tau=-0.05$, $P_{\rm null}=0.88$) or $\overline{\Gamma}_{\rm obs}$
($\tau=-0.14$, $P_{\rm null}=0.65$). On the other hand, Fig.~\ref{fig:fig6}a
suggests that  the line's EW anti-correlates with the source's luminosity: the
strength of the line decreases with increasing X-ray luminosity ($\tau=-0.52$,
$P_{\rm null}=0.09$). This result is not statistically significant, most
probably due to the small number of objects in the sample. However,  this
``equivalent width -- luminosity" anti-correlation that we observe \citep[the
so-called Iwasawa-Taniguchi effect,][]{iwasawataniguchi:1993}  is well
established for nearby AGN \citep[see e.g.][and references
therein]{bianchiea:2007}. The solid line in Fig.~\ref{fig:fig6}a indicates the
best power-law fit to the data. The best-fit slope of $-0.23 \pm 0.06$ is
consistent with the slope of $-0.17\pm0.03$ found by \citet[][]{bianchiea:2007}.

\begin{table}
\caption{The mean EW of the iron line and its standard deviation. N$_{\rm 2\sigma}$ is the number of
spectra with a $2\sigma$ detection of an iron line.}
\centering
\begin{tabular}{l c c c c}
\hline
Target      & N$_{\rm 2\sigma}$ & $\overline{\rm EW}$ & $\chi^2$ & $F_{\rm var}$\\
            &           & (eV)            &          & (\%)\\
\hline
Fairall 9   & -   & -           & -   & -  \\
Ark 564     & -   & -           & -   & -  \\
Mrk 766     & -   & -           & -   & -  \\
NGC 4051    & 13  & 682$\pm$351 & 41  & -  \\
NGC 3227    & 45  & 460$\pm$182 & 141 & -  \\
NGC 5548    & 19  & 248$\pm$184 & 101 & 37$\pm$14 \\
NGC 3783    & 103 & 329$\pm$72  & 52  & -  \\
NGC 5506    & 177 & 242$\pm$63  & 88  & -  \\
MCG-6-30-15 & 47  & 498$\pm$335 & 260 & 38$\pm$8 \\
NGC 3516    & 39  & 502$\pm$258 & 129 & 25$\pm$7 \\
\hline
\end{tabular}
\label{tab:tab4} 
\end{table}

\subsection{The spectral variability of each source}

\begin{figure}
\centering
\includegraphics[height=6.0cm,bb=174 71 480 438,clip,angle=-90]{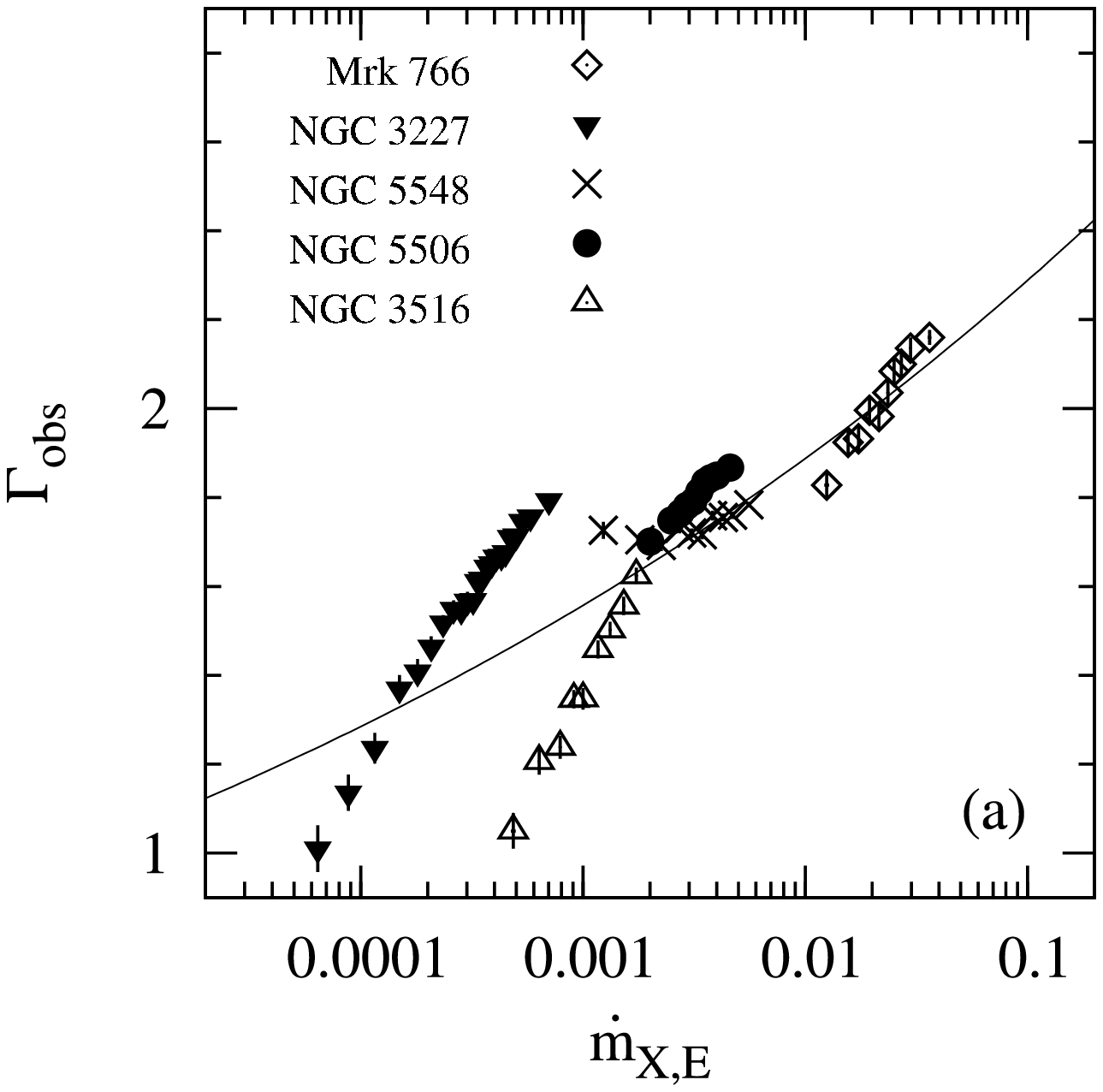}\\
\includegraphics[height=6.0cm,bb=174 71 545 438,clip,angle=-90]{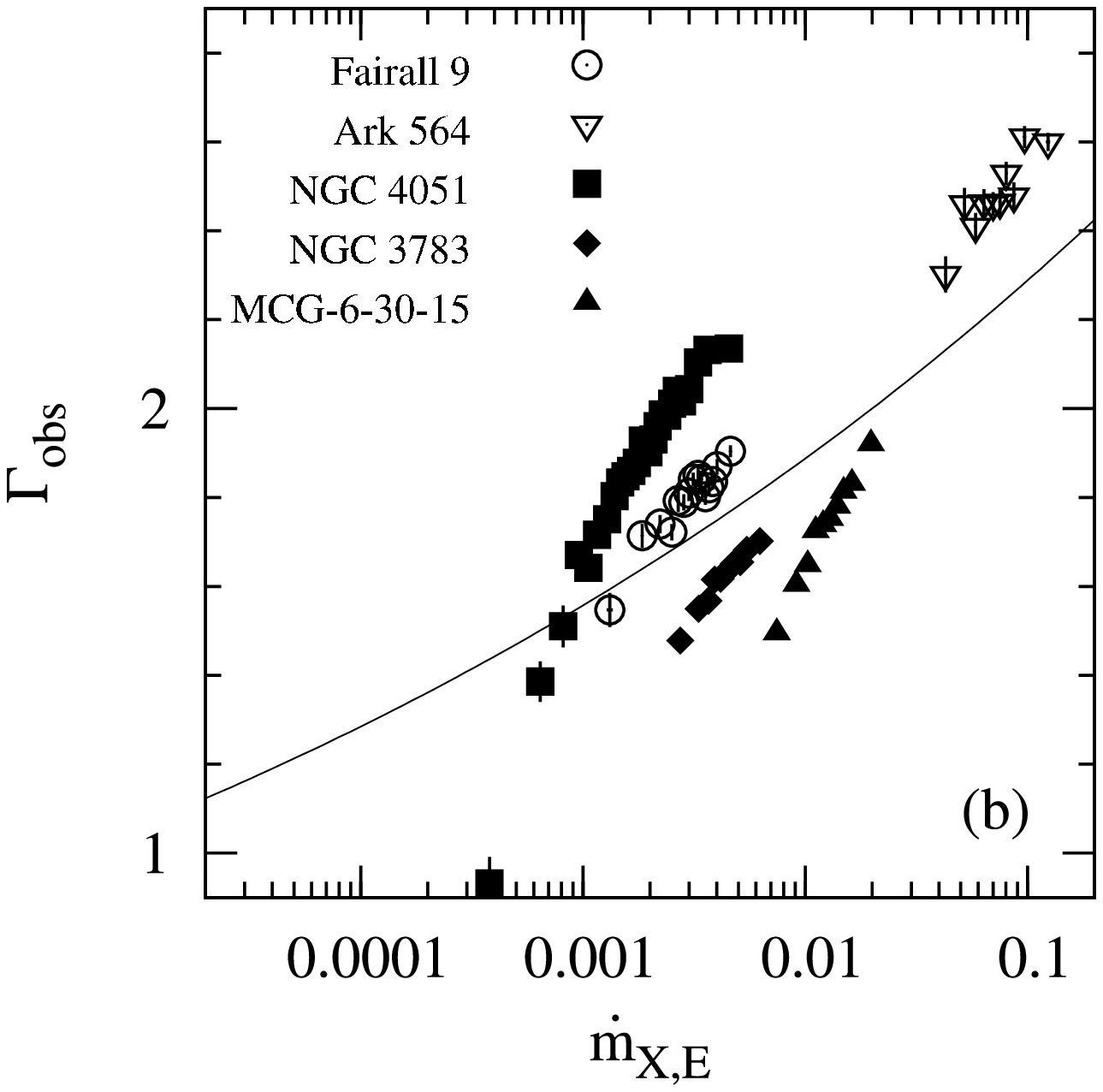}
\caption{The binned ``spectral slope - flux" data,  as in Fig.~\ref{fig:fig7},
plotted with the best-fit function to the data plotted in Fig.~\ref{fig:fig5}c.
The sources were split arbitrarly between panels (a) and (b) for clarity of
presentation.}
\label{fig:fig8}
\end{figure}

Crosses in panels (a--j) of Fig.~\ref{fig:fig7} indicate the individual
$(\Gamma_{\rm obs}, F_{\rm 2-10})$ points for each AGN in the sample. To better
illustrate the long-term, ``observed spectral slope -- flux" relation  in these
objects, we  grouped the $(\Gamma_{\rm obs}, F_{\rm 2-10})$ data  in flux bins
and calculated the (unweighted) mean flux and photon index in each bin. Each bin
contains at least 20 measurements, and the number of bins is roughly equal for
all sources. Filled circles in Fig.~\ref{fig:fig7} indicate the (average
spectral slope, average $F_{\rm 2-10}$) points. Both the binned and
un-binned data points in Fig.~\ref{fig:fig7} suggest that, in almost all
objects, the spectrum softens with increasing 2--10 keV flux.

In Fig.~\ref{fig:fig7}k we plot the binned ``spectral slope -- flux" data for
NGC~3516 using the best fit results of model M6 (filled circles) together with
those when we used the best fit results of model M1 (open circles; this is the
model that shows the largest discrepancy in $\overline{\Gamma}_{\rm obs}$ with
the statistically acceptable model M6 for this source). We fitted a power-law
model to the M1 and M6 data plotted in this panel,  and the best fit slopes
differ by 0.08$\pm 0.02$, with the M1 slope being flatter. Therefore, the {\it
exact} shape of the ``$\Gamma-F_{\rm 2-10}$" plots in Fig.~\ref{fig:fig7} does
depend on the choice of the model used to fit the individual spectra of the
sources. However, the overall trend stays the same in both cases: the spectrum
steepens with increasing flux.

In fact, the spectrum steepens with increasing flux in almost the {\it same} way
in all sources. The opened circles in Fig.~\ref{fig:fig7}l indicate the binned 
``$\Gamma_{\rm obs}-F_{\rm 2-10}$" data for all sources. We did not use
different symbols for each source in order to emphasize the fact that, in all
sources, the spectral shape varies with flux in a similar way. The only
exception is NGC~5548. Filled squares in Fig.~\ref{fig:fig7}l indicate the
``$\Gamma_{\rm obs}-F_{\rm 2-10}$" data for this source. The spectral slope
appears to stay roughly constant although the 2--10 keV flux varies by almost an
order of magnitude. A positive correlation between spectral slope and X-ray flux
may exist at the highest luminosity bins, but there is also a hint of an
anti-correlation at the lowest luminosity bins.

The positive correlation between  $\Gamma_{\rm obs}$ and $F_{\rm 2-10}$ can be
translated into a correlation between $\Gamma_{\rm obs}$ and $\dot{m}_{\rm
X,E}$, similarly to that between their {\it average} spectral slope and {\it
average} accretion rate (compare Fig.~\ref{fig:fig5}c), but not necessarily the
same. Figure~\ref{fig:fig8} shows the binned $(\Gamma_{\rm obs}, \dot{m}_{\rm
X,E})$ data for all sources plotted together with the best-fit power-law model
to the $(\overline{\Gamma}_{\rm obs}, \overline{\dot{m}}_{\rm X,E})$ data from
Fig.~\ref{fig:fig5}c. It is clear that, in at least some cases, such power-law
cannot fit well the individual ``observed spectral slope -- accretion rate"
relations.

To quantify the comparison between the average and the individual ``observed
spectral slope - accretion rate" relations,  we fitted the ($\Gamma_{\rm obs},
\dot{m}_{\rm X,E})$ data plotted in Fig.~\ref{fig:fig8} with the same power-law
model we used to fit the data shown in  Fig.~\ref{fig:fig5}c. The model fits
well the data of Akn 564 ($\beta_{\rm Akn564}=0.10\pm 0.01, \chi^2=15/8~d.o.f.$,
$p_{\rm null}=0.06$), NGC~5506  ($\beta_{\rm NGC5506}=0.12\pm 0.01,
\chi^2=8.1/8~d.o.f.$, $p_{\rm null}=0.43$),  Mrk~766 ($\beta_{\rm Mrk766}=0.16\pm
0.01, \chi^2=6/8~d.o.f.$, $p_{\rm null}=0.75$), and  NGC~3516 ($\beta_{\rm
NGC3516}=0.32\pm 0.02, \chi^2=5.8/7~d.o.f.$, $p_{\rm null}=0.56$).  Interestingly,
$\beta_{\rm Akn564}$ and $\beta_{\rm NGC5506}$ are similar to the best fit slope
in the case of the ($\overline{\Gamma}_{\rm obs},\overline{\dot{m}}_{\rm X,E})$
plot.

\section{Discussion}

The main aim of this work was to determine the shape of the X--ray continuum,
and study its variability, in 10 X--ray luminous AGN which have been observed
regularly with \rxte\ since 1996. We used a large number of spectra, which were
taken regularly over a period of many years. As we argued in the introduction, 
since this time period is much larger than the characteristic time scale in
these objects, it is possible that we have sampled the full range of their
spectral variations. We therefore believe that we have determined,  as
accurately as it is possible at present, the  ``observed spectral slope -- flux"
relation for the AGN in the sample. In Section 4.1 we discuss the implications
from the average ``observed spectral slope - accretion rate" correlation
reported in Section 3.1.

In Section 2.2.5 we argued that despite the limited spectral resolution of the
PCA on board \rxte, the $\Gamma_{\rm obs}$ values we derived from the  model
fitting of the individual spectra of each source are representative of their
actual X--ray spectral shape. However, this conclusion does not necessarily
imply that $\Gamma_{\rm obs}=\Gamma_{\rm intr}$ ($\Gamma_{\rm intr}$ being the
{\it intrinsic} slope of the X--ray continuum). Due to the short exposure of the
individual \rxte\ observations and the limited resolution of the PCA on board
\rxte, we cannot constrain the properties of any warm absorbing material that
may affect the spectra above 3 keV. Consequently, if there exists such a warm
absorbing material in the vicinity of our X--ray sources, we would expect
$\Gamma_{\rm obs}<\Gamma_{\rm intr}$. Similarly, the possible presence of a
significant reflection component (which we did not take into account in our
models) will also affect the observed X--ray spectral shape. As a result, we
would expect again the spectral shape of the continuum to {\it appear} harder
than it actually is, i.e. $\Gamma_{\rm obs}<\Gamma_{\rm intr}$. 

Although all these mechanisms may operate in AGN, one of our main results is
that the ``$\Gamma_{\rm obs}-$flux" relation is similar to all the objects in
the sample, despite the fact that their BH mass and accretion rate are quite
different. This common behaviour argues that their spectral variability is
mainly driven by the same mechanism in all of them. In the following sections we
demonstrate that, using various reasonable assumptions, the plots shown in
Figs.~\ref{fig:fig7}--\ref{fig:fig8} can be used to test various models for the
AGN spectral variability that have been proposed in the last few years. 

\subsection{The average X--ray spectral shape in AGN and Comptonization
models}

One of the main results of our study is that the AGN in the sample do not have
the same average spectral slope. Instead we found that $\overline{\Gamma}_{\rm
obs}$ correlates positively with $\overline{\dot{m}}_{\rm X,E}$, that is AGN
with higher accretion rate show softer X-ray spectra as well. This is in
agreement with  recent studies which also suggested a positive correlation
between  spectral slope and accretion rate in AGN
\citep[e.g.][]{porquetea:2004,bian:2005,shemmerea:2006,saezea:2008} and Galactic
black hole binaries \citep[GBHs]{wugu:2008}.

\citet[][]{papadakisea:2009} argued that the difference in the
$\overline{\Gamma}_{\rm obs}$ of the objects reflects a real difference in their
intrinsic spectral slopes and it is not due to either the presence of a
reflection component of different amplitude  in the spectra of the sources or
the effects of strong absorption in the 2--10 keV band. Furthermore, the 
reality of the average ``observed spectral slope -- accretion rate" correlation 
we observed can also be justified from its similarity with the
``$\Gamma-$accretion rate" relation that has been observed in luminous quasars
\citep[i.e. PG  quasar luminosity and higher][]{shemmerea:2006}, since the
energy spectrum in  such sources is not expected to be significantly affected by
either absorption or reflection effects. We therefore believe that this
correlation most probably reflects a ``true/intrinsic" correlation between the
intrinsic spectral slope and accretion rate in AGN.

It is widely believed that hard X--rays from AGN are produced by Comptonization, i.e.
by multiple up-scattering of seed soft photons by hot electrons in a corona
located close to the black hole. The Comptonization process generally produces
power-law X--ray spectra. The main factor which determines the resulting X--ray
spectrum is the so-called Compton amplification factor, $A=(L_{\rm diss}+L_{\rm
s})/L_{\rm s}$, where  $L_{\rm diss}$ is the power dissipated in the corona and
$L_{\rm s}$ is the intercepted soft luminosity. According to
\citet{beloborodov:1999} and \citet*{malzacea:2001}, $\Gamma \approx
a(A-1)^{-b}$, where $a=2.15$--2.30, and  $b=0.07$--0.10 in the case of AGN, in
which the energy of the input soft photons is of the order of a few eVs.

If hard X--rays from AGN are indeed generated by the Comptonization process, then a
possible explanation for the ``spectral slope - accretion rate" correlation
shown in Fig.~\ref{fig:fig5}c is that the ratio $L_{\rm diss}/L_{\rm s}$ in these objects
decreases proportionally with increasing accretion rate. Indeed, if $L_{\rm
diss}/L_{\rm s}\propto \dot{m}^{-1}$, then the $\Gamma \propto (A-1)^{-b}$
relation implies that $\Gamma \propto \dot{m}^b$, with $b=0.07$--0.1, as
observed.

The ratio $L_{\rm diss}/L_{\rm s}$ depends mostly on the geometry of the
accretion flow, and the geometry may depend on the accretion rate. One
possibility is that the cold disc is disrupted in the inner region
\citep[e.g.][]{esinea:1997}. As $\dot{m}$ increases, the inner radius of the
disc may approach the radius of the innermost stable circular orbit, and so more
soft photons are supplied to the hot inner flow causing an increase in $L_{\rm
s}$. As a consequence, $L_{\rm diss}/L_{\rm s}$ should decrease and the X-ray
spectrum should soften accordingly  \citep*[e.g.][]{zdziarskiea:1999}. However,
in this case, we will have to accept that the inner radius of the disc  in AGN
has not reached the innermost radius of the last circular stable orbit, even in
the case of objects such as Ark~564 which probably accretes at almost its 
Eddington limit. 

Another possibility is  that the coronal plasma is moving away from the disc and
emits beamed X--rays \citep[][]{malzacea:2001}. In this case, the Compton
amplification factor depends on the plasma velocity, $\beta$. As this velocity
increases, $A$ should also increase and the spectrum should harden. If this
picture is correct, then our results imply that the plasma velocity should
decrease with increasing accretion rate, which is somewhat opposite to what one
would expect. Furthermore, if the X-rays are beamed away from the disc, one
would expect the EW of the iron line to decrease with increasing plasma velocity
as well. Hence, one would expect the EW to increase with increasing accretion
rate. However, we do not observe a significant correlation between the line's EW
and accretion rate for the objects in the sample.

The ``observed spectral slope -- accretion rate" relation could also be
explained, quantitatively, if we assume that  either the amount of reflection
decreases or  the effects of absorption become weaker as the accretion rate
increases. In the first case, we should also expect to detect an
anti-correlation between $\overline{\rm EW}$ and accretion rate, which is not the
case. In the second case, if the warm absorber is located far away from the
central source and the weakening of the absorption is due to the decrease in its
opacity, by increasing its ionization parameter for example, we should expect
$\overline{\Gamma}_{\rm obs}$ to correlate with the source luminosity, which is
not the case. It is more difficult to  anticipate the ``observed spectral slope
-- accretion rate" relation in the case when the absorber originates from an
outflowing disc wind. The reason is that no physical models predict at the
moment how does the accretion rate control the mass loss rate and the ionization
state of the wind, and hence its opacity.

\subsection{The spectral variations within each source}

\subsubsection{Intrinsic spectral slope variability} 

One possible explanation for the ``observed spectral slope -- flux" relations
shown in Fig.~\ref{fig:fig7} is that the observed flux variations correspond to
$\dot{m}$ variations in each source. We argued above that it is possible for the
Compton amplification factor, $A$, to decrease proportionally with increasing
$\dot{m}$ for different AGNs. Similarly, this factor may vary with $\dot{m}$
also in individual objects, Iin which case we would expect $\Gamma\propto F_{\rm
2-10}^b$, where $b\sim 0.07$--0.1 (see discussion in the previous section).
However, we found that a power-law model fits the ``$\Gamma$--$F_{\rm 2-10}$"
relation of just 4 AGN in the sample. Furthermore, only in the case of Ark~564
and NGC~5506 the best-fitting slope is consistent with the expected value of
$\sim 0.07$--0.1. Therefore, the flux related spectral variations, at least in
the other objects, most probably have a different origin.

\subsubsection{The case of variable absorption.}

\begin{figure}
\includegraphics[bb=28 395 523 760,clip,width=8cm]{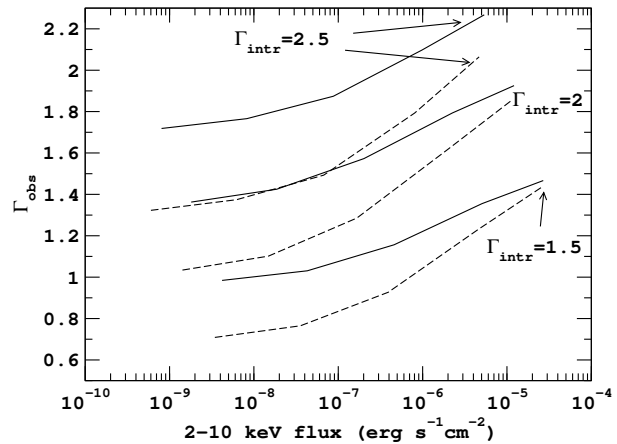}
\caption{The ``$\Gamma_{\rm obs}-F_{\rm 2-10}$" relation  for three $\Gamma_{\rm
intr}$ values in the case the X--ray continuum varies only in flux, and there
exists absorbing material whose ionization parameter varies proportionally with
the continuum flux ( solid and dashed lines correspond to  N$_{\rm H} = 5\times
10^{22}$ and N$_{\rm H} = 10^{23}$ cm$^{-2}$, respectively).}
\label{fig:fig9}
\end{figure}

Spectral variations can also be caused by variations in the column density,
covering fraction (say $f$), and/or the opacity of an absorber while
$\Gamma_{\rm intr}$ remains constant \citep[e.g.][]{turnerea:2007,millerea:2008}.
It is rather unlikely that variations of either the column density or the
covering factor are the main driver for the ``spectral slope -- flux" relation.
Under this hypothesis, the N$_{\rm H}$ or $f$ variations flatten the
observed spectral slope {\it and} and also reduce the continuum flux, as
observed. In other words, in this case the observed {\it flux} variations in AGN
should not be intrinsic, but they should be caused by the variations in the
column density or the covering factor of the absorber. However, many sources  in
the sample show significant flux variations on time scales as short as a few
hundred seconds. Therefore, the continuum flux variability in these objects 
must be intrinsic to a large extent, as neither N$_{\rm H}$ nor $f$ are expected
to vary on these time scales. 

On the other hand, it is possible that the variability of the ionization state
of the absorber, due to intrinsic variations of the continuum flux, can affect
the spectral shape of the continuum and result in ``spectral slope -- flux"
relations which are similar to the relations we observed: a decrease in the
continuum flux will decrease the ionization state of the absorber,  increasing
its opacity and hence flattening the observed spectrum. 

To investigate this possibility we considered the {\tt absori*powerlaw} model
in {\tt XSPEC}, where {\tt absori} is an ionised absorption model
\citep[][]{doneea:1992}. The main parameters of this model are the ionisation
parameter  $\xi$ and  the column density, $N_{\rm H}$, of the absorbing medium,
as well as $\Gamma_{\rm intr}$ and N$_{\rm PL}$, normalisation of the power-law component. We
created model spectra assuming $\Gamma_{\rm intr}=1.5$, 2 and 2.5, and  various 
values of N$_{\rm PL}$ (from 1 up to 5000). We then  considered three values of
column density, N$_{\rm H} = 10^{22}$, $5\times 10^{22}$ and 10$^{23}$ cm$^{-2}$
and, for each $\Gamma_{\rm intr}$, we assumed that the ionisation parameter of
the absorber is changing proportionally with N$_{\rm PL}$ (i.e.
$\xi\propto$N$_{\rm PL}$). 

We then used  {\tt XSPEC} to fit a simple PL model to the resulting spectra and
determined $\Gamma_{\rm obs}$ and $F_{\rm 2-10}$.  In Fig.~\ref{fig:fig9}
we plot  $\Gamma_{\rm obs}$ as a function of the observed flux  for the three
$\Gamma_{\rm intr}$ values we considered and in the case when N$_{\rm H} =
5\times 10^{22}$ and N$_{\rm H} = 10^{23}$ cm$^{-2}$ (solid and dashed lines,
respectively; for simplicity, we assumed, arbitrarily, that $\xi = $N$_{\rm
PL}$). As expected, the observed spectral slope flattens as the source
luminosity decreases. The $\Gamma_{\rm obs}$ change is more pronounced in the
case of the absorber with the larger absorbing column (the $\Gamma_{\rm obs}$
variations are of low amplitude in the case of the N$_{\rm H} = 10^{22}$
absorber, and for this reason we do not plot the respective models in Fig.~\ref{fig:fig9}).

We fitted the curves shown in Fig.~\ref{fig:fig9} to the data plotted in Fig.
\ref{fig:fig7}, by allowing them to shift both in the vertical (i.e. $y_{\rm
model}= \Gamma_{\rm obs}+y_{\rm shift}$), and the horizontal direction (i.e.
$x_{\rm model}=F_{\rm 2-10}\times x_{\rm shift}$). These model curves fitted
well the ``$\Gamma_{\rm obs}$ -- flux" data of NGC~5548 only. A N$_{\rm H} =
5\times 10^{22}$ cm$^{-2}$ absorber whose  ionisation parameter varies between
$\xi\sim 30-180$, can explain satisfactorily ($\chi^2=14.5/8~d.o.f.$) the
spectral variations we observed in this object. In all other cases, the observed
``spectral slope -- flux" relations are significantly steeper than any of the
model curves plotted in Fig.~\ref{fig:fig9}.

We conclude that opacity variations, due to changes in the continuum flux, of a
single absorber, whose column density remains constant, cannot explain the
observed spectral variations in most AGN in the sample. Of course, if there are
multiple layers of absorbing material, the resulting ``spectral slope -- flux"
model curves will be different. However, it is not easy to model such a
situation  without some apriori knowledge of how many absorbing layers there may
be, what their average ionization is, etc.

\subsubsection{The case of a constant reflection component} 

\begin{figure}
\includegraphics[bb=28 395 523 760,clip,width=8cm]{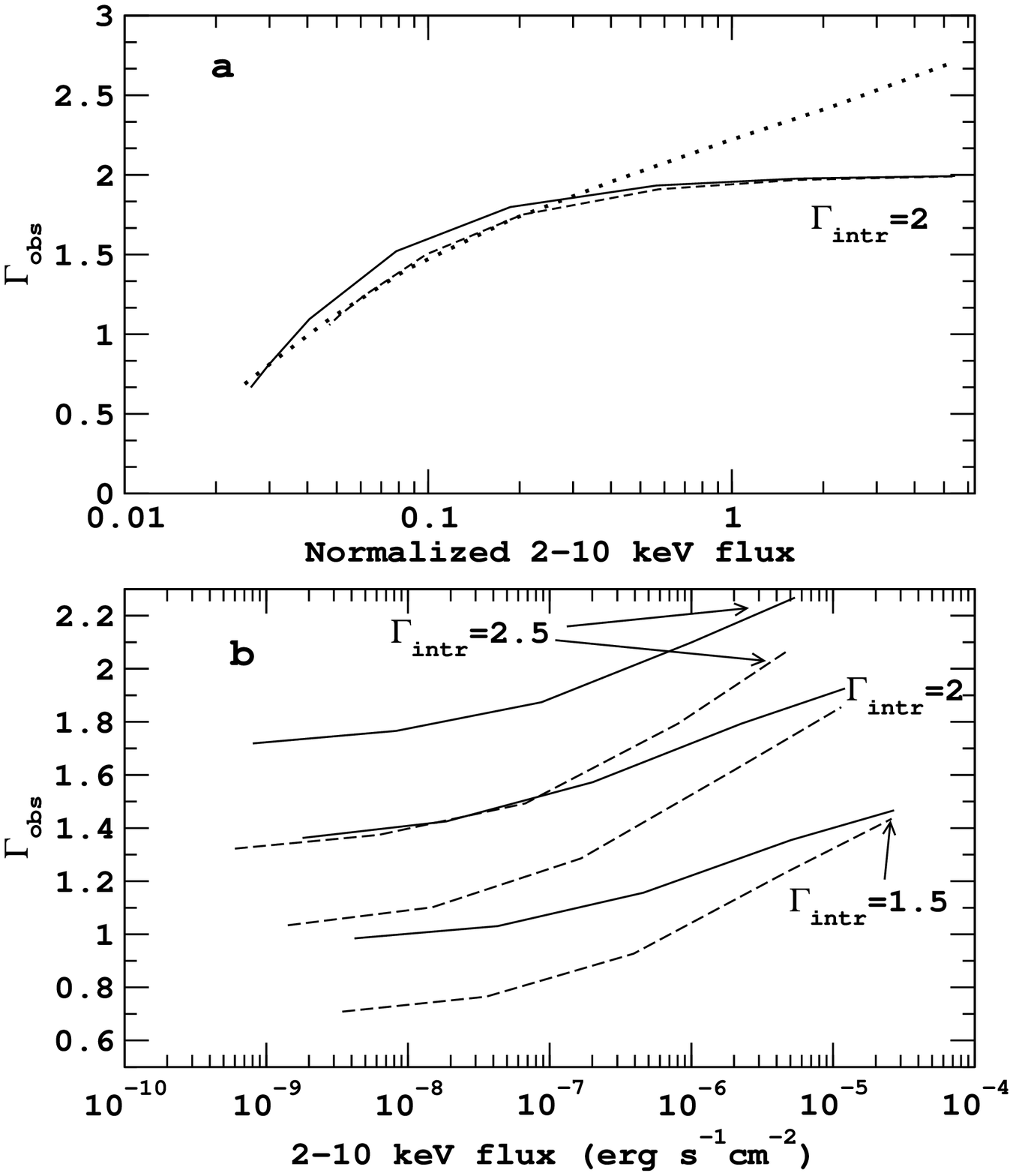}
\caption{$\Gamma_{\rm obs}$ plotted as a function of $F_{\rm 2-10}$
(normalized to the mean model flux) in the case of a power law with 
$\Gamma_{\rm intr}=2$ which varies only in normalization and a constant
reflection component in the case when the reflecting material is neutral (i.e.
$\log(\xi) = 0$; solid line) and ionized ($\log(\xi) = 3$; dashed line). The
dotted line indicates the ``spectral slope -- flux" relation in the case when
$\Gamma_{\rm intr}$ varies with flux  ($\Gamma_{\rm intr}\propto F_{\rm
2-10}^{0.08}$) and there also exists a constant reflection component (with
$R=1$).}
\label{fig:fig10}
\end{figure}

It is possible that the observed spectral variations are due to the combination
of a highly variable (in flux) power-law continuum  (with $\Gamma_{\rm
intr}$=constant) and a constant reflection component
\citep*[e.g.][]{taylorea:2003,pontiea:2006,miniuttiea:2007}. In this case, as
the amplitude of the continuum increases, the relative contribution of the
reflection component to the observed spectrum will decrease, and $\Gamma_{\rm
obs}$ will increase, reaching $\Gamma_{\rm intr}$ when the continuum amplitude
is large enough.

To investigate this possibility, we considered a ``power-law plus constant
reflection" model. To calculate the reflected component we used the {\tt pexriv}
code by \citet[][]{magdziarzzdziarski:1995}, available in {\tt XSPEC}. The main
model parameters are: (i) the X-ray photon index of the intrinsic power-law,
$\Gamma_{\rm intr}$, (ii) the ionisation parameter of the reflecting medium,
$\xi$, and (iii) the amplitude of reflection, $R$. We computed model spectra
assuming $\Gamma_{\rm intr}=2$, various  values of the power law normalization,
N$_{\rm PL}$ (from $3\times10^{-5}$ up to 0.1), while the normalization of the
reflected component, N$_{\rm refl}$, was kept fixed (the {\tt pexriv} model was
defined, arbitrarily, so that $|R|=0.1$ and N$_{\rm refl} = 0.1$, where $N_{\rm
refl}$ means normalization of the power-law continuum that is reflected). Since
the power-law was defined explicitly in our model, we used {\tt pexriv} with
$R<0$, which outputs only the reflected component. 

We then used {\tt XSPEC} to fit a power-law to the model spectra in the 2--10
keV band and determined model $\Gamma_{\rm obs}$. In  Fig.~\ref{fig:fig10} we
plot $\Gamma_{\rm obs}$ as a function of the total 2--10 keV flux (normalized to
the mean model flux) in the case when the reflecting material is neutral (i.e.
$\log(\xi) = 0$; solid line) and ionized ($\log(\xi) = 3$; dashed line), and the
slope of the power law is fixed at $\Gamma_{\rm intr} = 2$. As expected, as the
flux decreases, the relative strength of the reflecting component increases, and
$\Gamma_{\rm obs}$ becomes harder.

The shape of the model curves in the case of different $\Gamma_{\rm intr}$
values is identical to the ones shown in Fig.~\ref{fig:fig10} except for the
fact that, in the high flux limit, they saturate at their respective
$\Gamma_{\rm intr}$ value. Similarly, the shape of the model curves remains the
same if we increase the total flux while keeping the relative strength of the
power-law and the reflection component (and hence $\Gamma_{\rm obs}$) the same.
As a result the curves will only shift to a higher flux.

We fitted the curves shown in Fig.~\ref{fig:fig10} to the data plotted in
Fig.~\ref{fig:fig7} by allowing them to shift both in the vertical (i.e. $y_{\rm
model}= \Gamma_{\rm obs}+y_{\rm shift}$), and the horizontal direction (i.e.
$x_{\rm model}=F_{\rm 2-10,mod}\times x_{\rm shift}$; $F_{\rm 2-10,mod}$ is the
model 2--10 keV flux). The shift along the $y-$axis is necessary because
$\Gamma_{\rm intr}$ may not be equal to 2 in all objects. The best fit vertical
offset value, $y_{\rm bf}$, will determine the best fit intrinsic slope for each
object, $\Gamma_{\rm intr}=2+y_{\rm bf}$. On the other hand, the horizontal
offset shift is necessary, given the different flux levels of the objects, and
the fact that the strength of the  reflection component in each object, with
respect to that of the power-law continuum (i.e. the average $R$ in each object,
$\overline{R}$), is not known apriori. In fact, the  best fit horizontal offset
value, $x_{\rm bf}$, can be used to determine $\overline{R}$ as follows.

The reflection amplitude, $R$, determines the fraction of the X--ray photons
that are reflected by the disc, so that $R \propto F_{\rm refl}/F_{\rm PL}$
(where $F_{\rm refl}$ and $F_{\rm PL}$ are the fluxes of the reflected component
and continuum, respectively). In our case, the average source flux,
$\overline{F}_{\rm 2-10,obs}$, corresponds to a model PL flux of $F_{\rm
PL,mod}= (\overline{F}_{\rm 2-10,obs}/x_{\rm bf})-F_{\rm refl,mod}$. The model
curve was constructed by assuming, arbitrarily,  the constant reflection
component when $R=0.1$ and N$_{\rm PL}=0.1$, therefore  $\overline{R}=0.1\times
F_{\rm PL,mod}(0.1)/F_{\rm PL,mod}$, where $F_{\rm PL,mod}(0.1)$ is the 2--10
keV flux of the power law with N$_{\rm PL}=0.1$.

\begin{figure}
\includegraphics[bb=28 180 523 760,clip,width=8cm]{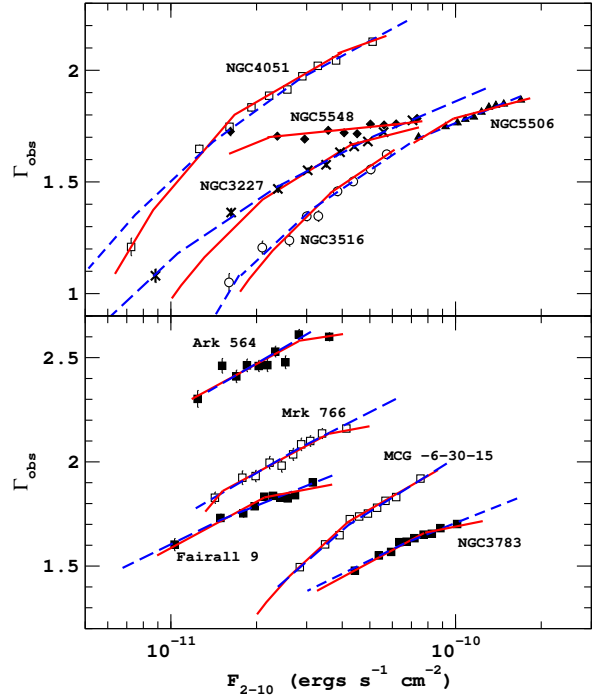}
\caption{The $\Gamma_{\rm obs}-F_{\rm 2-10}$ relations for the objects in the
sample together with the best fit curves for the models described in sections
4.2.3 (solid line) and 4.2.4 (dashed line).}
\label{fig:fig11}
\end{figure}

The model curves plotted in Fig.~\ref{fig:fig10} fitted well the data of all
objects, except for NGC~3227 ($\chi^2=55.1/8~d.o.f.$), NGC~5506
($\chi2=22.2/8~d.o.f.$), and NGC~5548 ($\chi2=40.4/8~d.o.f.$). In the case of
MCG~-6-30-15 and NGC~3516 it was reflection from ionized material that fitted
the data best. We plot the ``$\Gamma_{\rm obs}-F_{\rm 2-10}$" data for the
objects in the sample in Fig.~\ref{fig:fig11}. Best fit lines are indicated by
the solid lines.

The best model fits do not provide a statistically acceptable fit to  the data
of NGC~3227,  NGC~5506, and NGC~5548. In the case of  NGC~5548, the model
describes rather well the ``$\Gamma_{\rm obs}-$flux" data, except for the lowest
flux point (the best fit model line plotted in Fig.~\ref{fig:fig11} for NGC~5548
corresponds to reflection from ionized material). The disagreement between best
model fit and data points is more pronounced in the case of NGC~3227, especially
at low fluxes. This is a source where a possible  absorption event, which lasted
for a few months, has been observed in its \rxte\ light curve
\citep*[][]{lamerea:2003}. In this case, one would expect $\Gamma_{\rm obs}$ to
be harder than the model predictions at a given (low) flux, which is opposite to
what is observed. However, such an event should also shift the model line to
lower fluxes. Therefore it is difficult to reach final conclusions regarding the
quality of the model fit to the NGC~3227 data.

The best fit intrinsic spectral slopes  ranged from $\Gamma_{\rm intr}\sim 1.8$
in NGC~5548, to $\Gamma_{\rm intr}\sim 2.8$ in Ark 564. Furthermore, the best
fit $x_{\rm bf}$ values imply large average $R$ values: from $\overline{R}\sim
0.5$ in the case of NGC~5548 up to $\overline{R}\sim 15$  in the case of
NGC~3516. The median $\overline{R}$ is 4.7. We did not detect any correlation
between the resulting best-fit $\Gamma_{\rm intr}$ and $\overline{R}$ values. 

The reflection amplitude $R$ is essentially equal to the solid angle covered by
the reflecting material and is equal to 1 in the case of a point source on top
of a disc at large heights (i.e. in the case when $\Omega=2\pi$). Values as
large as $R\sim 5$ suggest either  a peculiar geometry (which is difficult to
envisage) or non-isotropic emission from the X--ray continuum in which case the
X--ray flux on the reflecting material is much larger than the flux which
reaches the observer \citep[see e.g.][for the case of an X--ray source which is
located very close to a rapidly rotating black hole, in which case relativistic
light bending effects are important]{miniuttifabian:2004}.

\subsubsection{The case of $\Gamma_{\rm intr}$ variations plus  a constant
reflection component}.

We considered next the possibility that the X--ray continuum in AGN consists of
a constant reflection component and a  power-law continuum with variable N$_{\rm
PL}$ {\it and} $\Gamma_{\rm intr}$. This is a combination of the two cases we
considered above: X--rays are produced by thermal Comptonization, the flux
variations (i.e. the N$_{\rm PL}$ variations) reflect intrinsic variations in
the accretion rate, and $\Gamma_{\rm intr}=2.7\dot{m}^{0.08}$ (this is the best
power-law model fit to the $\Gamma_{\rm intr}-\dot{m}$ relation plotted in
Fig.~\ref{fig:fig5}c). The X--ray source illuminates a reflecting material,
which effectively responds to the average X--ray continuum (i.e. to a
power-law of $\sim \overline{\Gamma}_{\rm intr}$ and $\overline{{\rm N}}_{\rm
PL}$). This could be for example the case of a reflector which is located away
from the X--ray source.

To investigate this possibility, we considered again the model {\tt
pexriv+powerlaw} in {\tt XSPEC}. This time, we  created model spectra
assuming 8 values of the power law normalization (from $10^{-4}$ up to 0.1) and a
spectral slope of $\Gamma_{\rm intr}=2.7$N$_{\rm PL}^{0.08}$ (the arbitrary
choice $\dot{m}=$N$_{\rm PL}$ was made for simplicity; it does not affect our
results, as long as the resulting model curves are allowed to shift freely along
the $x-$axis when we fit them to the data). Then, for each pair of values
(N$_{\rm PL},\Gamma_{\rm intr})$ we considered the respective reflection
component (from neutral material) in the case when $R=1$, and we added it to all 8 power-law spectra.
In this way, we constructed 8 sets of model curves (8 model spectra in each set).

We then used  {\tt XSPEC} to fit a simple PL model to the 8 spectra of each
model set and determined $\Gamma_{\rm obs}$ and the spectrum flux in the 2--10
keV band ($F_{\rm 2-10}$). The dotted line in Fig.~\ref{fig:fig10} indicates
$\Gamma_{\rm obs}$ in the case when we added to all 8 spectra the reflection
component (with $R=1$) from a $\overline{\Gamma}_{\rm intr}=2$ power-law continuum, plotted as a
function of $F_{\rm 2-10}$ (normalized to the mean model flux).

We fitted the 8 ``$\Gamma_{\rm obs}$ -- $F_{\rm 2-10}$" model curves to the observed ``spectral slope -- luminosity"
relations plotted in Fig.~\ref{fig:fig7} by allowing them to shift both in the vertical (i.e.
$y_{\rm model}= \Gamma_{\rm obs}+y_{\rm shift}$), and the horizontal direction
(i.e. $x_{\rm model}=x_{\rm shift}\times F_{\rm 2-10,mod}$). These model curves
fitted well the ``$\Gamma-\dot{m}$" data of all objects, except for NGC~4051
and NGC~5548.

The dashed lines in Fig.~\ref{fig:fig11} indicate the best fit curves for this
model. Interestingly, the model fits well the NGC~3227 data, a result which
implies that the flat $\Gamma_{\rm obs}$ at low fluxes may correspond to
intrinsically flatter spectra, and not an increase in the external absorption by
cold gas. In the case of NGC~4051, the fit is not statistically acceptable
($\chi^2=20.9/8~d.o.f.$, $p_{\rm null}=7.4\times 10^{-3}$) mainly because of the
lowest flux point. Furthermore, the model fails entirely to describe the
NGC~5548 data. This   result is obvious given the fact that the model curves in
this case have no flat part that could fit, even approximately, the observed
NGC~5548 data.

In principle, we should not allow a $y-$axis shift of the model curves in this
case. The intrinsic spectral slope is uniquely determined by the $\Gamma_{\rm
intr}-$N$_{\rm PL}$ relation we assumed. However, the assumed relation was
determined by the PL model fit to the ``average spectral slope - average
accretion rate" data plotted in Fig.~\ref{fig:fig5}c, hence there is an uncertainty
associated with its normalization and slope. The average best fit $y-$axis shift
for all objects is 0.13$\pm0.08$, i.e. it is consistent with zero. This result
implies that the relation $\Gamma_{\rm intr}\sim 2.7\dot{m}^{0.08}$ is in
agreement with both the average ``spectral slope - accretion rate" relation of
all objects and with the individual ``slope - luminosity" relations of each AGN.

The Ark~564 data are fitted best in the case when the constant reflection
component is the one which corresponds to $\overline{\Gamma}_{\rm intr}=2.45$.
The Mrk~766 and NGC~4051 data are fitted best in the case of constant reflection
component for $\overline{\Gamma}_{\rm intr}=2.25$ and 1.9, respectively. In all
other cases, the addition of a constant component for $\overline{\Gamma}_{\rm intr}=1.7$ is
needed to fit the data best.

\section{Conclusions}

We analyzed more than 7,500 \rxte\ PCA data of 10 nearby AGN with the goal of
studying their spectral variability properties. The objects were observed
regularly with \rxte\ over many years since 1996. Given the large number of
observations, spread over many years, we were able  to determine accurately
their average observed spectral slope, luminosity and accretion rate.  Using the
best model-fit results from the hundreds of spectra of each source, we also
determined their long-term ``observed spectral slope -- X--ray flux'' relation. 
This kind of long term variability analysis can not be accomplished with the 
rather short AGN observations currently performed by {\it Chandra}, {\it
XMM-Newton} or {\it Suzaku}, despite their higher effective area and superior 
spectral resolution. Our main results are as follows:

\begin{list}{\labelitemi}{\leftmargin=0em\itemsep=0.5em}

\item[1.] The average spectral slope is {\it not} the same in all objects. It
does not correlate either with BH mass or luminosity. However,
$\overline{\Gamma}$ correlates positively with the mass accretion rate: objects
with higher accretion rate have steeper spectra, and $\Gamma \simeq 2.7\dot{m}^{0.08}$. This result can be
explained if the Compton amplification factor, $A$, decreases proportionally
with the accretion rate in AGN.

\item[2.] We detected the iron K$\alpha$ line in many spectra of seven sources
in the sample. The average EW of the line anti-correlates with the average 
2--10 keV luminosity (i.e. the so-called Iwasawa-Taniguchi effect). Our
results are in agreement with recent results which are based on recent studies
of large samples of AGN \citep[][]{bianchiea:2007}.

\item[3.] The spectra of each source  become steeper with increasing flux. This
is a well known result. What we have shown in this work that this trend holds
over many years, and over almost the full range of flux variations that AGN
exhibit. We also found that the $\Gamma_{\rm obs}-F_{\rm 2-10}$ relations are
{\it similar} in all objects, except for NGC~5548. There are many reasons which can cause apparent
and/or intrinsic  variations in the X--ray spectrum of AGN. It is possible that
they all operate, to a different extent in various objects. However, the common
spectral variability pattern we detected implies that just a single mechanism is
responsible for the bulk of the observed spectral variations.

\item[4.] NGC~5548 displays limited spectral variations for its flux
variability. Although uncommon, this behaviour has also been observed in other
Type I Seyfert, PG 0804+761 \citep*[][]{papadakisea:2003}. This behaviour,
different to what is observed in most (Type I) AGN, raises the issue of different
spectral states in AGN, just like in GBHs. Investigation of this issue is
beyond the scope of the present work.

\item[5.] The scenario in which the spectral variability is caused by absorption
of X-rays by a {\it single} medium whose ionization parameter varies
proportionally to the continuum flux variations fails to account for the
observed spectral variations.

\item[6.] A ``power-law, with constant $\Gamma_{\rm intr}$ and variable flux,
plus a constant reflection component" can fit the observed ``$\Gamma -F_{\rm
2-10}$'' relations of most (but not all) objects in the sample. The flux of the
reflection component necessary to explain the observed spectral variability is
rather large, implying average reflection amplitudes of the order of $R\sim 5$.
If true, this result implies that the primary source of X-rays is located close
to a maximally rotating Kerr black hole, where relativistic effects (such as 
light bending) are expected to be strong. 

\item[7.] The observed ``$\Gamma -F_{\rm 2-10}$'' relations (except for
NGC~5548) can be explained if we assume that the power-law continuum varies both
in flux and shape as $\Gamma_{\rm intr}\propto F_{\rm 2-10}^{0.08}$ (as implied
by the {\it average} ``spectral slope - accretion rate" relation that we observed)
{\it and} the reflecting material responds to the average continuum spectrum,
with $R=1$.

\end{list}

\section*{Acknowledgements}
We acknowledge support by the EU grant MTKD-CT-2006-039965. MS also acknowledges
support by the the Polish grant N20301132/1518 from Ministry of Science and
Higher Education.

\bsp

\label{lastpage}

\end{document}